\documentclass[%
preprint,
 amsmath,amssymb,
 aps,
longbibliography
]{revtex4-1}

\usepackage{amsmath,amssymb}
\usepackage{graphicx,color}
\usepackage{bm}% bold math
\usepackage{ulem}

\begin{document}

% Use the \preprint command to place your local institutional report
% number in the upper righthand corner of the title page in preprint mode.
% Multiple \preprint commands are allowed.
% Use the 'preprintnumbers' class option to override journal defaults
% to display numbers if necessary
%\preprint{}

%Title of paper
%\title{Fluorescence via Reverse Intersystem Crossing\\
%from Higher Triplet States:\\
%Suppressed Radiative and Non-Radiative Processes\\
%between Higher and Lower Triplet States\\
%in a Bisanthracene Derivative}
\title{Fluorescence via Reverse Intersystem Crossing\\
from Higher Triplet States in a Bisanthracene Derivative}

% repeat the \author .. \affiliation  etc. as needed
% \email, \thanks, \homepage, \altaffiliation all apply to the current
% author. Explanatory text should go in the []'s, actual e-mail
% address or url should go in the {}'s for \email and \homepage.
% Please use the appropriate macro foreach each type of information

% \affiliation command applies to all authors since the last
% \affiliation command. The \affiliation command should follow the
% other information
% \affiliation can be followed by \email, \homepage, \thanks as well.
%\author{}
%\email[]{Your e-mail address}
%\homepage[]{Your web page}
%\thanks{}
%\altaffiliation{}
%\affiliation{}

\author{Tohru Sato}
\email[]{tsato@moleng.kyoto-u.ac.jp}
\affiliation{Department of Molecular Engineering,
Graduate School of Engineering, Kyoto University,
Nishikyo-ku, Kyoto 615-8510, Japan}
\affiliation{Unit of Elements Strategy Initiative
for Catalysts {\rm \&} Batteries, Kyoto University,
Nishikyo-ku, Kyoto 615-8510, Japan}

\author{Rika Hayashi}
\affiliation{Undergraduate School of Industrial Chemistry,
Faculty of Engineering, Kyoto University, Nishikyo-ku,
Kyoto 615-8510, Japan}

\author{Naoki Haruta}
\affiliation{Department of Molecular Engineering,
Graduate School of Engineering, Kyoto University,
Nishikyo-ku, Kyoto 615-8510, Japan}

\author{Yong-Jin Pu}
\affiliation{Department of Organic Device Engineering and
Research Center for Organic Electronics, Yamagata University,
4-3-16, Johnan, Yonezawa, 992-8510, Japan}
\affiliation{PRESTO (Sakigake), JST}

%Collaboration name if desired (requires use of superscriptaddress
%option in \documentclass). \noaffiliation is required (may also be
%used with the \author command).
%\collaboration can be followed by \email, \homepage, \thanks as well.
%\collaboration{}
%\noaffiliation

\date{\today}

\begin{abstract}
% insert abstract here
To elucidate the high external quantum efficiency
observed for organic light-emitting diodes
using a bisanthracene derivative, BD1, as the emitting molecule,
off-diagonal vibronic coupling constants (VCCs)
between the excited states of BD1, which govern non-radiative transition rates,
were calculated employing time-dependent density functional theory.
The VCCs were analysed based on the concept of vibronic coupling density.
The VCC calculations suggest
{\it a fluorescence via higher triplets (FvHT) mechanism}, which entails the conversion of
a T$_4$ exciton generated during electrical excitation
into an S$_2$ exciton via reverse intersystem crossing (RISC); moreover,
the S$_2$ exciton relaxes to a fluorescent S$_1$ exciton
because of large vibronic coupling between S$_2$ and S$_1$.
This mechanism is valid as long as the relaxation of triplet states higher than T$_1$
to lower states is suppressed.
The symmetry-controlled thermally activated delayed fluorescence
(SC-TADF) and inverted singlet and triplet (iST) structure,
which have been proposed in our previous studies,
are the special examples of the FvHT mechanism that need high molecular symmetry.
However, BD1 achieves the FvHT mechanism in spite of its asymmetrical structure.
A general condition for the suppression
of radiative and non-radiative transitions
in molecules with pseudo-degenerate electronic structures such as BD1
is discussed.
A superordinate concept, {\it fluorescence via RISC},
which includes TADF, SC-TADF, iST structure, and FvHT is also proposed.
\end{abstract}

% insert suggested PACS numbers in braces on next line
% insert suggested keywords - APS authors don't need to do this
%\keywords{}

%\maketitle must follow title, authors, abstract, \pacs, and \keywords
\maketitle

% body of paper here - Use proper section commands
% References should be done using the \cite, \ref, and \label commands
%\section{}
% Put \label in argument of \section for cross-referencing
%\section{\label{}}
%\subsection{}
%\subsubsection{}

%%%%%%%%%%%%%%%%%%%%%%%%%%%%%%%%%%%%%%%%%%%%%%%%%%%%%%%%%%%%%%%%%%%%%
\section{Introduction\label{Sec:Introduction}}
%%%%%%%%%%%%%%%%%%%%%%%%%%%%%%%%%%%%%%%%%%%%%%%%%%%%%%%%%%%%%%%%%%%%%
Thermally activated delayed fluorescence (TADF) has attracted 
significant attention as the emission mechanism in molecules used
in organic light-emitting diodes (OLEDs)\cite{Adachi2014_060101}.
Although the phenomenon of TADF has been known for a long time,
its application in OLEDs was first reported by Endo {\it et al.}
in 2009\cite{Endo2009_4802}.
TADF OLEDs utilize fluorescence via reverse intersystem crossing (RISC)
from the triplet state, T$_1$,
generated during electrical excitation, as well as
fluorescence from the singlet excited state, S$_1$,
generated during electrical excitation.

In order to make RISC possible in a molecule,
the energy difference between S$_1$ and T$_1$, $\Delta E_{\rm ST}$, 
must be small enough that the RISC energy barrier can be overcome through thermal excitation.
$\Delta E_{\rm ST}$ can be written as
\begin{equation}
\Delta E_{\rm ST} 
= 2J 
= 
2\int\int 
  \psi_{\rm HO}^\ast(\boldsymbol{r}_{1})\psi_{\rm LU}(\boldsymbol{r}_{1})
  \frac{1}{r_{12}}
  \psi_{\rm HO}(\boldsymbol{r}_{2})\psi_{\rm LU}^\ast(\boldsymbol{r}_{2})
d^3 \boldsymbol{r}_{1}
d^3 \boldsymbol{r}_{2},
\end{equation}
where $J$ is an exchange integral, and
$\psi_{\rm HO}(\boldsymbol{r})$ and $\psi_{\rm LU}(\boldsymbol{r})$ denote
the HOMO and LUMO, respectively. Based on this equation, Endo {\it et al.} proposed a design principle 
to reduce $\Delta E_{\rm ST}$\cite{Endo2009_4802}, i.e. candidates for TADF molecules must be
donor-acceptor systems with small overlap between the HOMO and LUMO.

Based on this design principle for emitting molecules,
a number of TADF OLEDs have exhibited 
very high external quantum efficiencies (EQEs).
For example, a phenoxazine derivative, PXZ-TRZ,
exhibits an EQE of 12.5 \% 
(photoluminescence quantum efficiency (PLQE): 65.7 \%)\cite{Tanaka2012_11392},
a carbazolyl dicyanobenzene derivative, 4CzIPN,
exhibits an EQE of 19.3 \% (PLQE: 94 \%)\cite{Uoyama2012_234},
a triazine derivative, CC2TA,
exhibits an EQE of 11 \% (PLQE: 62 \%)\cite{Lee2012_093306},
a spiro bifluorene derivative, Spiro-CN,
exhibits an EQE of 4.4 \% (PLQE: 27 \%)\cite{Nakagawa2012_9580}, and
an acridine derivative, ACRFLCN,
exhibits an EQE of 10.1 \% (PLQE: 67 \%)\cite{Mehes2012_11311}.
Recently, Kaji {\it et al.} reported a triazine derivative, DACT-II,
exhibiting an extremely high EQE of 41.5 \%
(PLQE:100 \%)\cite{Kaji2015_8476}.
These results demonstrate the success of this design principle.

However, this design principle also has certain drawbacks: 
(1) the small overlap between the HOMO and LUMO leads to suppression
of the oscillator strength\cite{Sato2013_012010,Uejima2014_14244},
and (2) TADF OLEDs exhibit broad emission wavelengths 
because of charge-transfer (CT) excitation.

In order to overcome these drawbacks, 
Sato {\it et al.} proposed
other concepts
for emitting molecules in OLEDs, viz.  
symmetry-controlled TADF (SC-TADF) and 
inverted singlet and triplet (iST) structure,
wherein fluorescence via RISC
from triplet states higher than T$_1$
is utilized based on the selection rules  
of transition dipole moment (TDM) and spin-orbit coupling (SOC)
\cite{Sato2015_870}.
The order of the preferable point groups for realizing SC-TADF and iST 
is as follows:
\begin{eqnarray}
D_{6h} > O_h > I_h & = & D_{4h} > D_{2h} > D_{3h} > T_{d} > C_i 
\nonumber
\\
&=& C_{2h} > D_{2d} = C_{4v} > D_{2} = C_{2v} > C_{3v} > C_s = C_1 
.
\label{Eq:OrderPointGroups}
\end{eqnarray}
These mechanisms are unlike TADF because they enable us to
(1) use candidates not belonging to the donor-acceptor type and
(2) induce RISC without thermal excitation.
Uejima {\it et al.} and Sato {\it et al.} have already designed and proposed iST molecules 
based on anthracene\cite{Sato2013_012010,Uejima2014_14244}
and perylene derivatives\cite{Sato2015_870}, respectively.

Even for asymmetric molecules,
RISC via higher T$_n$ is possible 
as long as undesirable interactions are suppressed.
Recently, a phenothiazin-benzothiadiazole derivative, PTZ-BZP, used as a fluorescent OLED 
exhibited a high EQE of 1.54 \% (PLQE:16 \%)\cite{Yao2014_2151}, which was attributed to fluorescence via RISC from T$_3$ 
based on the energy gap law. Sato also proposed that the high EQE in PTZ-BZP,
which is an asymmetric molecule, can be attributed 
to suppressed radiative and non-radiative transitions 
from triplet states higher than T$_1$ to lower triplet states 
because of small overlap densities 
in the pseudo-degenerate electronic structure
as well as the small energy gap 
between the relevant triplet and singlet states\cite{Sato2015_189}.

The overlap density is related to the rate constants of
radiative and non-radiative transitions as follows.
The rate constant of the radiative transition
between electronic states $m$ and $n$
depends on the square of TDM, ${\boldsymbol{\mu}_{mn}}$,
defined as
\begin{equation}
\boldsymbol{\mu}_{mn}
:=
\int\cdots\int
\Psi_{m}^{\ast}(\boldsymbol{R}_{0},\boldsymbol{r})
\left( \sum_i -e\boldsymbol{r}_i \right)
\Psi_{n}(\boldsymbol{R}_{0},\boldsymbol{r})
d^4\boldsymbol{x}_{1}\cdots d^4\boldsymbol{x}_{N},
\label{Eq:TDMDef}
\end{equation}
while that of the non-radiative transition
via vibrational mode $\alpha$
between vibronic states
depends on the square of off-diagonal vibronic coupling constant 
(VCC) $V^{mn}_\alpha$,
defined as
\begin{equation}
V^{mn}_\alpha
:=
\int\cdots\int
\Psi_{m}^{\ast}(\boldsymbol{R}_{0},\boldsymbol{r})
\left(
\frac{\partial \hat{H}}{\partial Q_\alpha}
\right)_{\boldsymbol{R}_{0}}
\Psi_{n}(\boldsymbol{R}_{0},\boldsymbol{r})
d^4\boldsymbol{x}_{1}\cdots d^4\boldsymbol{x}_{N},
\label{Eq:VCCDef}
\end{equation}
where $\hat{H}$ is a molecular Hamiltonian,
$Q_\alpha$ stands for a mass-weighted normal coordinate of mode $\alpha$,
$\boldsymbol{R}_{0}$ denotes a reference nuclear configuration,
$\boldsymbol{x}_{i}=(\boldsymbol{r}_{i},s_{i})$ 
with spatial coordinate $\boldsymbol{r}_{i}$ and spin coordinate $s_{i}$
for electron $i$, $e$ is the elementary charge,and
$\Psi_{m}$ and $\Psi_{n}$ are the $N$-electron wave functions 
of electronic states $m$ and $n$, respectively
\cite{Sato2013_012010,Uejima2014_14244}.
TDM ${\boldsymbol{\mu}_{mn}}$
and off-diagonal VCC $V^{mn}_\alpha$ represent
the strengths of radiative and non-radiative transitions,
respectively.
The relations of the radiative and non-radiative transition rate constants
to ${\boldsymbol{\mu}_{mn}}$ and $V^{mn}_\alpha$
are described in more detail in SEC. S1 of the SM
\footnote{See Supplemental Material
at http://link.aps.org/supplemental/xxx/PhysRevApplied.xxx for
radiative and non-radiative transition rates;
vibronic coupling density and transition dipole moment density;
optimized structures of BD1;
selection rules for $D_2$;
energy shifts of the triplet excited states during optimizations;
vibrational modes with strong couplings;
vibronic coupling density analyses;
general conditions for the disappearance of overlap densities
in pseudo-degenerate systems; and
imposition of the $D_{2h}$-symmetry constraint on BD1.}.

Both TDM ${\boldsymbol{\mu}_{mn}}$ and off-diagonal VCC $V^{mn}_\alpha$
are related to the overlap density 
$\rho^{mn}(\boldsymbol{r}_{i})$\cite{Sato2013_012010,Uejima2014_14244}.
The overlap density $\rho^{mn}$ is defined by
\begin{equation}
\rho^{mn}(\boldsymbol{r}_{i})
:=
N\int\cdots\int 
\Psi_{m}^{\ast}(\boldsymbol{R}_{0},\boldsymbol{r})
\Psi_{n}(\boldsymbol{R}_{0},\boldsymbol{r})
d^4\boldsymbol{x}_{1}\cdots d^4\boldsymbol{x}_{i-1} ds_{i} d^4\boldsymbol{x}_{i+1}\cdots
d^4\boldsymbol{x}_{N}
.
\label{Eq:OverlapDensity}
\end{equation}
Hereafter $\mathbf{r}_{i}$ is simply denoted as $\mathbf{r}$.
$\rho^{mn}(\boldsymbol{r})$ is sometimes called a transition density,
especially within the orbital approximation.
For example, in the case of the HOMO-LUMO transition,
it is equal to HOMO-LUMO overlap density.
The off-diagonal VCC $V_{\alpha}^{mn}$ can be
exactly expressed using the off-diagonal vibronic coupling density (VCD)
$\eta_{\alpha}^{mn}(\boldsymbol{r})$:
\begin{equation}
V_{\alpha}^{mn}=\int\eta_{\alpha}^{mn}(\boldsymbol{r})d^{3}\boldsymbol{r}
,
\label{Eq:VCCVCD}
\end{equation}
where the off-diagonal VCD is defined by
\begin{equation}
\eta_{\alpha}^{mn}(\boldsymbol{r})
:=
\rho^{mn}(\boldsymbol{r})\times v_{\alpha}(\boldsymbol{r})
,
\label{Eq:VCD}
\end{equation}
and the potential derivative $v_{\alpha}(\boldsymbol{r})$ is defined by
\begin{equation}
v_{\alpha}(\boldsymbol{r})
:=
\left(
\frac{\partial u(\boldsymbol{r})}{\partial Q_{\alpha}}
\right)_{\boldsymbol{R}_{0}}
,
\quad
u(\boldsymbol{r})
:=
\sum_{A=1}^{M}
-\frac{Z_{A}e^{2}}
{4\pi\epsilon_{0}\left|\boldsymbol{r} - \boldsymbol{R}_{A}\right|}
,
\label{Eq:PotentialDerivative}
\end{equation}
where $u(\boldsymbol{r})$ is the attractive potential 
of a single electron due to all nuclei, and
$\boldsymbol{R}_{A}$ and $Z_{A}$ are the position and charge, respectively of nucleus $A$. 
VCD $\eta_{\alpha}^{mn}(\boldsymbol{r})$ illustrates
the origin of VCC which gives rise to non-radiative transition
as a local picture.
The detailed derivation of VCD can be found in SEC. S2 of the SM.
TDM $\boldsymbol{\mu}_{mn}$ is also related to the overlap density:
\begin{equation}
\boldsymbol{\mu}_{mn} 
=
\int \boldsymbol{\tau}_{mn}(\boldsymbol{r}) d^3 {\boldsymbol{r}}
,
\label{Eq:TDMOverlapDensity}
\end{equation}
where the transition dipole moment density (TDMD)
$\boldsymbol{\tau}_{mn}(\boldsymbol{r})$ is defined by
\begin{equation}
\boldsymbol{\tau}_{mn}(\boldsymbol{r}) 
:= 
-e \boldsymbol{r} \rho^{mn}(\boldsymbol{r})
.
\label{Eq:TDMD}
\end{equation}
TDMD $\boldsymbol{\tau}_{mn}(\boldsymbol{r})$
illustrates the origin of TDM which causes radiative transition
as a local picture.
The detailed derivation of TDMD is shown in SEC. S2 of the SM.
Based on Eqs. \ref{Eq:VCCVCD}, \ref{Eq:VCD},
\ref{Eq:TDMOverlapDensity}, and \ref{Eq:TDMD},
both radiative and non-radiative transitions are suppressed,
when the overlap density $\rho^{mn}(\boldsymbol{r})$ is
so small
that VCD $\eta_{\alpha}^{mn}(\boldsymbol{r})$ and
TDMD $\boldsymbol{\tau}_{mn}(\boldsymbol{r})$ are small.

\begin{figure}[!h]
\centering
\includegraphics[scale=0.7]{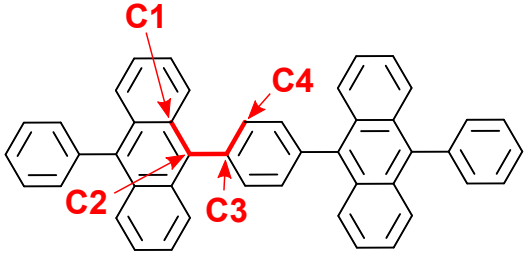}
\caption{Bisanthracene derivative, BD1.\label{Fig:BD1}}
\end{figure}

Hu {\it et al.} observed blue-light emission 
in OLEDs using bisanthracene derivatives
including
1,4-bis(10-phenylanthracene-9-yl)benzene (BD1)
(see FIG. \ref{Fig:BD1})\cite{Hu2014_2064}.
Since the observed PLQE of BD1 is 14\% in the neat film while the maximum EQE is 8.9\% in the doped film,
the emission is not conventional fluorescence using only singlet excitons.
In other words, triplet excitons must
contribute to the observed emission.
One possible mechanism to explain the observed high EQE is the triplet-triplet annihilation (TTA).
In the TTA mechanism, 
62.5\% of generated excitons can be up-converted into the singlet excited state at best.
Hu {\it et al.} have reported that,
for example, 
the highest EQE value is 5.6\% in the device employing a neat film of BD1 as an emitting layer, 
and the observed PLQE of the neat film is 14\%
\cite{Hu2014_2064}.
Using these values, the estimated upper-limit of the EQE is 1.8\%--3.5\%
if the outcoupling efficiency is assumed to be 20\%--40\%.
Therefore, we cannot explain the observed EQE on the basis of the TTA mechanism. 

In this study, we theoretically investigate the mechanism of light emission 
from an OLED using a bisanthracene derivative, BD1, as the emitting molecule 
based on the concept of VCD. 
We propose {\it a fluorescence via higher triplets (FvHT)} mechanism
to explain the high EQE in OLEDs using BD1, which is valid as long as undesirable radiative and non-radiative transitions are suppressed.
In addition, we
propose a general condition for the suppression of radiative and non-radiative transitions
in molecules with pseudo-degenerate electronic structures such as BD1.
In our previous work on PTZ-BZP, we have mentioned that 
the observed EQE is due to its pseudo degenerate electronic structure.
In this study, we perform the analyses for electronic wave functions 
of BD1 in detail to obtain general design principles for the realization of the present mechanism.
The present article consists of the following sections:
The method of calculation is described in SEC. \ref{Sec:Method}.
In SEC. \ref{SubSec:FC}, the results for the Franck--Condon (FC) states are discussed.
The optimized structures for the excited states are discussed in SEC. \ref{SubSec:ADStructures}.
The frontier orbitals and orbital overlap densities are discussed in SEC. \ref{SubSec:Orbital}.
In SEC. \ref{SubSec:SelectrionRules}, the selection rules for the couplings are described.
The excited states at the optimized structures for the excited states are discussed in SEC. \ref{SubSec:AD}.
The calculated VCCs are presented in SEC. \ref{SubSec:VCC}.
The obtained VCCs are discussed on the basis of the concept of VCD in SEC. \ref{SubSec:VCD}.
The disappearance of the overlap densities is discussed in SEC. \ref{SubSec:OverlapDensity}.
We also present a general condition for the suppression of the radiative and non-radiative transitions
in the SM.
We conclude this study in SEC. \ref{Sec:Conclusion}. 

%%%%%%%%%%%%%%%%%%%%%%%%%%%%%%%%%%%%%%%%%%%%%%%%%%%%%%%%%%%%%%%%%%%%%
\section{Method of Calculation\label{Sec:Method}}
%%%%%%%%%%%%%%%%%%%%%%%%%%%%%%%%%%%%%%%%%%%%%%%%%%%%%%%%%%%%%%%%%%%%%
The optimized structure of BD1 in the ground state was obtained.
The structure was confirmed to be the minimum energy structure using vibrational analysis.
Excited adiabatic (AD) states were obtained 
by carrying out geometry optimizations and calculating the electronic states 
at the optimized structures. 
The normal modes were also obtained by vibrational analyses.
Regarding the AD states, vibrational analyses were
carried out for S$_0$ at the optimized structures for the excited states.
These calculations were performed at the B3LYP/6-311+G(d,p) and TD-B3LYP/6-311+G(d,p) 
levels of theory for the ground and excited states, respectively.
In the excited state calculations,
ten singlet and ten triplet states were taken into consideration.
Off-diagonal VCCs between triplet states T$_m$--T$_n$, as well as 
singlet states S$_m$--S$_n$ were calculated. 
VCD analyses were carried out for strong coupling modes.
The electronic and vibrational states were calculated 
using Gaussian 09 Revision D.01\cite{g09}, while the VCC calculations and VCD analyses were performed using our in-house codes.

%%%%%%%%%%%%%%%%%%%%%%%%%%%%%%%%%%%%%%%%%%%%%%%%%%%%%%%%%%%%%%%%%%%%%
\section{Results and Discussion\label{Sec:ResultsDiscussin}}
%%%%%%%%%%%%%%%%%%%%%%%%%%%%%%%%%%%%%%%%%%%%%%%%%%%%%%%%%%%%%%%%%%%%%
\subsection{Franck--Condon Excited States\label{SubSec:FC}}
\begin{figure}[!h]
\centering
\begin{tabular}{cc}
\multicolumn{1}{l}{{\bf\large (a)}} &
\multicolumn{1}{l}{{\bf\large (b)}}\\
\includegraphics[scale=0.45]{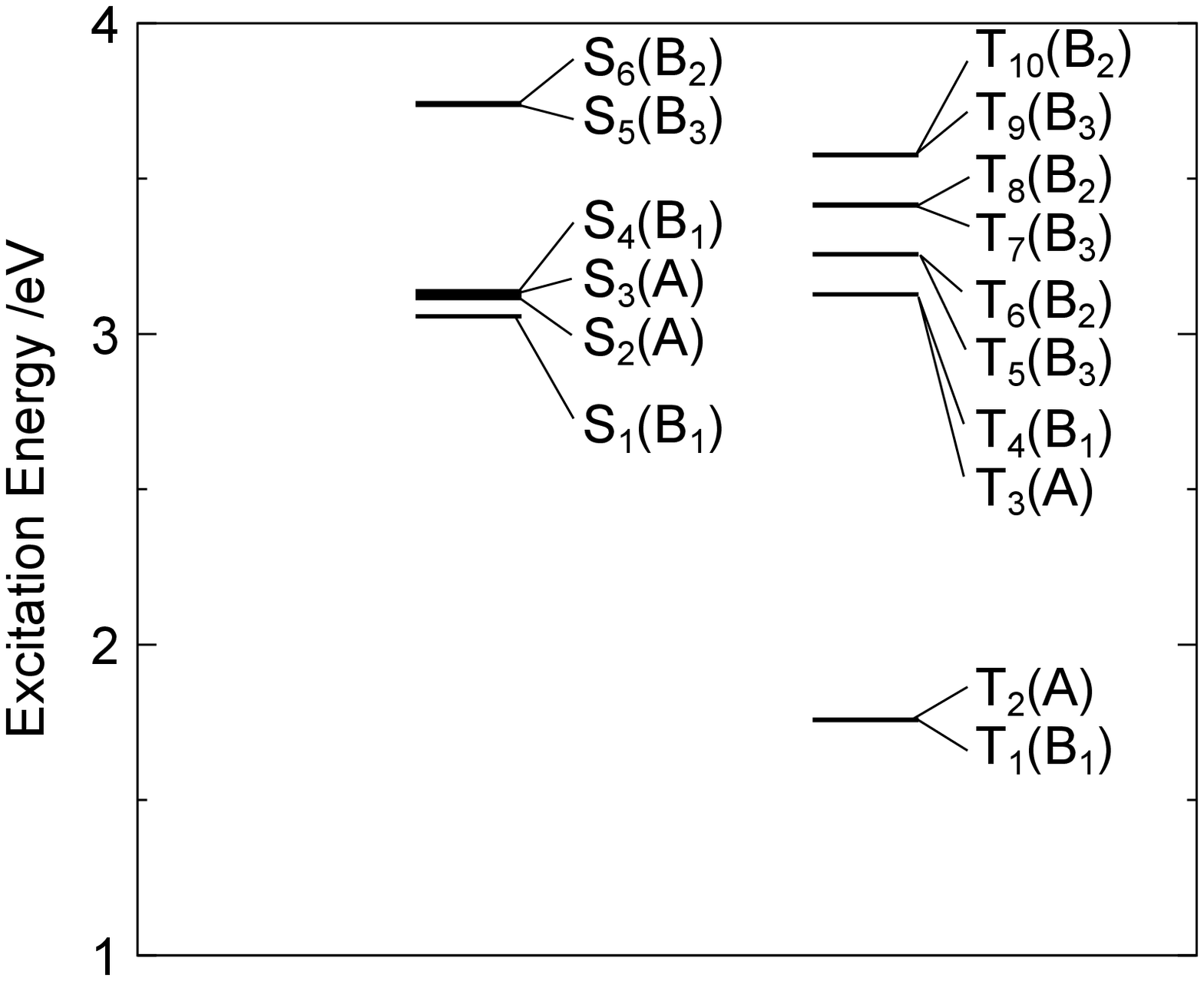}&
\includegraphics[scale=0.45]{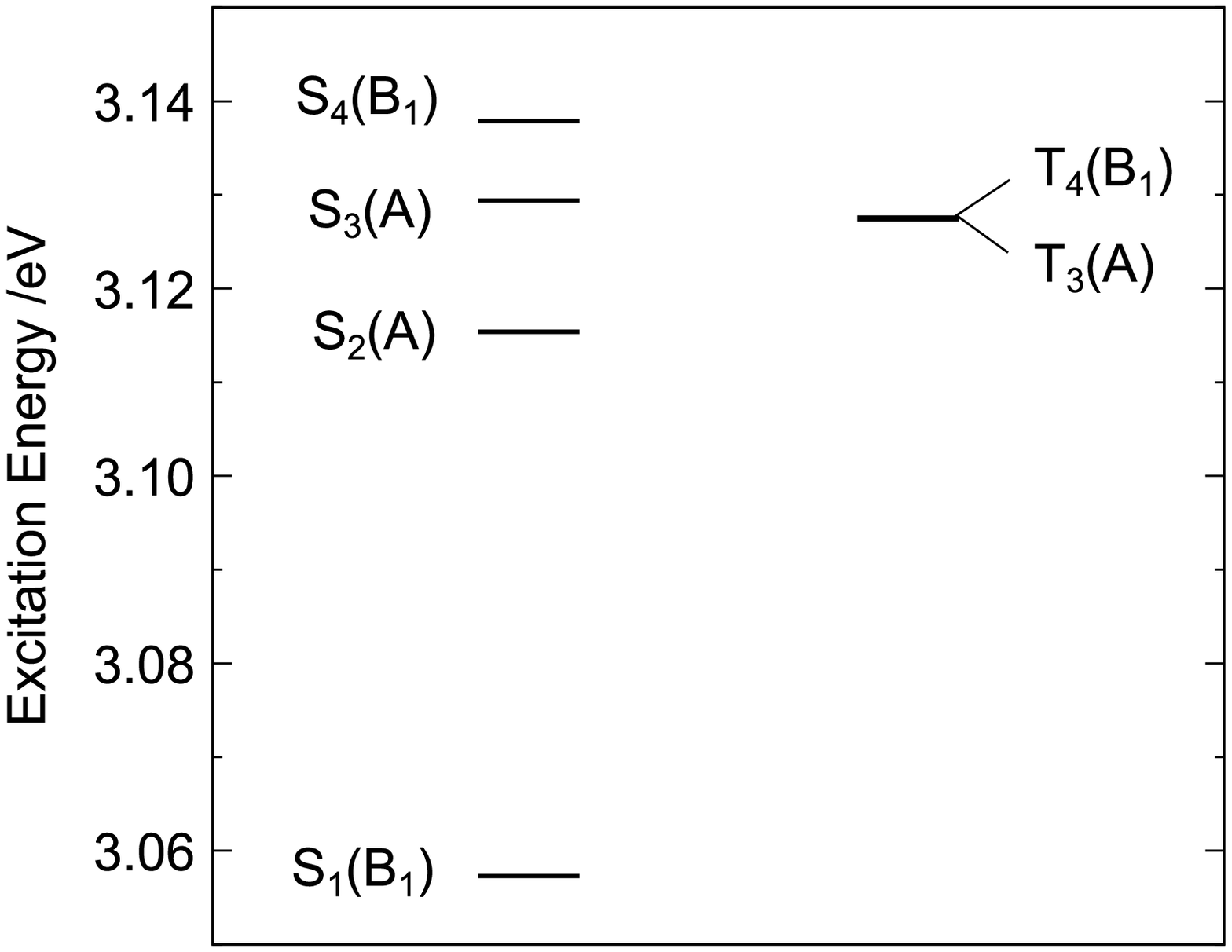}\\
\end{tabular}
\caption{(a) Energy levels of the excited states in BD1
at the optimized structure for $S_0$,
and (b) an enlarged view of the relevant levels.
\label{Fig:FCLevels}}
\end{figure}

%%%%%%%%%%%%%%%%%%%%%%%%%%%%%%%%%%%%%%%%%%%%%%%%%%%%%%%%%%%%%%%%%%%%%
The symmetry of the optimized structure for the ground state is $D_2$.
FC states were calculated at the optimized structure for S$_0$.
FIG. \ref{Fig:FCLevels} shows the energy levels of the FC states.
The energy levels of T$_3$ and T$_4$ are close to those of S$_2$, S$_3$, and S$_4$.
Therefore, the geometrical structures 
of T$_4$, T$_3$, S$_4$, S$_3$, S$_2$, and S$_1$ were optimized.

%%%%%%%%%%%%%%%%%%%%%%%%%%%%%%%%%%%%%%%%%%%%%%%%%%%%%%%%%%%%%%%%%%%%%
\subsection{Optimized Structures of the Excited states\label{SubSec:ADStructures}}
%%%%%%%%%%%%%%%%%%%%%%%%%%%%%%%%%%%%%%%%%%%%%%%%%%%%%%%%%%%%%%%%%%%%%
FIG. S1 in the SM\footnotemark[1]
shows the optimized structures of the relevant excited states, i.e. 
T$_3$, T$_4$, S$_2$, and S$_1$ as well as S$_0$.
All the optimized structures show $D_2$ symmetry.
The dihedral angles (C1--C2--C3--C4)
between the anthracene and benzene moieties (see FIG. \ref{Fig:BD1})
for the optimized structures are shown in  S2.
The dihedral angles of S$_0$ and S$_4$ are close to the right angle,
while those of T$_3$, T$_4$, and S$_1$ are small.

%%%%%%%%%%%%%%%%%%%%%%%%%%%%%%%%%%%%%%%%%%%%%%%%%%%%%%%%%%%%%%%%%%%%%
\subsection{Frontier Orbitals\label{SubSec:Orbital}}
\begin{figure}[!h]
\centering
\begin{tabular}{cc}
\multicolumn{1}{l}{{\bf\large (a)}} &
\multicolumn{1}{l}{{\bf\large (b)}}\\
\includegraphics[scale=0.12]{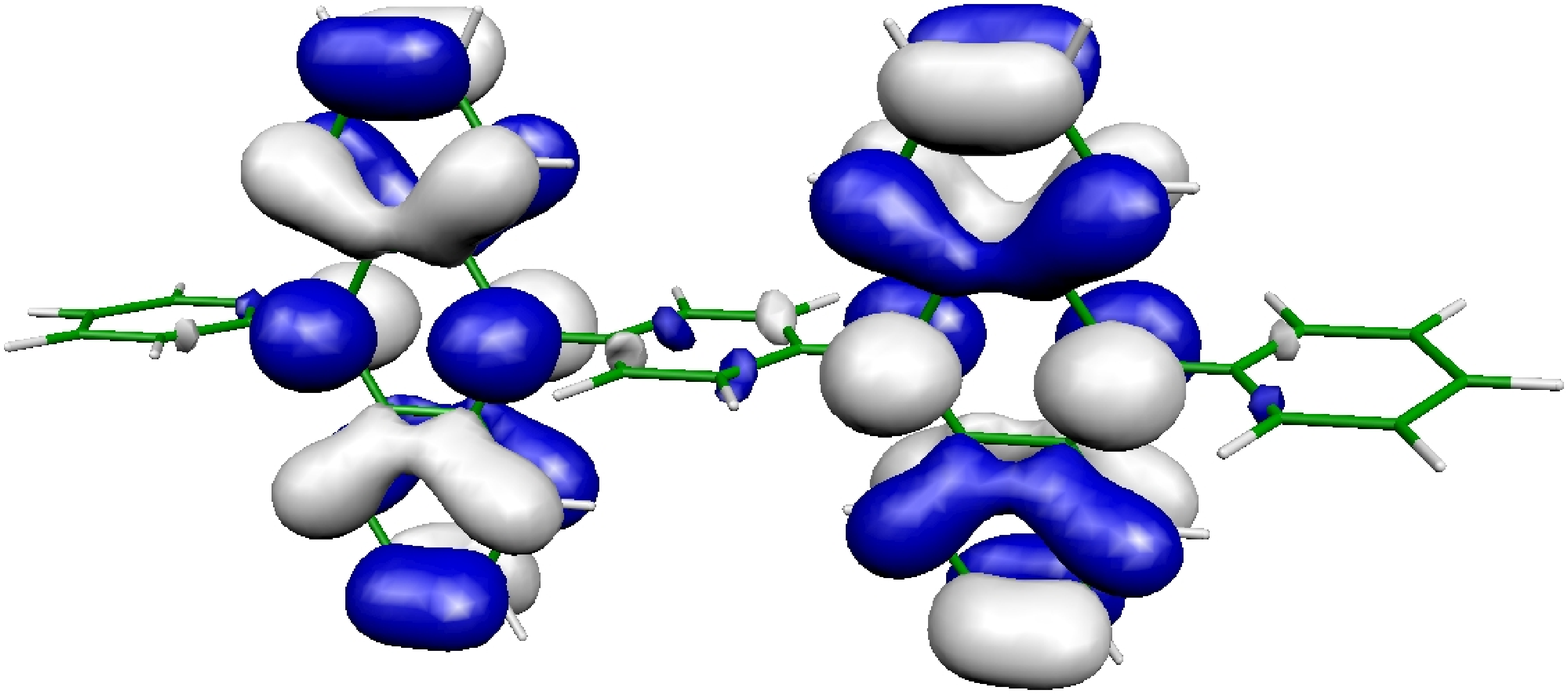} &
\includegraphics[scale=0.12]{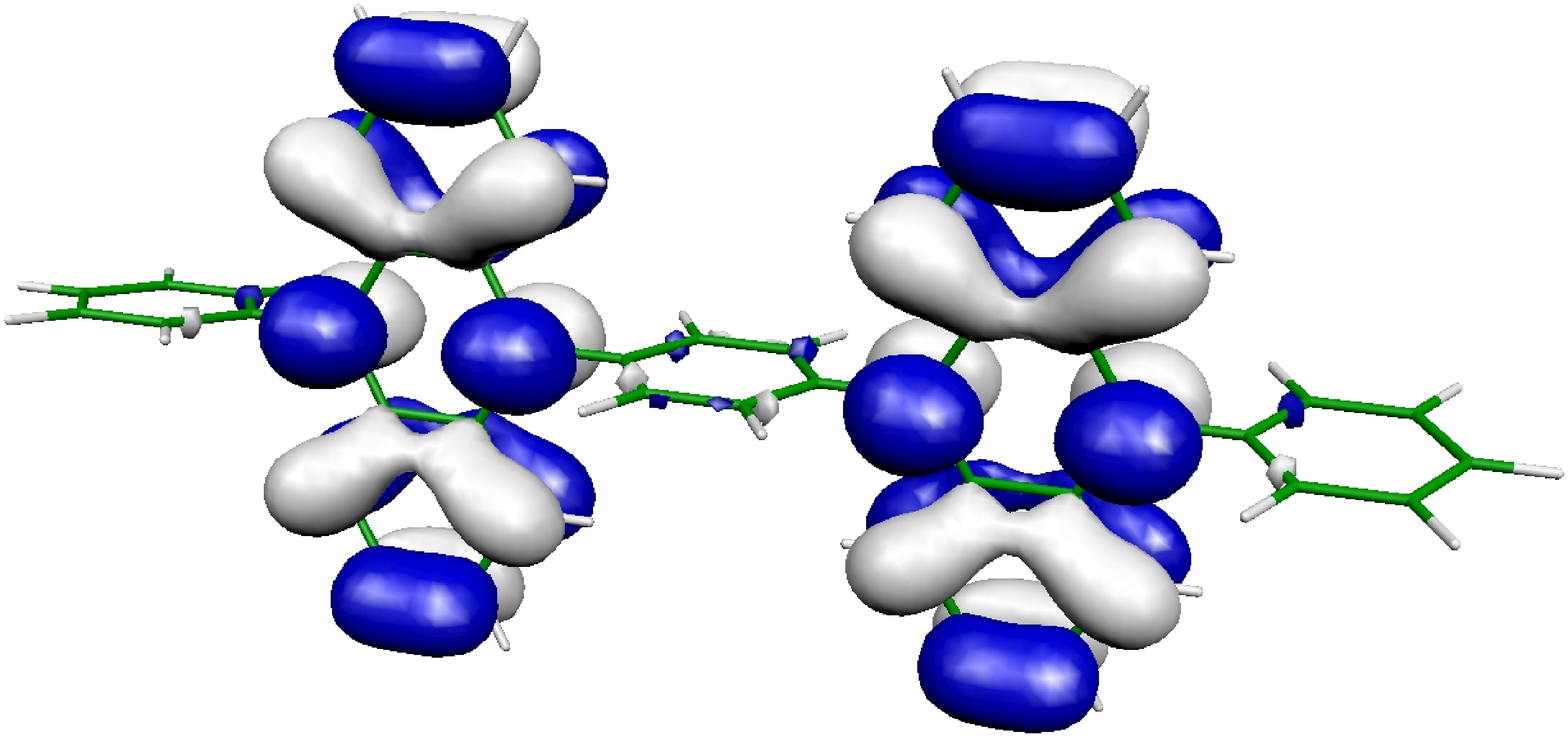} \\
%&\\
\multicolumn{1}{l}{{\bf\large (c)}} &
\multicolumn{1}{l}{{\bf\large (d)}}\\
\includegraphics[scale=0.12]{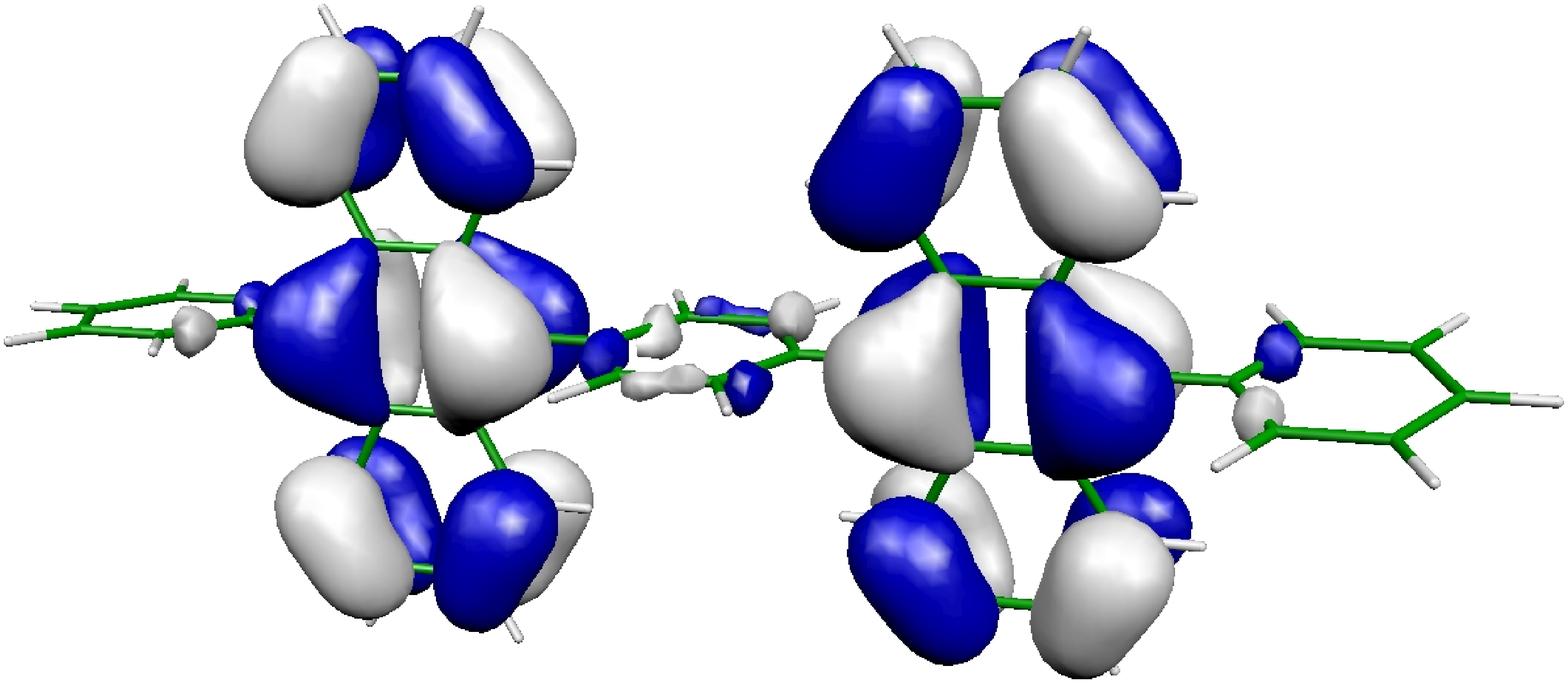} &
\includegraphics[scale=0.12]{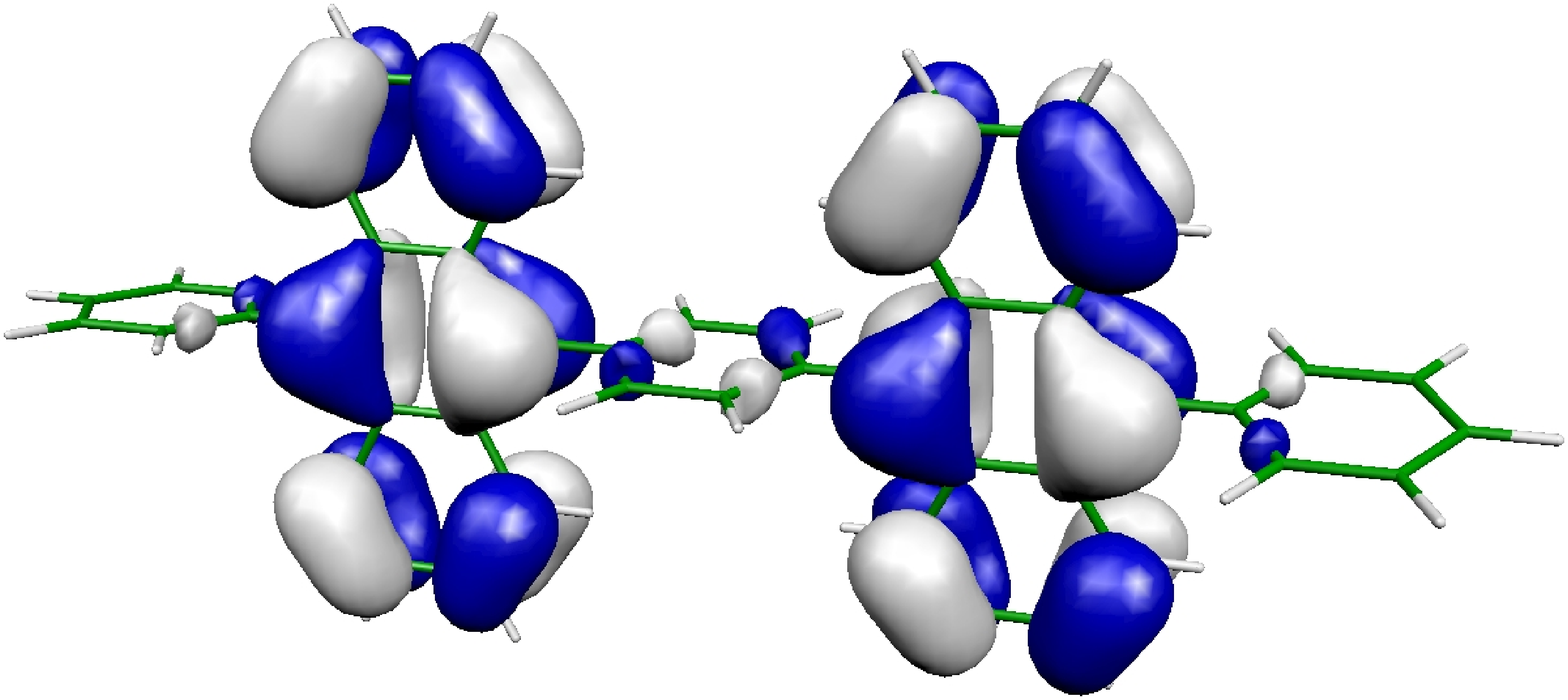} \\
\end{tabular}
\caption{Frontier orbitals of BD1:
(a) $B_2$ LUMO,
(b) $B_3$ NLUMO,
(c) $B_3$ HOMO, and
(d) $B_2$ NHOMO.
The isosurface value is $2.0 \times 10^{-2}$ a.u.
\label{Fig:S0Frontier}}
\end{figure}

\begin{figure}[!h]
\centering
\includegraphics[scale=0.5]{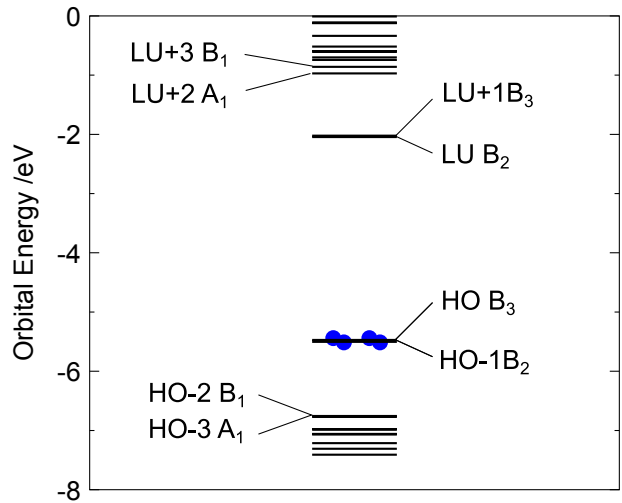}
\caption{Frontier orbital levels of BD1.
It should be noted that the HOMO and next HOMO
as well as
the LUMO and next LUMO are pseudo-degenerate.
\label{Fig:S0MOLevels}}
\end{figure}

%%%%%%%%%%%%%%%%%%%%%%%%%%%%%%%%%%%%%%%%%%%%%%%%%%%%%%%%%%%%%%%%%%%%%
Frontier orbitals and their energy levels at the optimized structure of the ground state 
are shown in FIGs. \ref{Fig:S0Frontier} and \ref{Fig:S0MOLevels}.
The NHOMO $\psi_{\textrm{NHO}}$ and HOMO $\psi_{\textrm{HO}}$
as well as the NLUMO $\psi_{\textrm{NLU}}$ and LUMO $\psi_{\textrm{LU}}$
are pseudo-degenerate.
From FIG. \ref{Fig:S0Frontier}, the frontier orbitals can be approximately represented as follows:
\begin{eqnarray}
\psi_{\textrm{LU}}\approx\frac{1}{\sqrt{2}}(\phi_{\textrm{LU}}(L) - \phi_{\textrm{LU}}(R)),
&\quad&
\psi_{\textrm{NLU}}\approx\frac{1}{\sqrt{2}}(\phi_{\textrm{LU}}(L) + \phi_{\textrm{LU}}(R))
,
\label{Eq:LUNLU}
\\
\psi_{\textrm{HO}}\approx\frac{1}{\sqrt{2}}(\phi_{\textrm{HO}}(L) + \phi_{\textrm{HO}}(R)),
&\quad&
\psi_{\textrm{NHO}}\approx\frac{1}{\sqrt{2}}(\phi_{\textrm{HO}}(L) - \phi_{\textrm{HO}}(R))
,
\label{Eq:HONHO}
\end{eqnarray}
where $\phi_{\textrm{HO/LU}}(L/R)$ denotes the fragment MOs consisting of the
HOMO/LUMO of the anthracene moiety Left(L)/Right(R).

%\sout{Neglecting the overlap between the different anthracene moieties, i.e.}
Since $\phi_{i}(L)$ and $\phi_j(R)$ are the fragment MOs, 
$\phi_{i}(L)\phi_j(R)\approx 0$ ($i,j=$HO, NHO, LU, and NLU).
Accordingly,
\begin{eqnarray}
|\psi_{\textrm{LU}}|^{2}
&\approx&
|\psi_{\textrm{NLU}}|^{2}
\label{Eq:SquareLUNLU}
\\
\psi_{\textrm{HO}}\psi_{\textrm{NLU}}
&\approx&
\psi_{\textrm{NHO}}\psi_{\textrm{LU}}
\label{Eq:OverlapHONLU-NHOLU}
\\
|\psi_{\textrm{HO}}|^{2}
&\approx&
|\psi_{\textrm{NHO}}|^{2}
\label{Eq:SquareHONHO}
\\
\psi_{\textrm{HO}}\psi_{\textrm{LU}}
&\approx&
\psi_{\textrm{NHO}}\psi_{\textrm{NLU}}
\label{Eq:OverlapHOLU-NHONLU}
\end{eqnarray}
%\begin{eqnarray}
%& &|\psi_{\textrm{LU}}|^{2}
%\approx
%\frac{1}{2}(|\phi_{\textrm{LU}}(L)|^{2} + |\phi_{\textrm{LU}}(R)|^{2})
%\approx
%|\psi_{\textrm{NLU}}|^{2}
%,
%\label{Eq:SquareLUNLU}
%\\
%& &
%\psi_{\textrm{HO}}\psi_{\textrm{NLU}}
%\approx
%\frac{1}{2}(
%  \phi_{\textrm{LU}}(L)\phi_{\textrm{HO}}(L)
%  +  
%  \phi_{\textrm{LU}}(R)\phi_{\textrm{HO}}(R))
%\approx
%\psi_{\textrm{NHO}}\psi_{\textrm{LU}}
%,
%\label{Eq:OverlapHONLU-NHOLU}
%\\
%& &
%|\psi_{\textrm{HO}}|^{2}
%\approx
%\frac{1}{2}(|\phi_{\textrm{HO}}(L)|^{2} + |\phi_{\textrm{HO}}(R)|^{2})
%\approx
%|\psi_{\textrm{NHO}}|^{2}
%,
%\label{Eq:SquareHONHO}
%\\
%& &
%\psi_{\textrm{HO}}\psi_{\textrm{LU}}
%\approx
%\frac{1}{2}(
%  \phi_{\textrm{LU}}(L)\phi_{\textrm{HO}}(L)
%  -
%  \phi_{\textrm{LU}}(R)\phi_{\textrm{HO}}(R))
%\approx
%\psi_{\textrm{NHO}}\psi_{\textrm{NLU}}
%.
%\label{Eq:OverlapHOLU-NHONLU}
%\end{eqnarray}
In addition,
%\begin{equation}
%\psi_{\textrm{HO}}\psi_{\textrm{NHO}}
%\approx
%\frac{1}{2}(
%  |\phi_{\textrm{HO}}(L)|^{2} - |\phi_{\textrm{HO}}(R)|^{2})
%,
%\label{Eq:OverlapHONHO}
%\end{equation}
%\begin{equation}
%\psi_{\textrm{LU}}\psi_{\textrm{NLU}}
%\approx
%\frac{1}{2}(
%  |\phi_{\textrm{LU}}(L)|^{2} - |\phi_{\textrm{LU}}(R)|^{2})
%.
%\label{Eq:OverlapLUNLU}
%\end{equation}
%Accordingly,
\begin{equation}
\psi_{\textrm{HO}}\psi_{\textrm{NHO}}
\ne
\psi_{\textrm{LU}}\psi_{\textrm{NLU}}
.
\label{Eq:OverlapHONHO-LUNLU}
\end{equation}
In order to analyze VCDs and TDMDs obtained
from the TD-DFT wave functions,
these relations of the frontier orbitals
will be used in SEC. \ref{SubSec:OverlapDensity}.

%%%%%%%%%%%%%%%%%%%%%%%%%%%%%%%%%%%%%%%%%%%%%%%%%%%%%%%%%%%%%%%%%%%%%
\subsection{Selection Rules\label{SubSec:SelectrionRules}}
%%%%%%%%%%%%%%%%%%%%%%%%%%%%%%%%%%%%%%%%%%%%%%%%%%%%%%%%%%%%%%%%%%%%%
Since all the optimized structures show $D_2$ symmetry,
the selection rules for TDM, SOC, and vibronic coupling (VC) within $D_2$ symmetry 
are discussed here.
TABLE S1 in the SM\footnotemark[1]
is the character table of the $D_2$ point group.
The direct products of the irreducible representations (irreps) are tabulated in TABLE S2.
From TABLE S1,
the components of the electric dipole operator,
$\hat{\mu}_{x}$, $\hat{\mu}_{y}$, $\hat{\mu}_{z}$ transform
according to the $B_3$, $B_2$, and $B_1$ irreps, respectively, and
the components of the orbital angular momentum operator,
$\hat{L}_{x}$, $\hat{L}_{y}$, $\hat{L}_{z}$ transform
according to the $B_3$, $B_2$, and $B_1$ irreps, respectively.
Based on TABLE S2, the selection rules for TDM, SOC, and VC were obtained  
and are listed in TABLEs S3, S4, and S5.
According to TABLEs S3 and S4,
an electric dipole transition or intersystem crossing 
between electronic states with the same irrep is  
symmetry forbidden.

TABLE S5 lists the selection rule for VC with $D_2$ symmetry.
The vibrational degrees of freedom in BD1 are decomposed
into irreps as follows:
\begin{equation}
\Gamma_{vib} = 54a_1 \oplus 52b_1 \oplus 58b_2 \oplus 58b_3
.
\label{Eq:Modes}
\end{equation}
According to TABLE S5,
the number of vibronically active modes accounts 
for 1/4 of all the modes
for every non-radiative transition $m\rightarrow n$.
For example, in non-radiative transition S$_1$ $\rightarrow$ S$_0$, 
the vibronically active modes are fifty two $b_1$ modes
because, as discussed later, the S$_1$ state transforms as $B_1$.

%%%%%%%%%%%%%%%%%%%%%%%%%%%%%%%%%%%%%%%%%%%%%%%%%%%%%%%%%%%%%%%%%%%%%
\subsection{Adiabatic Excited States\label{SubSec:AD}}
%%%%%%%%%%%%%%%%%%%%%%%%%%%%%%%%%%%%%%%%%%%%%%%%%%%%%%%%%%%%%%%%%%%%%
In this section, we discuss the excited states at their optimized geometries.
The T$_4$ state has lower energy than T$_3$ 
at the T$_n$ optimized structure ($n=3,4$), as shown
in FIG. S3 in SEC. S5 %S3
of the SM\footnotemark[1].
Hereafter, we will refer to the electronic state at a certain geometry
corresponding to the T$_n$ state at the FC state as T$_n$.

\begin{figure}[!h]
\centering
\begin{tabular}{cc}
\multicolumn{1}{l}{{\bf\large (a)}} &
\multicolumn{1}{l}{{\bf\large (b)}}\\
\includegraphics[scale=0.45]{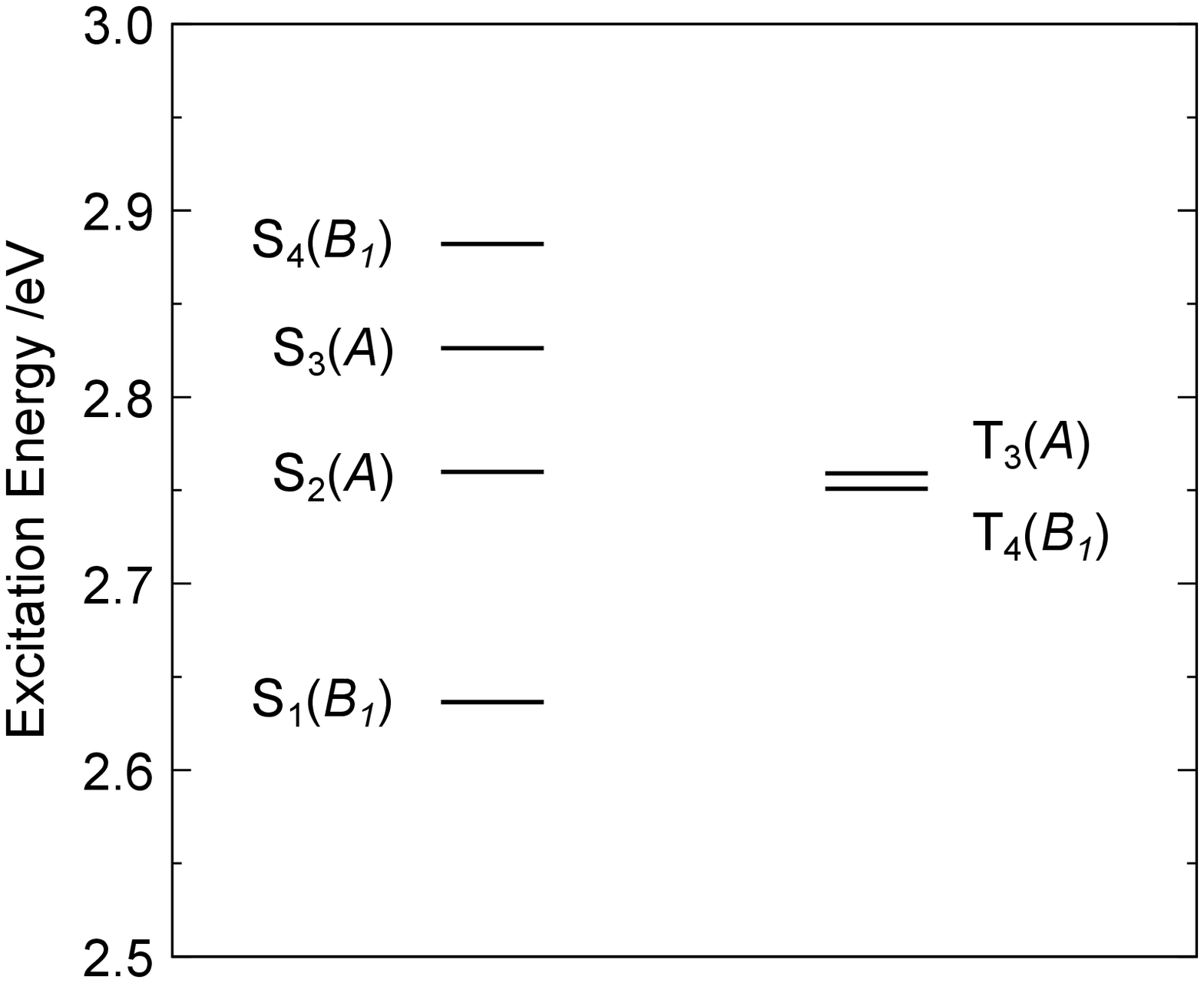} &
\includegraphics[scale=0.45]{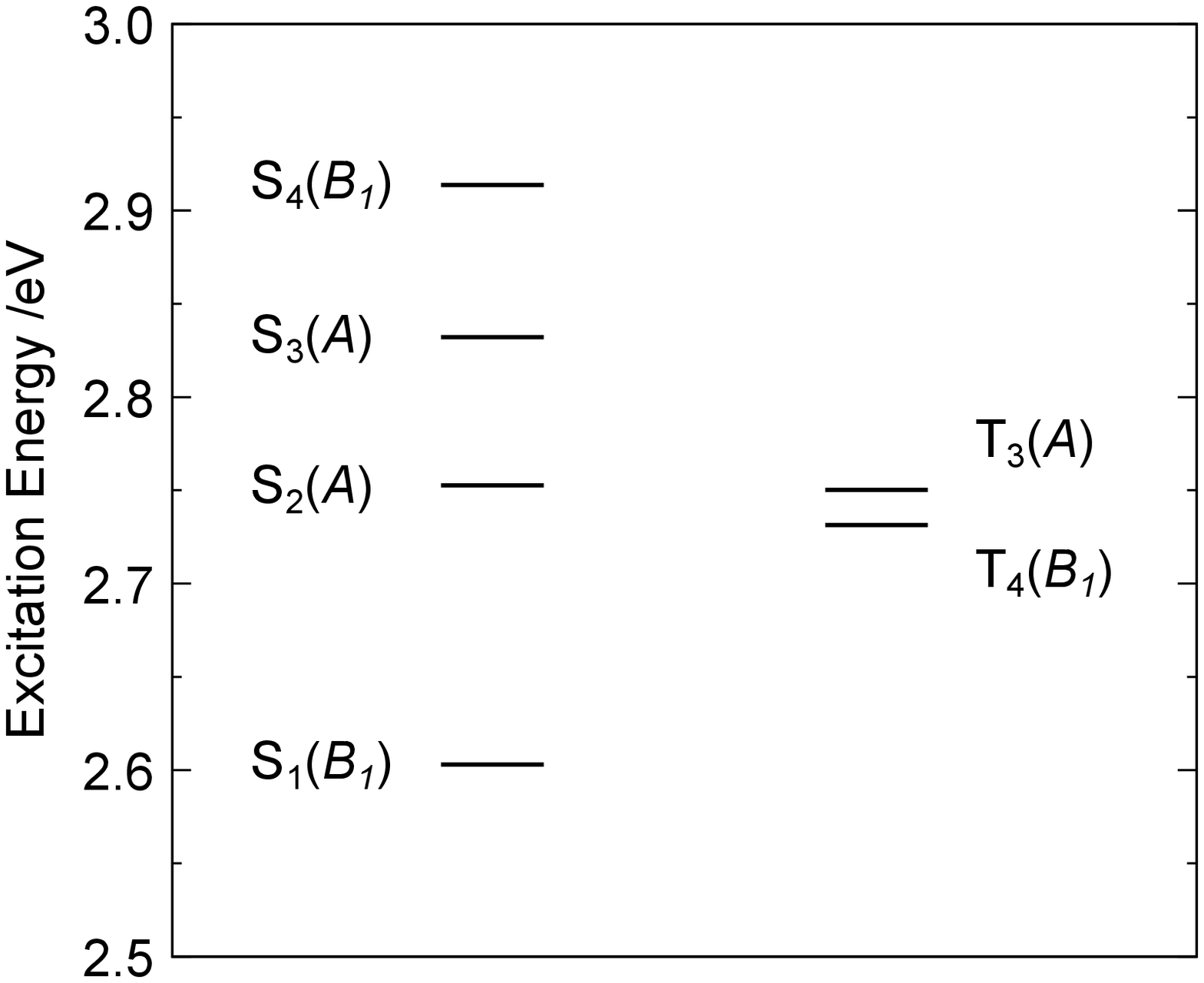}\\
\end{tabular}
\caption{Energy levels of excited states at the optimized structures
for (a) T$_3$ and (b) T$_4$.
\label{Fig:EnergyLevels@T3T4}}
\end{figure}

FIG. \ref{Fig:EnergyLevels@T3T4} shows the energy levels of the excited states
at the optimized structures for T$_3$ and T$_4$.
TABLE \ref{Table:ExcitedStates@T3} lists
the triplet excited states at the optimized structure for T$_3$.
The irrep of T$_3$ is $A$.
Although T$_3$ is close to S$_2$ with $\Delta E_{T_3-S_2}$ = 0.8 meV,
RISC between T$_3$ and S$_2$ is symmetry forbidden.
The electric dipole transition T$_3$ $\rightarrow$ T$_4$ and
internal conversion (IC) T$_3$ $\rightarrow$ T$_4$ via $b_1$ modes are allowed.
These interactions are large
because of a large overlap density between T$_3$ and T$_4$,
as discussed later.
Therefore, a T$_3$ exciton is immediately converted into a T$_4$ exciton.
\begin{table}[!h]
\centering
\caption{Triplet excited states at the optimized structure for T$_3$.
\label{Table:ExcitedStates@T3}}
\begin{tabular}{lccl}
\hline\hline
             &\multicolumn{2}{c}{Excitation Energy}& Major Configuration\\
             & eV & nm &(CI Coefficient) \\
\hline
% slightly changed because of LUMO@T3 = LUMO@S0 * (-1)
T$_4$($B_1$) & 2.7508 & 450.72 & HO-1$\rightarrow$LU+1(0.524),HO$\rightarrow$LU(-0.455)\\
T$_3$($A$)   & 2.7591 & 449.37 & HO-1$\rightarrow$LU(-0.501),HO$\rightarrow$LU+1(0.496)\\
T$_2$($A$)   & 1.4403 & 860.84 & HO$\rightarrow$LU+1(0.502),HO-1$\rightarrow$LU(0.497)\\
T$_1$($B_1$) & 1.4252 & 869.93 & HO$\rightarrow$LU(-0.539),HO-1$\rightarrow$LU+1(-0.458)\\
\hline\hline
\end{tabular}
\end{table}

TABLE \ref{Table:ExcitedStates@T4} lists
the triplet excited states at the optimized structure for T$_4$.
T$_4$ is close to S$_2$ with $\Delta E_{S_2 - T_4}$ = 21 meV, and
RISC between T$_4$ and S$_2$ is symmetry allowed
because the T$_4$ and S$_2$ states belong to
the $B_1$ and $A$ irreps, respectively.
Although the electric dipole transition T$_4$ $\rightarrow$ T$_1$ is symmetry forbidden,
that between T$_4$ and T$_2$ is symmetry allowed.
In addition, IC T$_4$ $\rightarrow$ T$_2$ with the help of $b_1$ modes and 
IC T$_4$ $\rightarrow$ T$_1$ with the help of $a$ modes
are symmetry allowed.
However, 
the transitions T$_4$ $\rightarrow$ T$_2$ and T$_4$ $\rightarrow$ T$_1$ are suppressed
because of small overlap densities of T$_4$ with T$_2$ and T$_1$,
as discussed later.
Therefore, a T$_4$ exciton can be up-converted into an S$_2$ exciton
with thermal excitation.
Accordingly, the T$_3$ and T$_4$ excitons generated via electrical excitation 
are up-converted into S$_2$ excitons via RISC from the T$_4$ state.
\begin{table}[!h]
\centering
\caption{Triplet excited states at the optimized structure for T$_4$.
\label{Table:ExcitedStates@T4}}
\begin{tabular}{lccl}
\hline\hline
          &\multicolumn{2}{c}{Excitation Energy}& Major Configuration \\
          & eV & nm & (CI coefficient) \\
\hline
% slightly changed because of LUMO@T4 = LUMO@S0 * (-1)
T$_3$($A$)  & 2.7503 & 450.81 & HO-1$\rightarrow$LU(-0.501),HO$\rightarrow$LU+1(0.495)\\
T$_4$($B_1$) & 2.7314 & 453.93 & HO-1$\rightarrow$LU+1(0.522),HO$\rightarrow$LU(-0.436)\\
T$_2$($A$)  & 1.4590 & 849.77 & HO$\rightarrow$LU+1(0.501),HO-1$\rightarrow$LU(0.495)\\
T$_1$($B_1$) & 1.4345 & 864.29 & HO$\rightarrow$LU(-0.549),HO-1$\rightarrow$LU+1(-0.444)\\
\hline\hline
\end{tabular}
\end{table}

\begin{figure}[!h]
\centering
\begin{tabular}{cc}
\multicolumn{1}{l}{{\bf\large (a)}} &
\multicolumn{1}{l}{{\bf\large (b)}}\\
\includegraphics[scale=0.45]{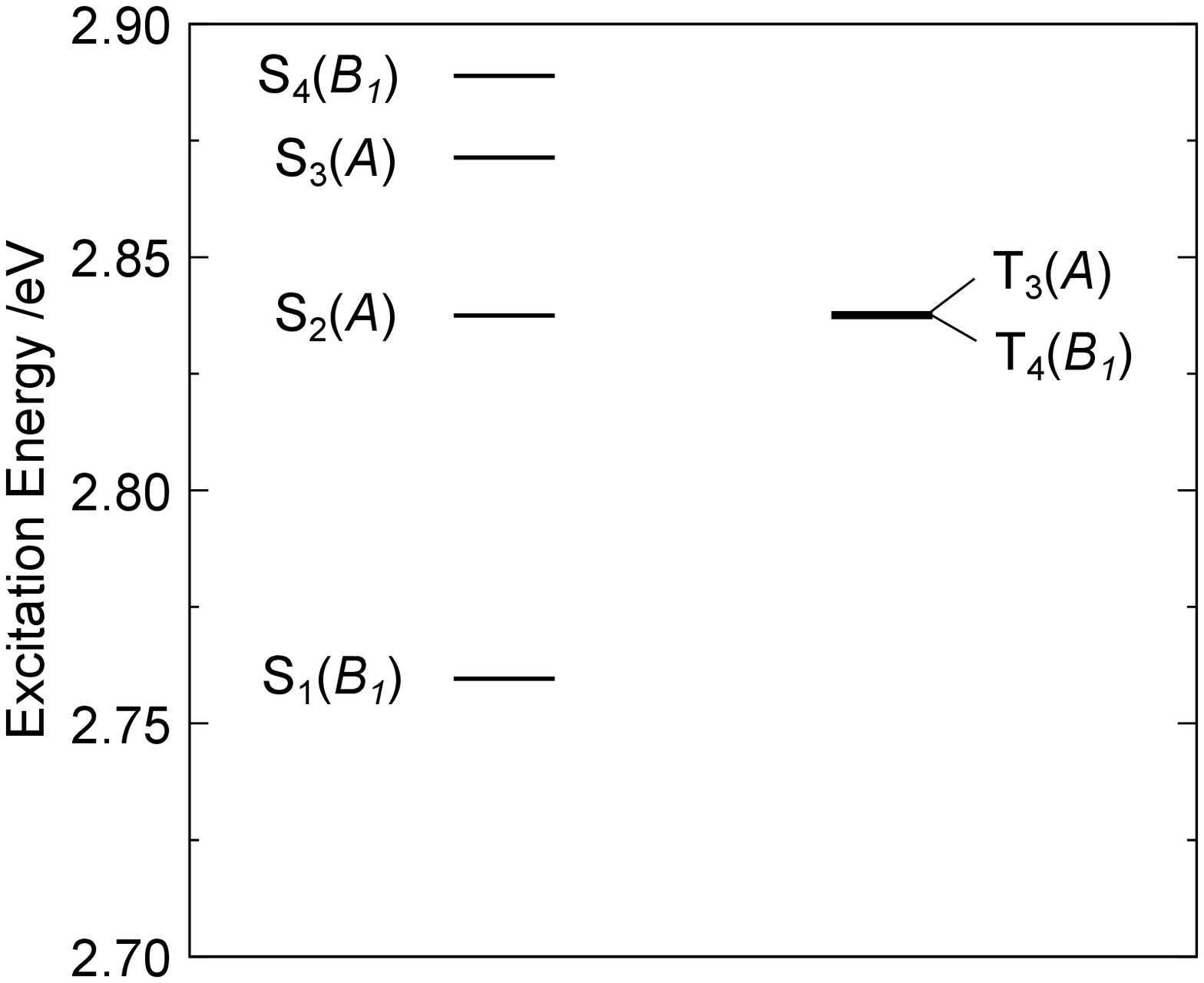}&
\includegraphics[scale=0.45]{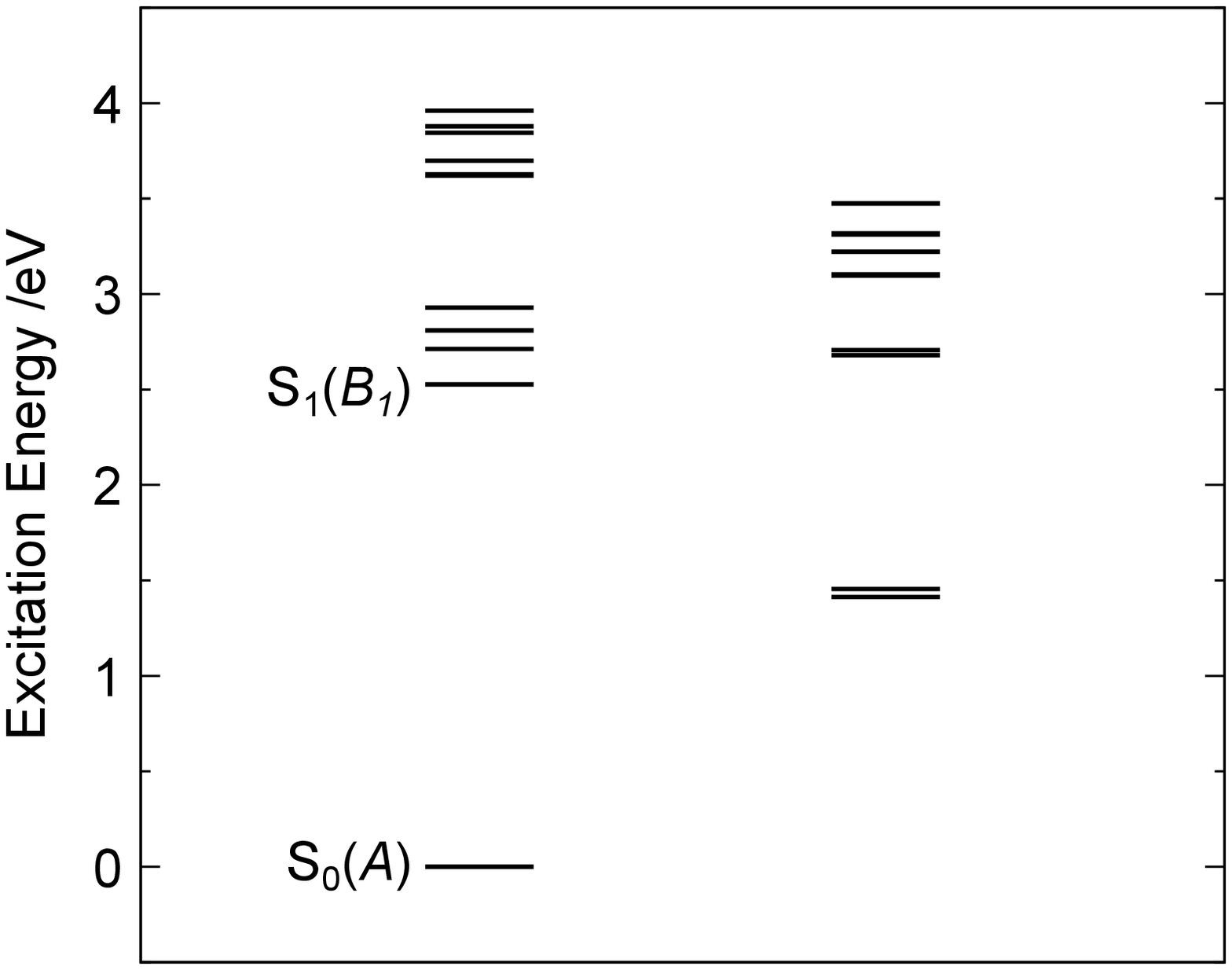}\\
\end{tabular}
\caption{Energy levels of the excited states at the optimized structures
for (a) S$_2$ and (b) S$_1$.
\label{Fig:EnergyLevels@S2S1}}
\end{figure}

FIG. \ref{Fig:EnergyLevels@S2S1} shows the energy levels 
of the excited states at the optimized structures for S$_2$ and S$_1$.
The singlet states at the optimized structure for S$_2$
are tabulated in TABLE \ref{Table:ExcitedStates@S2}.
IC S$_2$ $\rightarrow$ S$_0$ via $a$ modes and IC S$_2$ $\rightarrow$ S$_1$ via $b_1$ modes
are symmetry allowed.
As we will show later, the transition probability of IC between S$_2$ and S$_0$ is small,
while that between S$_2$ and S$_1$ is large.
The electric dipole transition S$_2$ $\rightarrow$ S$_0$ is symmetry forbidden.
On the other hand,
the electric dipole transition S$_2$ $\rightarrow$ S$_1$ is symmetry allowed, and
the TDM is thus large because of a large overlap density between S$_2$ and S$_1$.
Therefore, an S$_2$ exciton is relaxed into the S$_1$ state.
\begin{table}[!h]
\centering
\caption{Singlet excited states at the optimized structure for S$_2$.
\label{Table:ExcitedStates@S2}}
\begin{tabular}{lcccl}
\hline\hline
             &\multicolumn{2}{c}{Excitation Energy}&& Major configuration \\
             & eV & nm & Osc. & (CI coefficient) \\
\hline
% LUMO@S2 = LUMO@S0
S$_4$($B_1$) & 2.8889  & 429.18  & 0.1041 & HO-1$\rightarrow$LU+1(0.698)\\
S$_3$($A$)   & 2.8714  & 431.80  & 0.0000 & HO-1$\rightarrow$LU(0.591),HO$\rightarrow$LU+1(0.379)\\
S$_2$($A$)   & 2.8375  & 436.95  & 0.0000 & HO$\rightarrow$LU+1(0.593),HO-1$\rightarrow$LU(-0.382)\\
S$_1$($B_1$) & 2.7596  & 449.28  & 0.3131 & HO$\rightarrow$LU(0.698)\\
\hline\hline
\end{tabular}
\end{table}

The singlet states at the optimized structure for S$_1$
are listed in TABLE \ref{Table:ExcitedStates@S1}.
S$_1$ belongs to the $B_1$ irrep and is the fluorescent state,
as indicated by the oscillator strengths $f$
listed in TABLE \ref{Table:ExcitedStates@S1}.
\begin{table}[!h]
\centering
\caption{Singlet excited states at the optimized structure for S$_1$.
\label{Table:ExcitedStates@S1}}
\begin{tabular}{lcccl}
\hline\hline
             &\multicolumn{2}{c}{Excitation Energy}&& Major configuration \\
             & eV & nm & Osc. & (CI coefficient) \\
\hline
% slightly changed because of LUMO@S1 = LUMO@S0 * (-1)
S$_4$($B_1$) & 2.9295  & 423.23  & 0.0981 & HO-1$\rightarrow$LU+1(0.704)\\
S$_3$($A$)   & 2.8099  & 441.24  & 0.0000 & HO-1$\rightarrow$LU(-0.546),HO$\rightarrow$LU+1(-0.442)\\
S$_2$($A$)   & 2.7124  & 457.09  & 0.0000 & HO$\rightarrow$LU+1(0.547),HO-1$\rightarrow$LU(-0.444)\\
S$_1$($B_1$) & 2.5267  & 490.70  & 0.4800 & HO$\rightarrow$LU(-0.705)\\
\hline\hline
\end{tabular}
\end{table}

%%%%%%%%%%%%%%%%%%%%%%%%%%%%%%%%%%%%%%%%%%%%%%%%%%%%%%%%%%%%%%%%%%%%%
\subsection{Off-Diagonal Vibronic Coupling Constants\label{SubSec:VCC}}
\begin{figure}[!h]
\centering
\begin{tabular}{ccc}
\multicolumn{1}{l}{{\bf\large (a)}} &
\multicolumn{1}{l}{{\bf\large (b)}} &
\multicolumn{1}{l}{{\bf\large (c)}}\\
\includegraphics[scale=0.27]{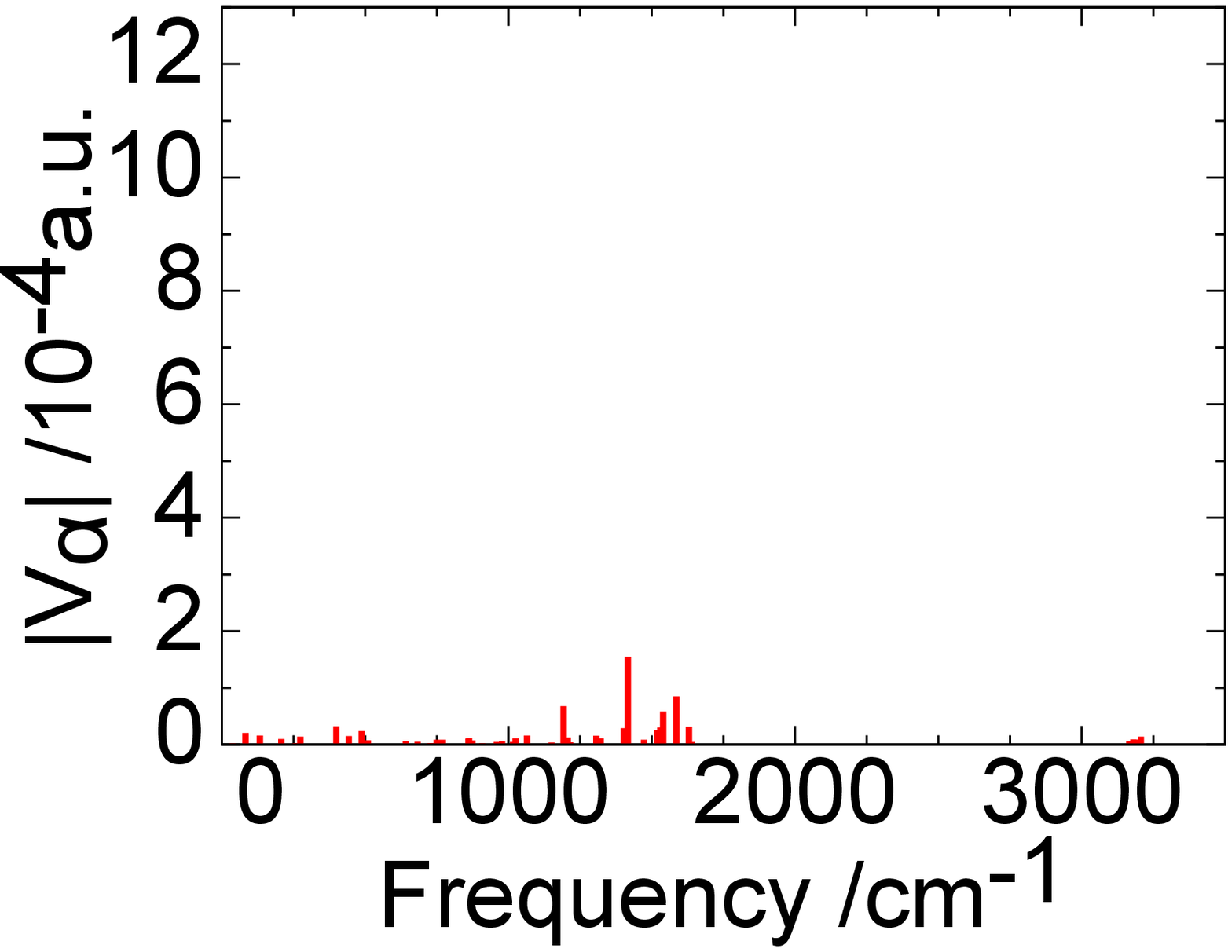} &
\includegraphics[scale=0.27]{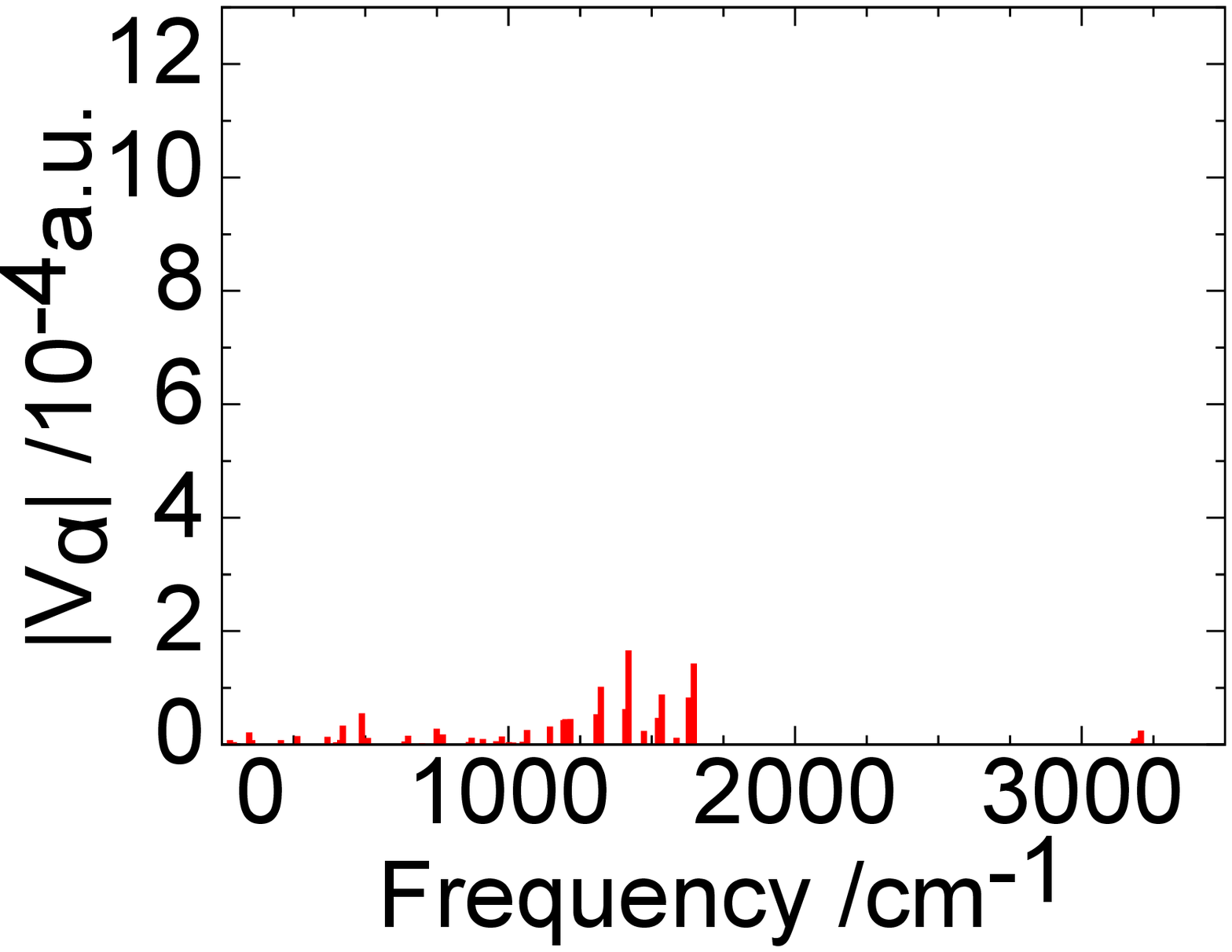} &
\includegraphics[scale=0.27]{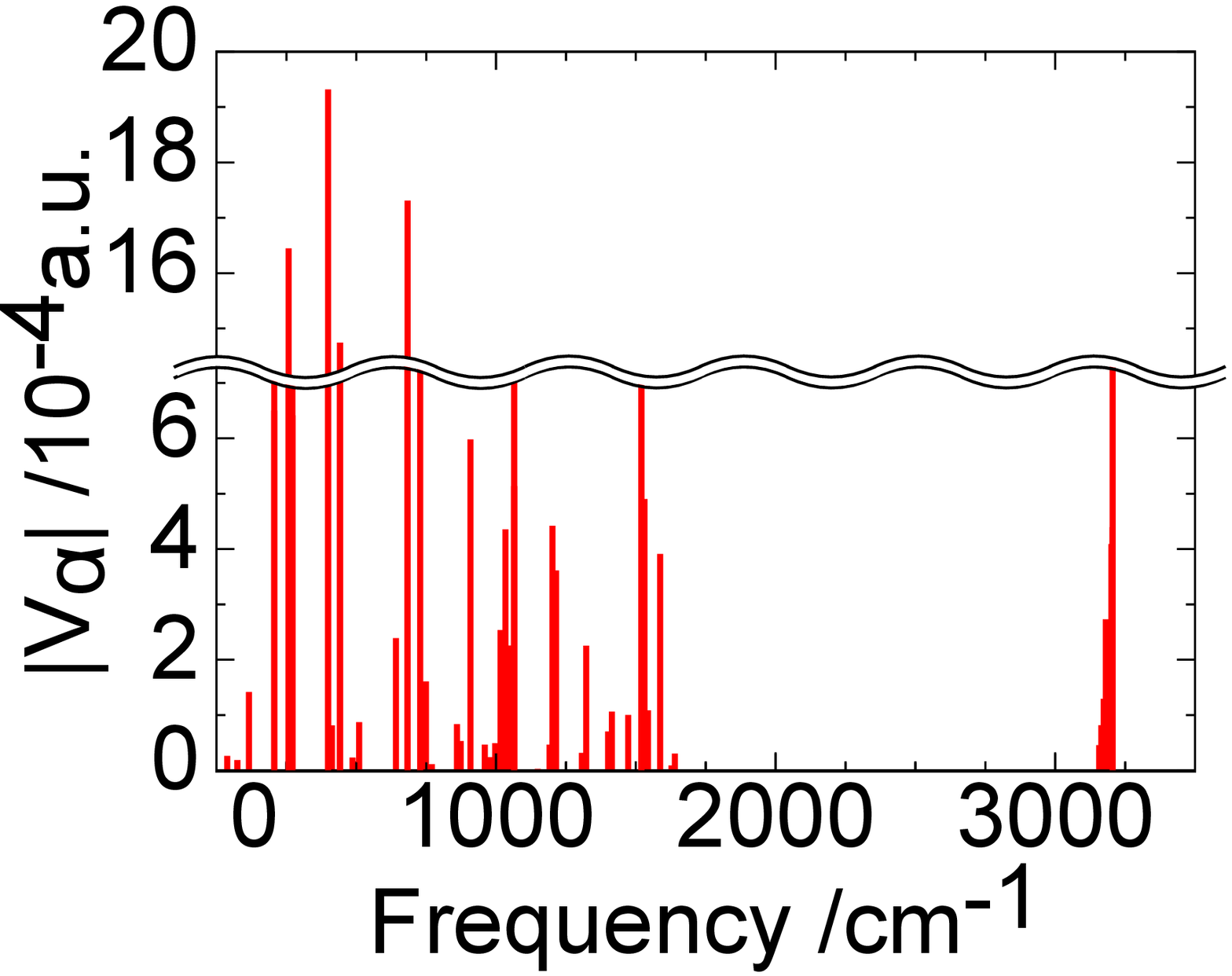} \\
 & & \\
\multicolumn{1}{l}{{\bf\large (d)}} &
\multicolumn{1}{l}{{\bf\large (e)}} &
\multicolumn{1}{l}{{\bf\large (f)}}\\
\includegraphics[scale=0.27]{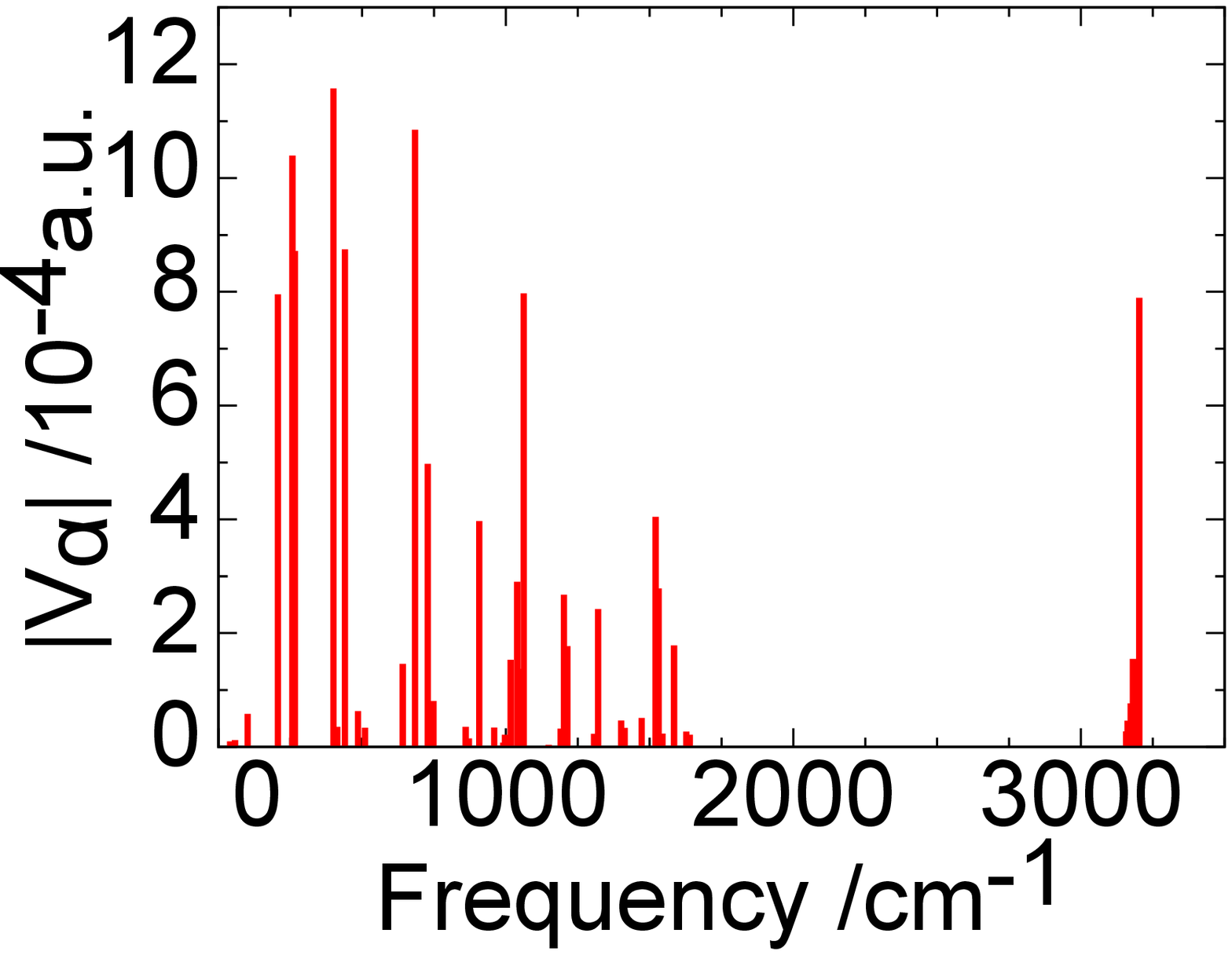} &
\includegraphics[scale=0.27]{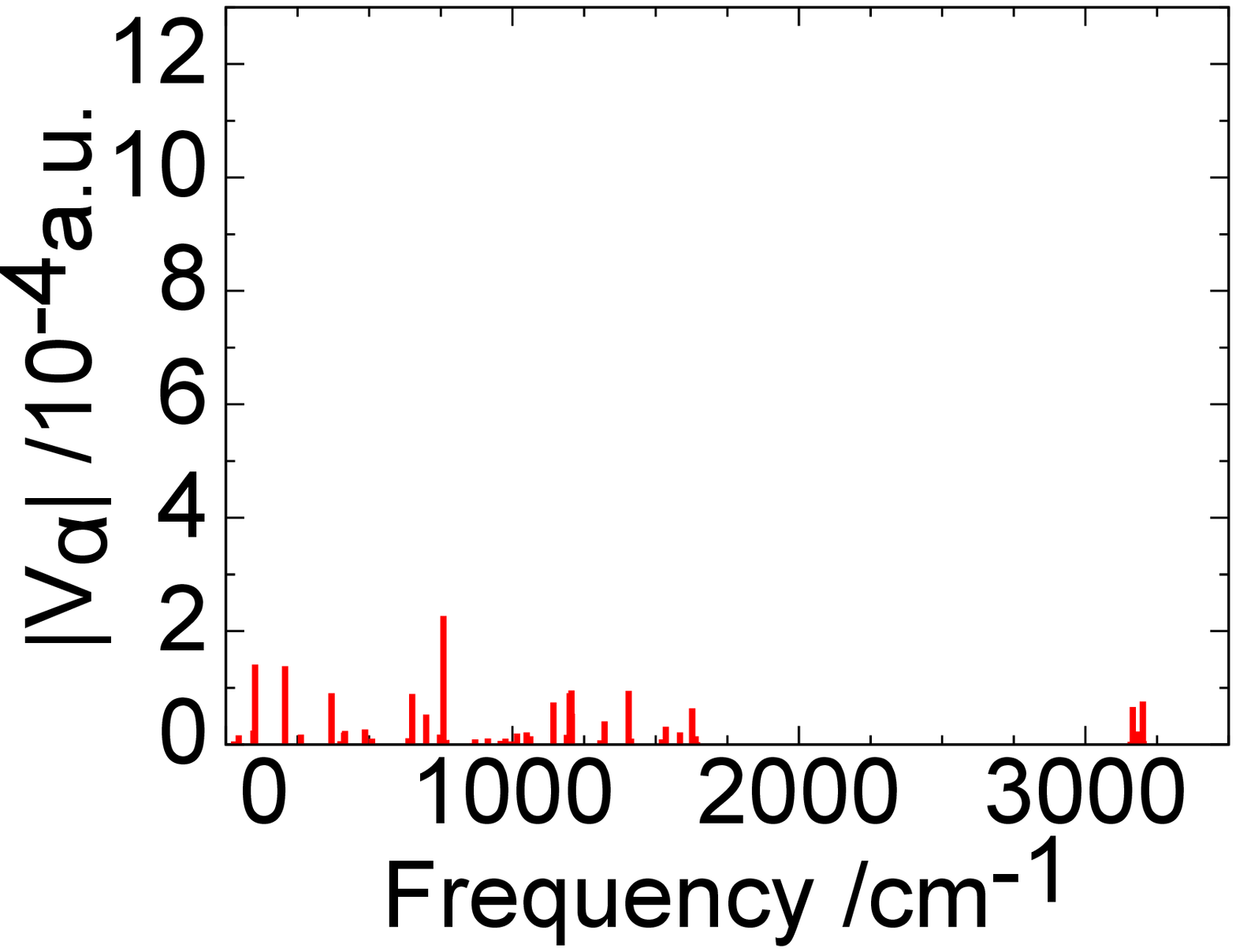} &
\includegraphics[scale=0.27]{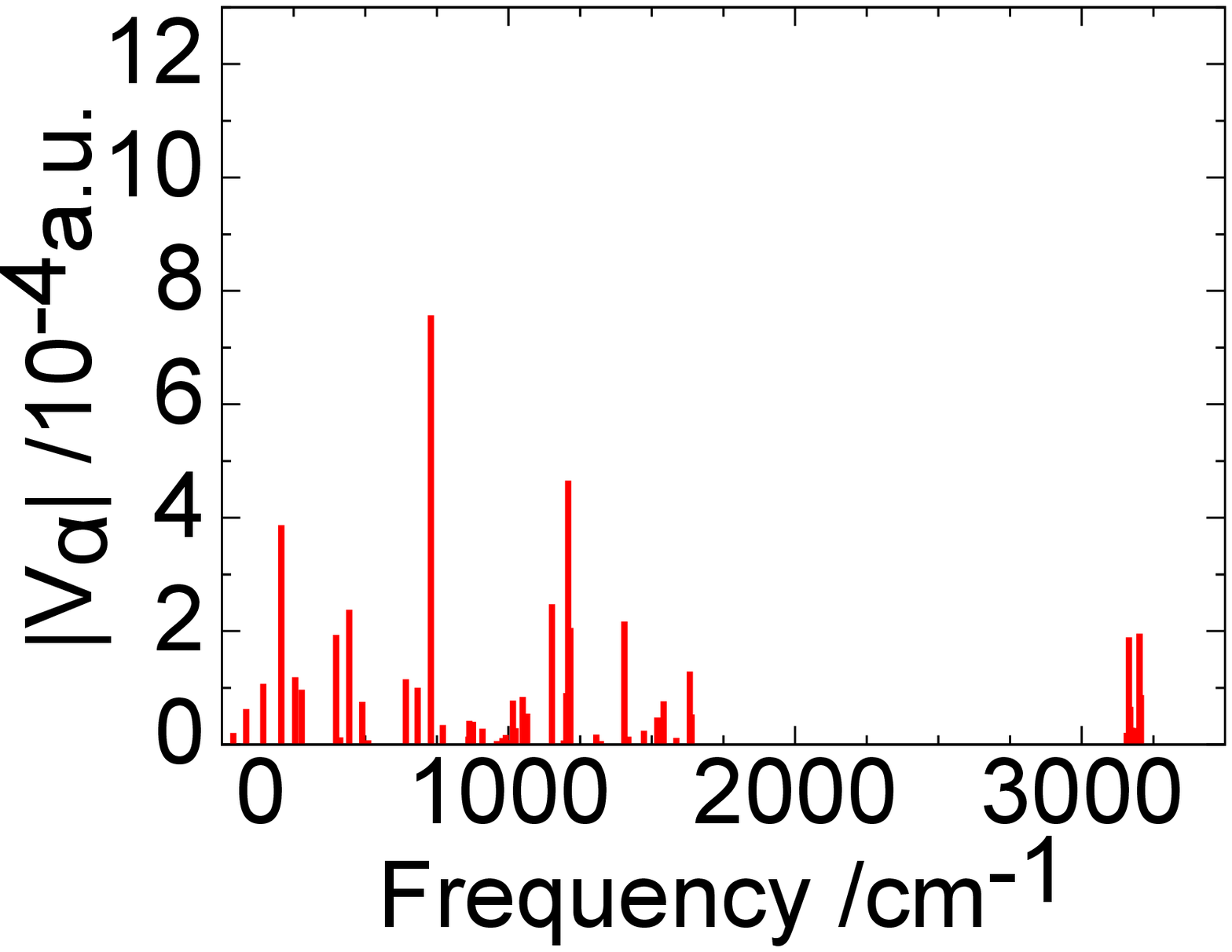} \\
\end{tabular}
\caption{Off-diagonal VCCs of
(a) T$_4$@T$_4$ $\rightarrow$ T$_2$@T$_4$,
(b) T$_4$@T$_4$ $\rightarrow$ T$_1$@T$_4$,
(c) T$_3$@T$_3$ $\rightarrow$ T$_4$@T$_3$,
(d) S$_2$@S$_2$ $\rightarrow$ S$_1$@S$_2$,
(e) S$_2$@S$_2$ $\rightarrow$ S$_0$@S$_2$, and
(f) S$_1$@S$_1$ $\rightarrow$ S$_0$@S$_1$.
@T$_n$/@S$_n$ denote the geometry used in optimization
for the T$_n$/S$_n$ state.
\label{Fig:OffDiagonalVCC}}
\end{figure}

%%%%%%%%%%%%%%%%%%%%%%%%%%%%%%%%%%%%%%%%%%%%%%%%%%%%%%%%%%%%%%%%%%%%%
The calculated off-diagonal VCCs are shown in FIG. \ref{Fig:OffDiagonalVCC}.
Vibrational modes with strong couplings are shown in SEC. S6 %S4
of the SM\footnotemark[1].
From FIG. \ref{Fig:OffDiagonalVCC},
the transition probabilities of the ICs are in the following order:
\begin{equation}
\textrm{T}_3 \rightarrow \textrm{T}_4  >
\textrm{S}_2 \rightarrow \textrm{S}_1  \gg
\textrm{S}_1 \rightarrow \textrm{S}_0  \gg
\textrm{S}_2 \rightarrow \textrm{S}_0  >
\textrm{T}_4 \rightarrow \textrm{T}_1  >
\textrm{T}_4 \rightarrow \textrm{T}_2
.
\label{Eq:ICProbability}
\end{equation}

\begin{figure}[!h]
\centering
\includegraphics[width=0.5\hsize]{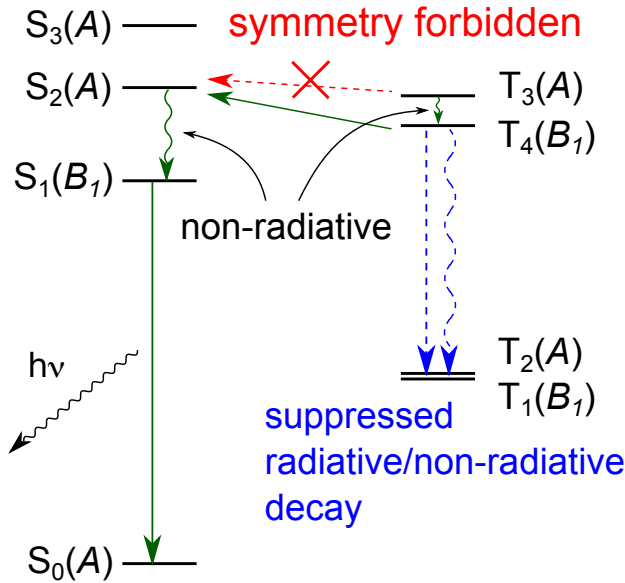}
\caption{Scheme of the excited state dynamics in BD1.
Horizontal solid arrows indicate RISC,
vertical straight arrows indicate radiative transitions, 
and wavy arrows denote non-radiative transitions.
\label{Fig:Excited-StateDynamics}}
\end{figure}

Based on the discussion in SEC. \ref{SubSec:AD} and Eq. \ref{Eq:ICProbability},
the scheme of excited state dynamics is depicted in FIG. \ref{Fig:Excited-StateDynamics}.
For the present FvHT mechanism,
ICs such as
$\textrm{S}_2 \rightarrow \textrm{S}_0$,  
$\textrm{T}_4 \rightarrow \textrm{T}_1$, and
$\textrm{T}_4 \rightarrow \textrm{T}_2$
should be suppressed.
From FIG. \ref{Fig:OffDiagonalVCC} and Eq. \ref{Eq:ICProbability},
the transition probabilities of these undesirable non-radiative processes 
are small.
On the other hand,
the non-radiative transition probabilities
of the ICs S$_2$ $\rightarrow$ S$_1$ and T$_3$ $\rightarrow$ T$_4$ are large.
In addition, electric dipole transition S$_2$ $\rightarrow$ S$_0$
and RISC T$_3$ $\rightarrow$ S$_2$ are symmetry forbidden.
Therefore, the transitions S$_2$ $\rightarrow$ S$_1$
and T$_3$ $\rightarrow$ T$_4$ are preferable.
A T$_3$ exciton is converted into a T$_4$ exciton to yield a S$_2$ exciton
via RISC, and the S$_2$ exciton is then converted into a fluorescent S$_1$ exciton.
Accordingly, we can conclude that
both T$_3$ and T$_4$ excitons are effectively converted 
into a fluorescent S$_1$ exciton. 
It should also be noted that the transition probability 
of IC S$_1$ $\rightarrow$ S$_0$ is small enough for a S$_1$ exciton to emit fluorescence.

%%%%%%%%%%%%%%%%%%%%%%%%%%%%%%%%%%%%%%%%%%%%%%%%%%%%%%%%%%%%%%%%%%%%%
\subsection{Vibronic Coupling Density Analyses\label{SubSec:VCD}}
%%%%%%%%%%%%%%%%%%%%%%%%%%%%%%%%%%%%%%%%%%%%%%%%%%%%%%%%%%%%%%%%%%%%%
In SEC. S7 %S5
of the SM\footnotemark[1],
the VCD analyses for T$_3$--T$_4$, T$_4$--T$_2$, T$_4$--T$_1$,
S$_2$--S$_1$, S$_2$--S$_0$, and S$_1$--S$_0$ are shown.
Note that the isosurface values are different.
As was discussed in SEC. \ref{SubSec:VCC},
the off-diagonal VCCs for T$_3$--T$_4$ and S$_2$--S$_1$ are large.
As shown in FIGs. S10 and S13 in SEC. S7 %S5
of the SM\footnotemark[1],
the VCDs for these VCs are localized on the anthracene moieties, which can be attributed to the large distributions of overlap densities
on the anthracene moieties.
On the other hand,
the off-diagonal VCCs for T$_4$--T$_2$, T$_4$--T$_1$, and S$_2$--S$_0$ are small.
As shown in s S11, S12, and S14,
the VCDs for these VCs are small, which is attributed to the disappearance of overlap densities.
The mechanisms of disappearance are discussed
in SEC. \ref{SubSec:OverlapDensity}.

%%%%%%%%%%%%%%%%%%%%%%%%%%%%%%%%%%%%%%%%%%%%%%%%%%%%%%%%%%%%%%%%%%%%%
\subsection{Overlap Densities\label{SubSec:OverlapDensity}}
\begin{figure}[!h]
\centering
\begin{tabular}{ccc}
\multicolumn{1}{l}{{\bf\large (a)}} &
\multicolumn{1}{l}{{\bf\large (b)}} &
\multicolumn{1}{l}{{\bf\large (c)}}\\
\includegraphics[scale=0.15]{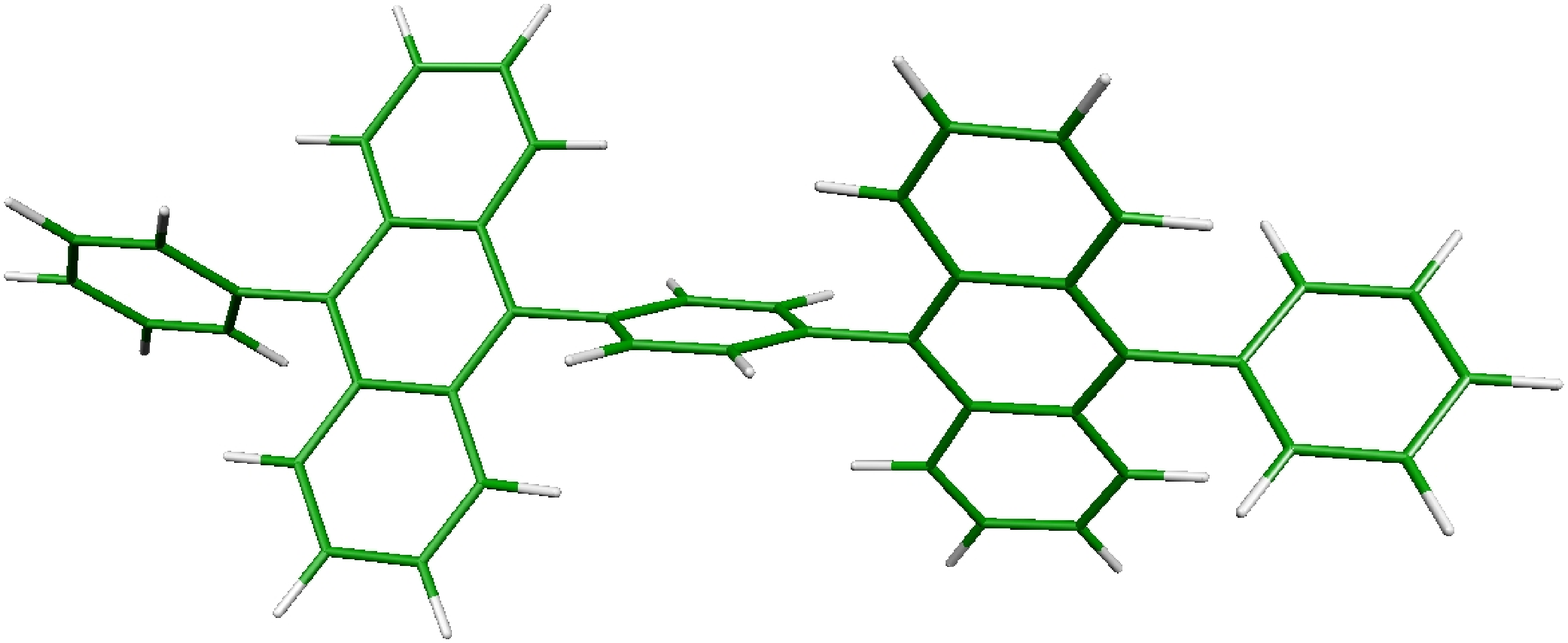} &
\includegraphics[scale=0.15]{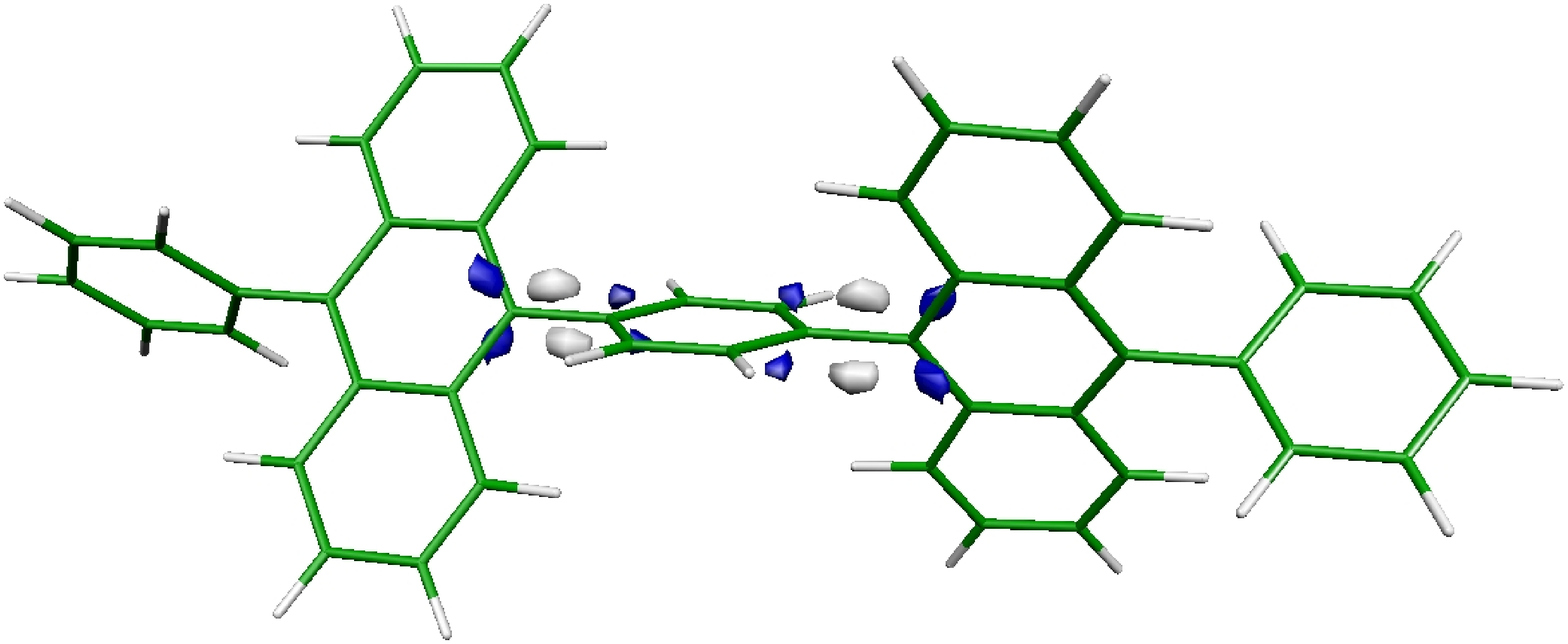} &
\includegraphics[scale=0.15]{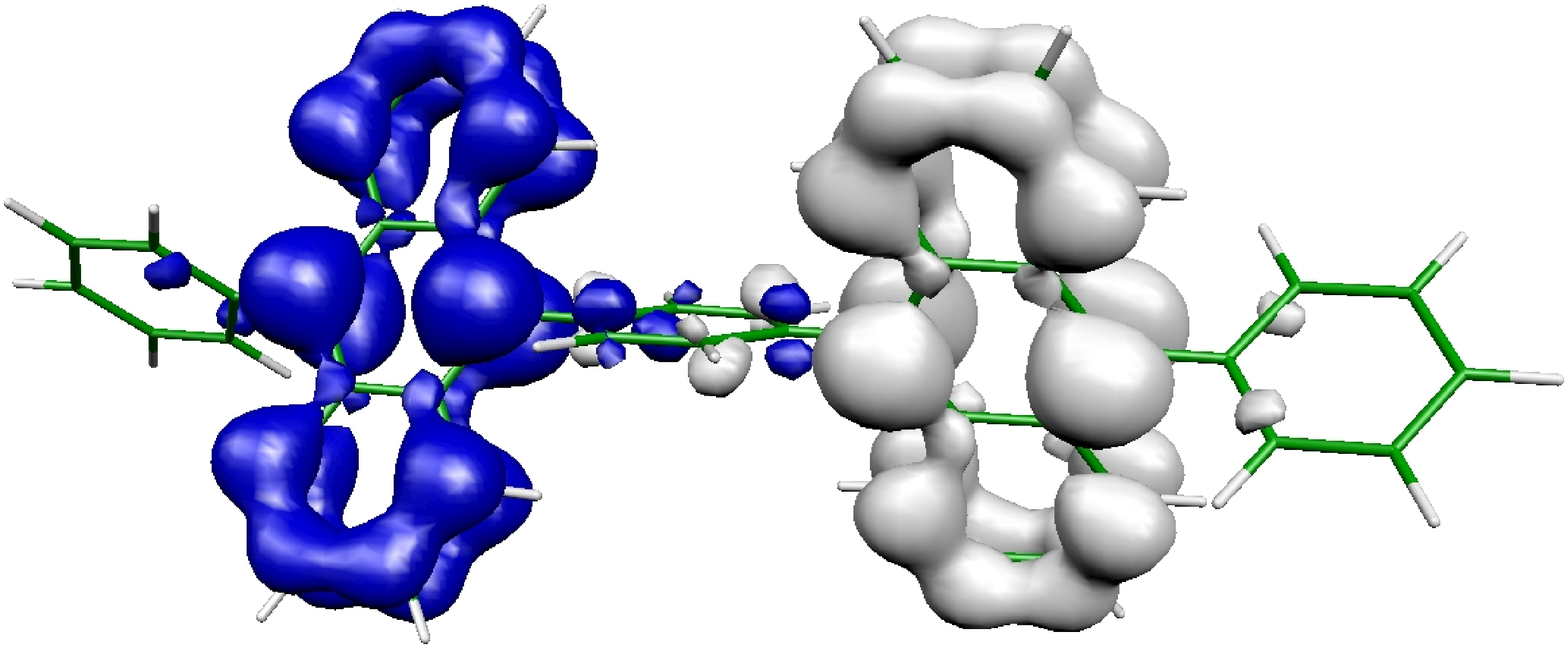} \\
 & & \\
\multicolumn{1}{l}{{\bf\large (d)}} &
\multicolumn{1}{l}{{\bf\large (e)}} &
\multicolumn{1}{l}{{\bf\large (f)}}\\
\includegraphics[scale=0.15]{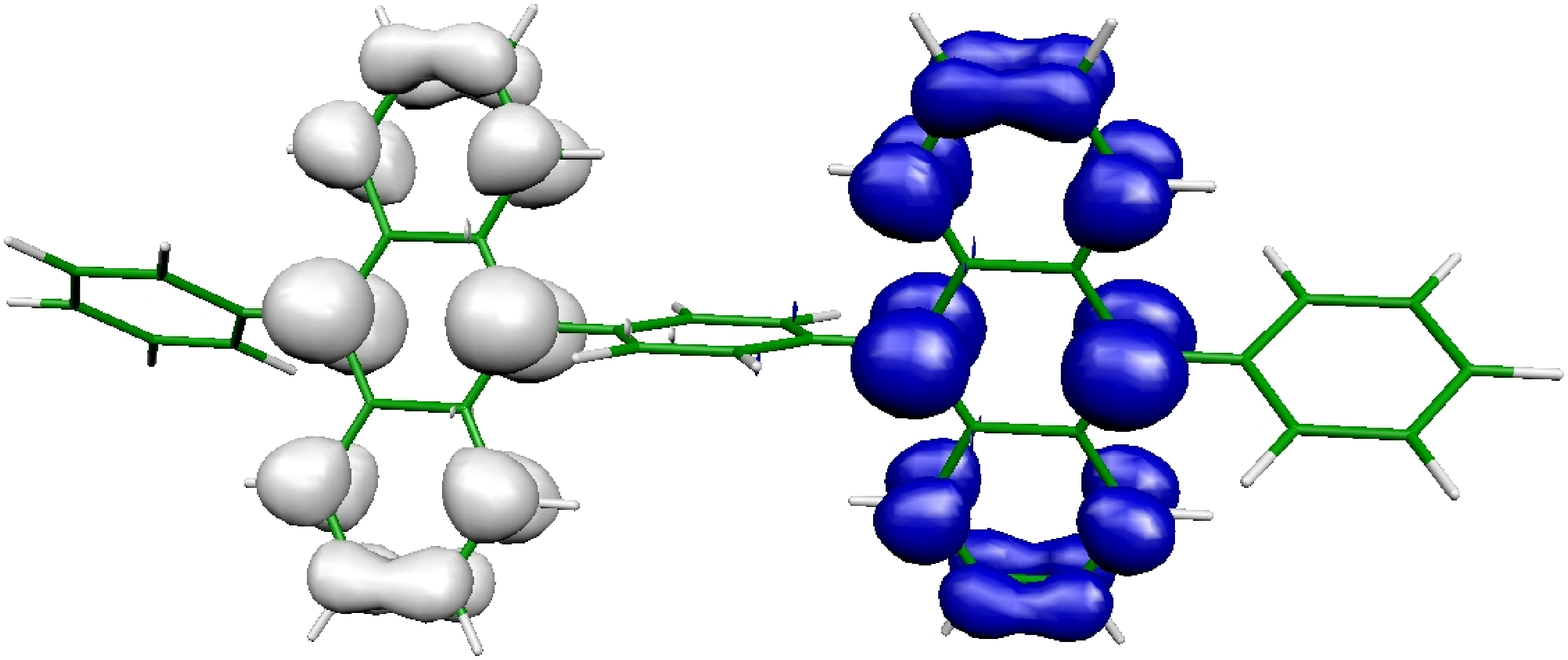} &
\includegraphics[scale=0.15]{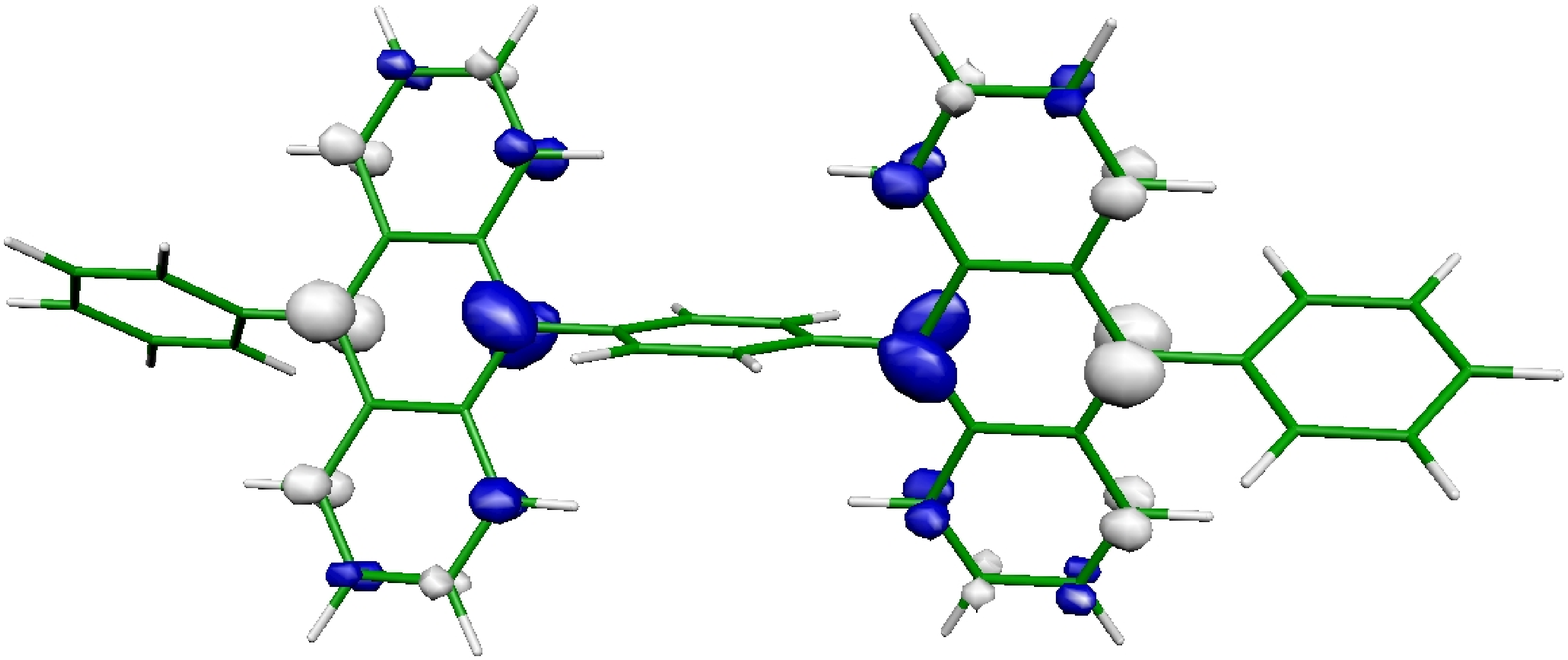} &
\includegraphics[scale=0.15]{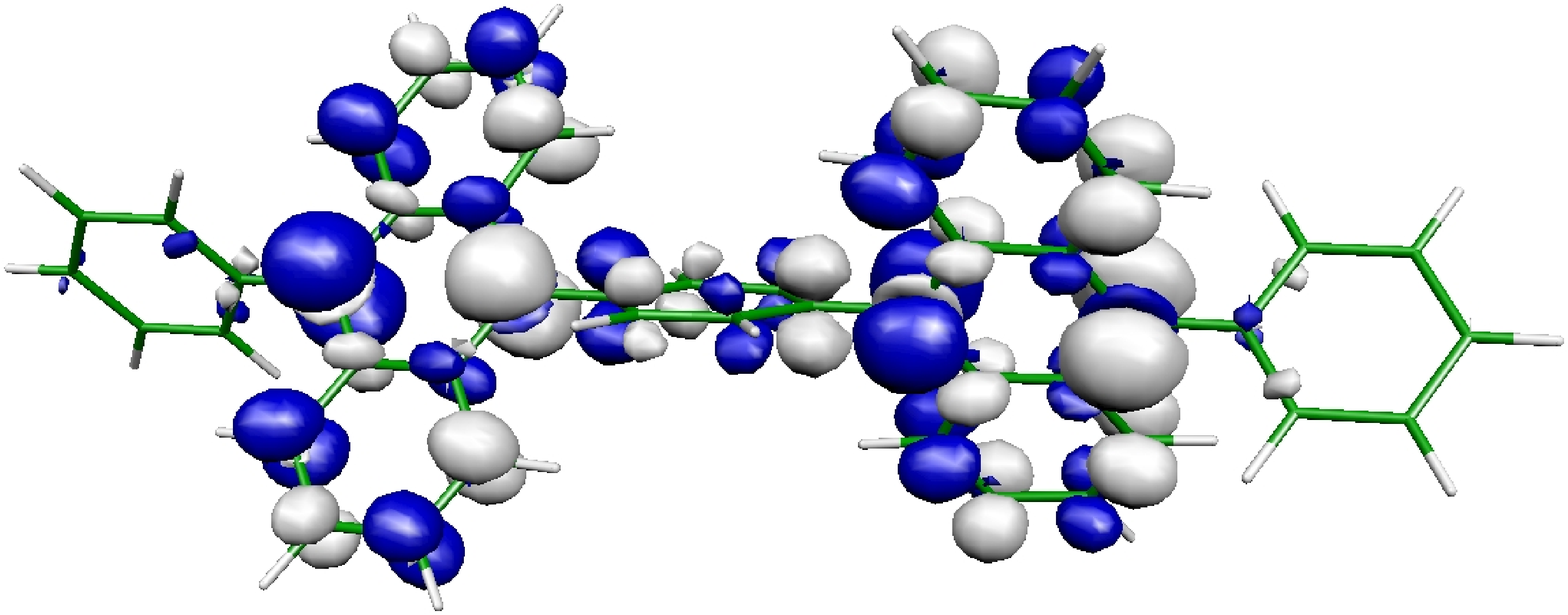} \\
\end{tabular}
\caption{Overlap densities for
(a) T$_4$@T$_4$--T$_2$@T$_4$,
(b) T$_4$@T$_4$--T$_1$@T$_4$,
(c) T$_3$@T$_3$--T$_4$@T$_3$,
(d) S$_2$@S$_2$--S$_1$@S$_2$,
(e) S$_2$@S$_2$--S$_0$@S$_2$, and
(f) S$_1$@S$_1$--S$_0$@S$_1$,
respectively.
All isosurface values are the same, i.e. $1.0 \times 10^{-3}$ a.u.
\label{Fig:OverlapDensities}}
\end{figure}

%%%%%%%%%%%%%%%%%%%%%%%%%%%%%%%%%%%%%%%%%%%%%%%%%%%%%%%%%%%%%%%%%%%%%
FIG. \ref{Fig:OverlapDensities} shows the overlap densities having the same isosurface values.
Among them, the overlap densities for T$_4$--T$_2$,
T$_4$--T$_1$, and S$_2$--S$_0$ disappear.
In this section, we discuss the disappearance of these overlap densities. 

A TD-DFT wave function is given by
\begin{equation}
\Psi_{n}=\sum_{i\in \text{occ}}\sum_{r\in \text{unocc}}C_{i}^{r} \Phi_{i}^{r}
,
\label{Eq:TDDFTWavefunction}
\end{equation}
where $\Phi_{i}^{r}$ represents the electronic configuration of a single-electron excitation 
from occupied orbital $i$ to vacant orbital $r$, and $C_{i}^{r}$ represents the CI coefficient.

The overlap density between two excited states $\Psi_{m}$ and $\Psi_{n}$, $\rho^{mn}$, is given by
\begin{equation}
\rho^{mn}
=
\sum_{i\in \text{occ}}\sum_{r\in \text{unocc}}
\sum_{j\in \text{occ}}\sum_{s\in \text{unocc}}
D_{j}^{s\ast}C_{i}^{r}\rho(\Phi_{j}^{s},\Phi_{i}^{r})
,
\label{Eq:OverlapDensityConfigurations}
\end{equation}
where $D_{j}^{s}$ is the CI coefficient of another excited state $m$,
and $\rho(\Phi_{j}^{s},\Phi_{i}^{r})$ denotes the overlap density between the two configurations.

$\Phi_0$ represents a ground state electron configuration, and its overlap density $\rho_0$ is given by
\begin{equation}
\rho(\Phi_0,\Phi_0)=:\rho_0
.
\label{Eq:OverlapGroundConfiguration}
\end{equation}
The overlap densities for various configurations are summarized as follows,
\begin{equation}
\rho(\Phi_i^r,\Phi_i^r)=\rho_0 - |\psi_i|^2 + |\psi_r|^2, \quad
\rho(\Phi_0,\Phi_i^r)=\psi_i^\ast\psi_r, \quad
\rho(\Phi_i^r,\Phi_j^r)=\psi_i^\ast\psi_j \quad (i\ne j),
\label{Eq:OverlapConfigurations1}
\end{equation}
\begin{equation}
\rho(\Phi_i^r,\Phi_i^s)=\psi_r^\ast\psi_s \quad (r\ne s), \quad
\rho(\Phi_i^r,\Phi_j^s)=0 \quad (i\ne j, r\ne s)
,
\label{Eq:OverlapConfigurations2}
\end{equation}
where $\psi$ represents a molecular orbital.

According to TABLEs
\ref{Table:ExcitedStates@T3},
\ref{Table:ExcitedStates@T4},
\ref{Table:ExcitedStates@S2}, and
\ref{Table:ExcitedStates@S1},
the approximate wave functions for the relevant excited states can be written as follows:
\begin{eqnarray}
\Psi_a&=&C_1^{a}\Phi_{\text{HO}}^{\text{LU}} + C_2^{a}\Phi_{\text{NHO}}^{\text{NLU}} 
%\approx c \Phi_{\text{HO}}^{\text{LU}} - c \Phi_{\text{NHO}}^{\text{NLU}} 
,
\quad (\textrm{T}_4@\textrm{T}_4)
\label{Eq:Psia}
\\
\Psi_b&=&C_3^{b}\Phi_{\text{HO}}^{\text{NLU}} + C_4^{b}\Phi_{\text{NHO}}^{\text{LU}} 
%\approx c \Phi_{\text{HO}}^{\text{NLU}} - c \Phi_{\text{NHO}}^{\text{LU}} 
,
\quad (\textrm{S}_2@\textrm{S}_2)
\label{Eq:Psib}
\\
\Psi_c&=&C_1^{c}\Phi_{\text{HO}}^{\text{LU}} + C_2^{c}\Phi_{\text{NHO}}^{\text{NLU}}
%\approx c \Phi_{\text{HO}}^{\text{LU}} + c \Phi_{\text{NHO}}^{\text{NLU}}
,
\quad (\textrm{T}_1@\textrm{T}_4)
\label{Eq:Psic}
\\
\Psi_{d}&=&\Phi_{\textrm{HO}}^{\textrm{LU}}
,
\quad (\textrm{S}_1@\textrm{S}_2)
\label{Eq:Psid}
\\
\Psi_e&=&C_3^{e}\Phi_{\text{HO}}^{\text{NLU}} + C_4^{e}\Phi_{\text{NHO}}^{\text{LU}} 
%\approx c \Phi_{\text{HO}}^{\text{NLU}} + c \Phi_{\text{NHO}}^{\text{LU}}
,
\quad (\textrm{T}_2@\textrm{T}_4)
\label{Eq:Psie}
\\
\Psi_{0}&=&\Phi_{0}
,
\quad (\textrm{S}_0@\textrm{S}_2)
\label{Eq:Psi0}
\end{eqnarray}
where the set of the CI coefficients is assumed to satisfy the following relation:
\begin{equation}
C_1^{a}\approx -C_2^{a}\approx C_3^{b}\approx -C_4^{b}
\approx C_1^{c}\approx C_2^{c} \approx C_3^{e}\approx C_4^{e} \approx c
.
\label{Eq:CIConditions}
\end{equation}
In addition, we should recall the conditions of orbital overlap densities
among the frontier orbitals discussed in SEC. \ref{SubSec:Orbital}.
In general, these conditions for excited wave functions can be satisfied 
in systems with pseudo-degenerate excited electronic states.

In order to elucidate the reason for the disappearance
of the overlap densities shown in FIG. \ref{Fig:OverlapDensities},
we discuss the overlap densities between
approximate excited wave functions.

\noindent
(Case 1: $\Psi_a$ and $\Psi_e$) This case corresponds to T$_4$--T$_2$.
The overlap density between $\Psi_a$ and $\Psi_e$
is given by
\begin{eqnarray}
\rho^{ae}
&=&
%C_1^{a}C_3^{e}\rho(\Phi_{\text{HO}}^{\text{LU}},\Phi_{\text{HO}}^{\text{NLU}})
%+
%C_1^{a}C_4^{e}\rho(\Phi_{\text{HO}}^{\text{LU}},\Phi_{\text{NHO}}^{\text{LU}})
%+
%C_2^{a}C_3^{e}\rho(\Phi_{\text{NHO}}^{\text{NLU}},\Phi_{\text{HO}}^{\text{NLU}})
%+
%C_2^{a}C_4^{e}\rho(\Phi_{\text{NHO}}^{\text{NLU}},\Phi_{\text{NHO}}^{\text{LU}})
%\nonumber
%\\
%&=&
(C_1^{a}C_3^{e} + C_2^{a}C_4^{e})\psi_{\text{LU}}\psi_{\text{NLU}}
+
(C_1^{a}C_4^{e} + C_2^{a}C_3^{e})\psi_{\text{HO}}\psi_{\text{NHO}}
.
\end{eqnarray}
From the condition for the CI coefficients
(Eq. \ref{Eq:CIConditions}),
\begin{equation}
C_1^{a}C_3^{e} + C_2^{a}C_4^{e}
\approx
%c^2 - c^2
%=
0
,
\quad
C_1^{a}C_4^{e} + C_2^{a}C_3^{e}
\approx
%c^2 - c^2
%=
0
.
\end{equation}
Thus, the overlap density between $\Psi_a$ and $\Psi_e$ is cancelled out.
Note that the disappearance of overlap density originates from the condition of the CI coefficients.

\noindent
(Case 2: $\Psi_a$ and $\Psi_c$) This case corresponds to T$_4$--T$_1$.
The overlap density between $\Psi_a$ and $\Psi_c$
is given by
\begin{eqnarray}
\rho^{ac}
&=&
%C_1^{a}C_1^{c}\rho(\Phi_{\text{HO}}^{\text{LU}},\Phi_{\text{HO}}^{\text{LU}})
%+
%C_1^{a}C_2^{c}\rho(\Phi_{\text{HO}}^{\text{LU}},\Phi_{\text{NHO}}^{\text{NLU}})
%+
%C_2^{a}C_1^{c}\rho(\Phi_{\text{NHO}}^{\text{NLU}},\Phi_{\text{HO}}^{\text{LU}})
%+
%C_2^{a}C_2^{c}\rho(\Phi_{\text{NHO}}^{\text{NLU}},\Phi_{\text{NHO}}^{\text{NLU}})
%\nonumber
%\\
%&=&
C_1^{a}C_1^{c}\left(\rho_0 - |\psi_{\text{HO}}|^2 + |\psi_{\text{LU}}|^2\right)
+
C_2^{a}C_2^{c}\left(\rho_0 - |\psi_{\text{NHO}}|^2 + |\psi_{\text{NLU}}|^2\right)
.
\end{eqnarray}
From the condition for the CI coefficients,
\begin{eqnarray}
\rho^{ac}
&\approx&
%c^2 \left(\rho_0 - |\psi_{\text{HO}}|^2 + |\psi_{\text{LU}}|^2\right)
%-
%c^2 \left(\rho_0 - |\psi_{\text{NHO}}|^2 + |\psi_{\text{NLU}}|^2\right)
%\nonumber
%\\
%&=&
c^2 \left( |\psi_{\text{NHO}}|^2 - |\psi_{\text{HO}}|^2 + |\psi_{\text{LU}}|^2 - |\psi_{\text{NLU}}|^2\right)
.
\end{eqnarray}
From the condition for the orbital densities,
Eqs. \ref{Eq:SquareLUNLU} and \ref{Eq:SquareHONHO}
\begin{equation}
\rho^{ac}
\approx
0.
\end{equation}
Note that the disappearance of overlap density originates from
the condition of the orbital densities of the frontier orbitals
as well as that of the CI coefficients.

\begin{figure}[!h]
\centering
\begin{tabular}{cc}
\multicolumn{1}{l}{{\bf\large (a)}} &
\multicolumn{1}{l}{{\bf\large (b)}}\\
\includegraphics[scale=0.15]{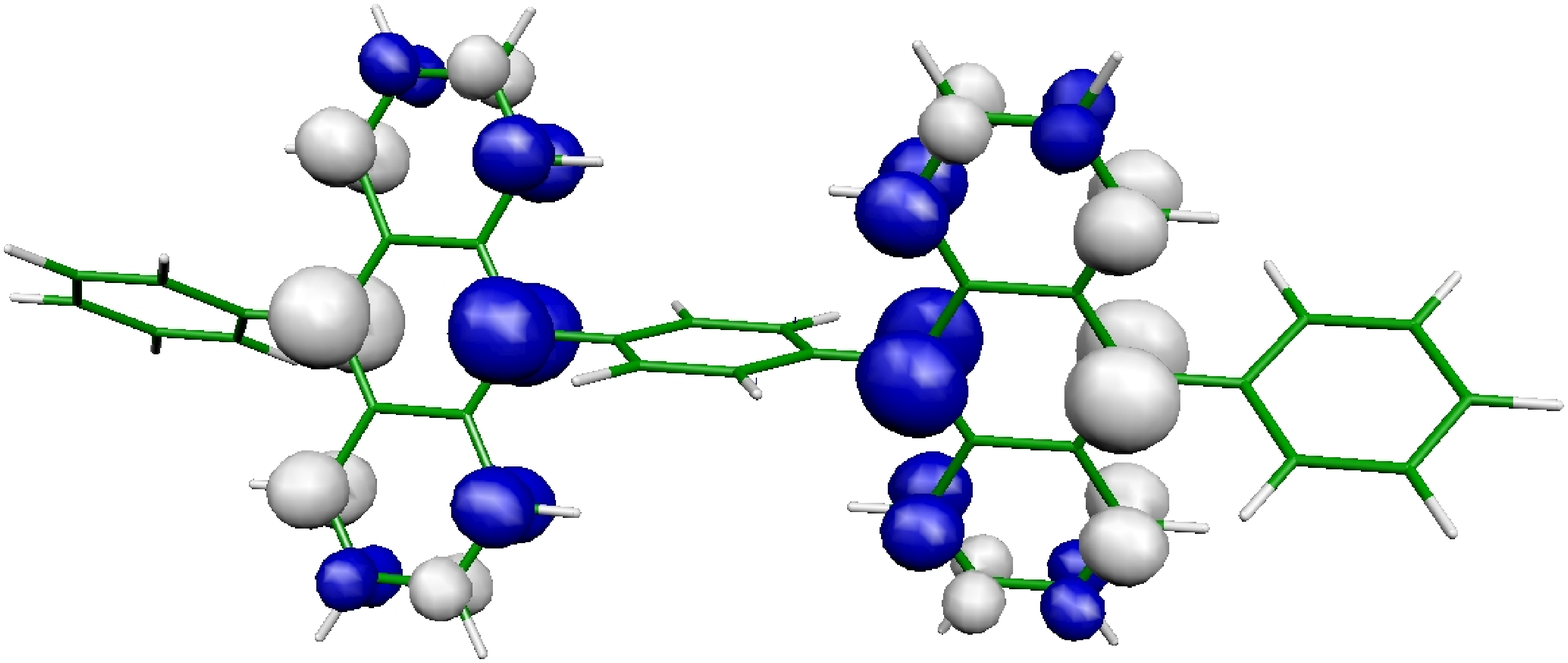}  &
\includegraphics[scale=0.15]{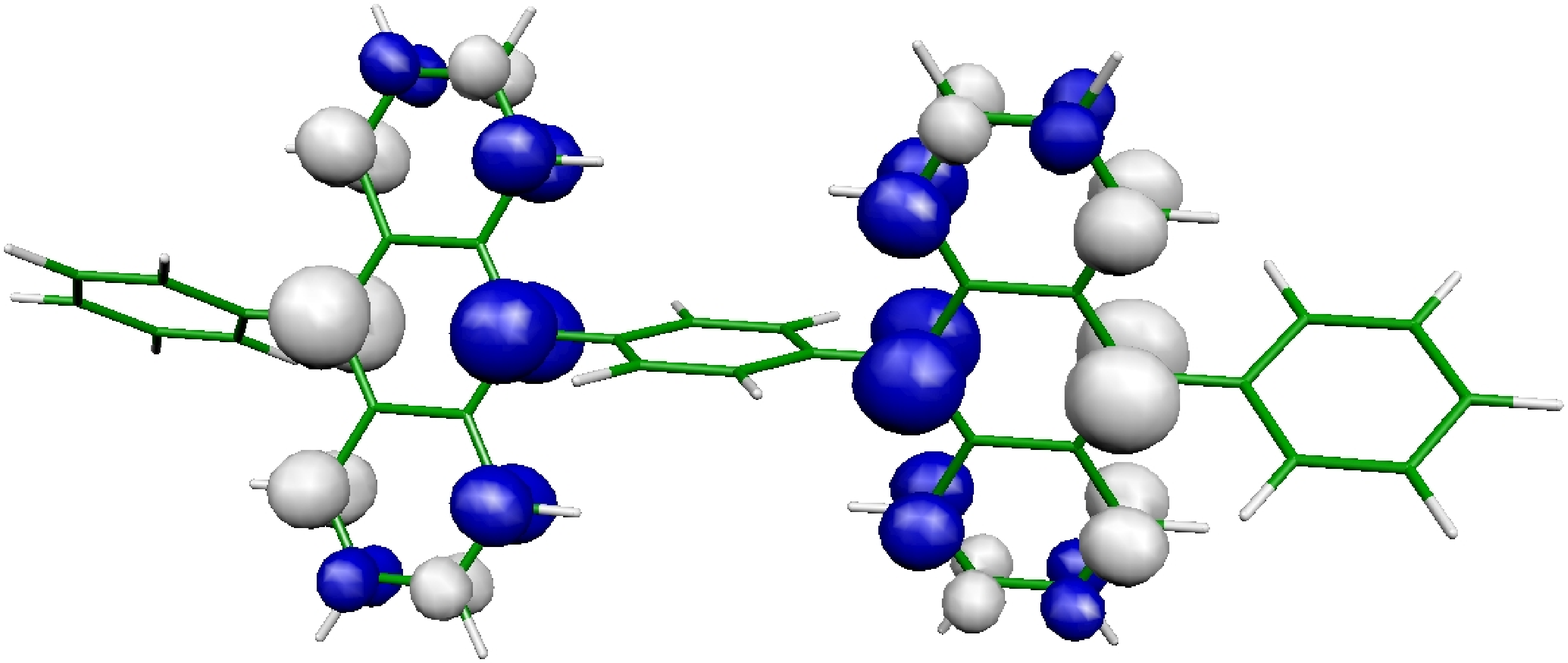}\\
\end{tabular}
\caption{Orbital overlap densities,
(a) $\psi_{\rm HO}\psi_{\rm NLU}$ and (b) $\psi_{\rm NHO}\psi_{\rm LU}$,
at the optimized structure for S$_2$.
The isosurface value is $1.0 \times 10^{-3}$ a.u.
\label{Fig:OrbitalOverlapDensity}}
\end{figure}

\noindent
(Case 3: $\Psi_{b}$ and $\Psi_{0}$) This case corresponds to S$_2$--S$_0$.
The overlap density between $\Psi_{b}$ and $\Psi_{0}$
is given by
\begin{equation}
\rho^{b0}
%=
%C_{3}^{b}\rho(\Phi_{\textrm{HO}}^{\textrm{NLU}},\Phi_{0})
%+
%C_{4}^{b}\rho(\Phi_{\textrm{NHO}}^{\textrm{LU}},\Phi_{0})
\approx
c(\psi_{\textrm{HO}}\psi_{\textrm{NLU}} - \psi_{\textrm{NHO}}\psi_{\textrm{LU}})
.
\end{equation}
According to Eq. \ref{Eq:OverlapHONLU-NHOLU}
and FIG. \ref{Fig:OrbitalOverlapDensity},
$\rho^{b0}$ is cancelled out.
Note that the disappearance of overlap density originates from
the condition of the orbital overlap densities of the frontier orbitals
as well as that of the CI coefficients.

\noindent
(Case 4: $\Psi_{b}$ and $\Psi_{d}$) This case corresponds to S$_2$--S$_1$.
According to Eq. \ref{Eq:OverlapHONHO-LUNLU},
the overlap density between $\Psi_{b}$ and $\Psi_{d}$
is given by
\begin{equation}
\rho^{bd}
%=
%C_{3}^{b}\rho(\Phi_{\textrm{HO}}^{\textrm{NLU}},\Phi_{\textrm{HO}}^{\textrm{LU}})
%+
%C_{4}^{b}\rho(\Phi_{\textrm{NHO}}^{\textrm{LU}},\Phi_{\textrm{HO}}^{\textrm{LU}})
\approx
c(\psi_{\textrm{NLU}}\psi_{\textrm{LU}} - \psi_{\textrm{NHO}}\psi_{\textrm{HO}})
\ne 0.
\end{equation}
Thus, $\rho^{bd}$ is not cancelled out.

General cases are discussed in SEC. S8 %S6
of the SM\footnotemark[1].

These reduced overlap densities, which originate from the pseudo degeneracy, 
are responsible for the suppression of
undesirable radiative and non-radiative transitions, 
T$_3$ $\rightarrow$ T$_2$, T$_3$ $\rightarrow$ T$_1$, and S$_2$ $\rightarrow$ S$_0$, for the FvHT mechanism
because TDM and VCC depend on overlap density.

%%%%%%%%%%%%%%%%%%%%%%%%%%%%%%%%%%%%%%%%%%%%%%%%%%%%%%%%%%%%%%%%%%%%%
\section{Conclusion\label{Sec:Conclusion}}
%%%%%%%%%%%%%%%%%%%%%%%%%%%%%%%%%%%%%%%%%%%%%%%%%%%%%%%%%%%%%%%%%%%%%
To elucidate the high EQE observed in OLEDs using BD1,
we calculated the off-diagonal VCCs and performed VCD analyses.
The findings based on the calculations and analyses are as follows:
\begin{enumerate}
\item The large off-diagonal VCCs between T$_3$--T$_4$
cause the non-radiative transition from T$_3$ to T$_4$.
\item As the overlap densities between T$_4$--T$_2$/T$_4$--T$_1$
are small, radiative/non-radiative processes are suppressed.
\item RISC from T$_4$ to S$_2$ is symmetry allowed and the energy gap between them is small 
($\Delta E_{S_2 - T_4}=21$ meV).
\item Owing to the large overlap density between S$_2$--S$_1$, 
radiative/non-radiative relaxation occurs from S$_2$ to S$_1$,
which results in emission from S$_1$.
\item The undesirable radiative/non-radiative transitions
can be suppressed by utilizing the pseudo-degeneracy.
Therefore, RISC via T$_4$ can be the dominant pathway.
\end{enumerate}

We employed the overlap densities calculated from $N$-electron wave functions
to take the multi-configurational property of the excited states into consideration.
This property cannot be captured using only a transition density (orbital overlap density).

We proposed a {\it fluorescence via higher triplets (FvHT)} emitting mechanism for OLEDs based on a bisanthracene derivative, BD1.
This mechanism is valid 
as long as all transitions from T$_m$ ($m>1$) to all lower T$_n$
($m>n \ge 1$)
are suppressed.
In BD1, we found that this condition is satisfied 
because of its pseudo-degenerate electronic structure.
The excited electronic structure of BD1 
is different from that of PTZ-BZP 
in which the LUMO and NLUMO are pseudo-degenerate, 
but the HOMO and NHOMO are not.
We also discussed the general conditions for the disappearance
of the overlap densities  in the pseudo-degenerate system.
The general conditions are applicable not only for transitions in a molecule,
but also for exciton migrations in the solid phase.

The concepts of iST and SC-TADF are based on the selection rules for a molecular symmetry group. 
In contrast, the design principle based on the FvHT mechanism allows the use of asymmetric molecules.
In fact, the molecular structure of BD1 shows $D_2$ symmetry and is thus not suitable as an iST nor SC-TADF molecule,
as expressed in Eq. \ref{Eq:OrderPointGroups}.
(If the dihedral angle of BD1 were to be the right angle, 
BD1 could show $D_{2h}$ symmetry and
exhibit iST or SC-TADF. See SEC. S9 %S7
of the SM\footnotemark[1].) 

Finally, we propose a superordinate concept, {\it fluorescence via RISC (FvRISC)}
from T$_1$ or higher triplet states. 
This concept includes TADF, SC-TADF, iST, and FvHT. 
The concept of FvRISC enables us 
to overcome the singlet exciton formation ratio 
of 25 \% for electrical excitations
and to realize highly efficient OLEDs.

\begin{acknowledgments}
% put your acknowledgments here.
We thank Prof. Hirofumi Sato.
Numerical calculations were partly performed at the
Supercomputer Laboratory of Kyoto University and at the
Research Center for Computational Science, Okazaki,
Japan. This study was also supported by a Grant-in-Aid
for Scientific Research (C) (15K05607) from the Japan
Society for the Promotion of Science (JSPS).
\end{acknowledgments}

% Create the reference section using BibTeX:
\bibliography{manuscript}

\end{document}

\maketitle
\footnotetext{
 \it 
 $^{a}$~
 Department of Molecular Engineering,
 Graduate School of Engineering,
 Kyoto University,
 Nishikyo-ku,
 Kyoto 615-8510,
 \vspace{-0.2cm}\\
 Japan;
 Tel:  +81-75-383-2544;
 Fax:  +81-75-383-2799;
 E-mail: tsato@moleng.kyoto-u.ac.jp
 }
\footnotetext{
 \it 
 $^{b}$~
 Unit of Elements Strategy Initiative for Catalysts \& Batteries, 
 Kyoto University,
 Nishikyo-ku,
 Kyoto 615-8510,
 Japan 
 }
\footnotetext{
 \it
 $^{c}$~
 Undergraduate School of Industrial Chemistry, Faculty of Engineering,
 Kyoto University,
 Nishikyo-ku, 
 Kyoto 615-8510,
 Japan
 }
\footnotetext{
 \it
 $^{d}$~
 Department of Organic Device Engineering and
 Research Center for Organic Electronics,
 Yamagata University,
 4-3-16,
 \vspace{-0.2cm}\\
 Johnan,
 Yonezawa,
 992-8510,
 Japan
 }
\footnotetext{
 \it
 $^{e}$~
 PRESTO (Sakigake), JST
 }
\clearpage
\tableofcontents

\clearpage
%%%%%%%%%%%%%%%%%%%%%%%%%%%%%%%%%%%%%%%%%%%%%%%%%%%%%%%%%%%%%%%%%%%%%
\section{Radiative and Non-Radiative Transition Rates}
%%%%%%%%%%%%%%%%%%%%%%%%%%%%%%%%%%%%%%%%%%%%%%%%%%%%%%%%%%%%%%%%%%%%%

We consider transition from the initial vibronic state
$\left| \Phi_{mi} \right> = \left| \Psi_m \right> \left| \chi_i \right>$
to the final one
$\left| \Phi_{nj} \right> = \left| \Psi_n \right> \left| \chi_j \right>$,
where $\left| \Psi_m \right>$ and $\left| \chi_i \right>$ denote
the initial electronic and vibrational states, and
$\left| \Psi_n \right>$ and $\left| \chi_j \right>$ stand for
the final electronic and vibrational states, respectively.
Radiative and non-radiative transition rate constants
between these vibronic states can be obtained as per
the Fermi's golden rule, which describes a transition rate constant
between quantum states in a general manner,
as described in our previous papers\cite{Sato2013_012010,Uejima2014_14244}.
Summing radiative transition rate constants over all the vibrational states,
the radiative transition rate constant $k_{r}$
from the initial electronic state $\left| \Psi_m \right>$
to the final one $\left| \Psi_n \right>$ is given by
\begin{equation}
k_{r}
=
\int_{0}^{\infty}d\omega \frac{4\omega^3}{3c^3}\sum_{i,j} P_{mi}(T)
|\boldsymbol{\mu}_{mn}|^2 |\langle\chi_{mi}|\chi_{nj}\rangle|^{2}
\delta(\hbar\omega - E_{mi} + E_{nj}),
\end{equation}
where $\omega$ denotes an angular frequency of an emitted photon,
$c$ is the speed of light,
$P_{mi}(T)$ stands for the statistical weight of
$\left| \Phi_{mi} \right>$ at the temperature $T$,
$\boldsymbol{\mu}_{nm}$ is the transition dipole moment
between $\left| \Psi_m \right>$ and $\left| \Psi_n \right>$,
and $\hbar$ is the reduced Planck Constant.
On the other hand, the non-radiative transition rate constant $k_{nr, \alpha}$
via mode $\alpha$ is given by
\begin{equation}
k_{nr, \alpha}
=
\frac{2\pi}{\hbar}\sum_{i,j} P_{mi}(T)
|V_{mn,\alpha}|^{2}|\langle\chi_{mi}|Q_{\alpha}|\chi_{nj}\rangle|^{2}
\delta(E_{mi} - E_{nj}),
\end{equation}
where $V_{\alpha}^{mn}$ is the off-diagonal vibronic coupling constant
between $\left| \Psi_m \right>$ and $\left| \Psi_n \right>$
with respect to mode $\alpha$, and $Q_{\alpha}$ denotes
a mass-weighted normal coordinate of mode $\alpha$.

\clearpage
%%%%%%%%%%%%%%%%%%%%%%%%%%%%%%%%%%%%%%%%%%%%%%%%%%%%%%%%%%%%%%%%%%%%%
\section{Vibronic Coupling Density and Transition Dipole Moment Density}
%%%%%%%%%%%%%%%%%%%%%%%%%%%%%%%%%%%%%%%%%%%%%%%%%%%%%%%%%%%%%%%%%%%%%

Suppose that an operator $\hat{O}$
consists of one-electron operators $\hat{o}$
without any differential operators,
\begin{equation}
\hat{O} = \sum_{i=1}^{N} \hat{o}(\boldsymbol{r}_i),
\end{equation}
where $\boldsymbol{r}_i$ denotes
the spatial coordinate of electron $i$.
A matrix element of $\hat{O}$ is given by\cite{Sato2013_012010}
\begin{eqnarray}
O_{mn} &=& \int\cdots\int \Psi_{m}^{\ast} \hat{O} \Psi_{n}
d^4\boldsymbol{x}_{1}\cdots d^4\boldsymbol{x}_{N}
\nonumber\\
&=& \sum_{i=1}^{N} \int \left[ \int\cdots\int {\Psi_m}^\ast \Psi_n
d^4\boldsymbol{x}_{1}\cdots d^4\boldsymbol{x}_{i-1}
ds_{i} d^4\boldsymbol{x}_{i+1}\cdots d^4\boldsymbol{x}_{N} \right]
\hat{o}(\boldsymbol{r}_i) d^3\boldsymbol{r}_i
\nonumber\\
&=& \sum_{i=1}^{N} \int \left[ \frac{1}{N}
\rho^{mn}(\boldsymbol{r}_i) \right] \hat{o}(\boldsymbol{r}_i) d^3\boldsymbol{r}_i
\nonumber\\
&=& \int \rho^{mn}(\boldsymbol{r}) \hat{o}(\boldsymbol{r}) d^3\boldsymbol{r},
\end{eqnarray}
where $\boldsymbol{x}_{i}=(\boldsymbol{r}_{i},s_{i})$
with spatial coordinate $\boldsymbol{r}_{i}$ and spin coordinate $s_{i}$
for electron $i$.
$\rho^{mn}(\boldsymbol{r}) \times \hat{o}(\boldsymbol{r})$
is a density form of the matrix element $O_{mn}$.
It should be noted that
any approximation is not employed in this derivation.

An electric dipole moment operator $\hat{\boldsymbol{\mu}}$
is an example of $\hat{O}$:
\begin{equation}
\hat{\boldsymbol{\mu}} := \sum_{i=1}^{N} -e\boldsymbol{r}_i.
\end{equation}
Therefore, a transition dipole moment $\boldsymbol{\mu}_{mn}$
between electronic states $m$ and $n$ is given by
the integral of a transition dipole moment density
$\boldsymbol{\tau}_{mn}(\boldsymbol{r}) 
= \rho^{mn}(\boldsymbol{r}) \times
\left(- e \boldsymbol{r} \right)$
\cite{Sato2013_012010,Uejima2014_14244}.

An off-diagonal vibronic coupling constant $V^{mn}_\alpha$ is
\begin{eqnarray}
V^{mn}_\alpha
&=& \int\cdots\int
\Psi_{m}^{\ast} \left(
\frac{\partial U_{ne}}{\partial Q_\alpha}
\right)_{\boldsymbol{R}_{0}} \Psi_{n}
d^4\boldsymbol{x}_{1}\cdots d^4\boldsymbol{x}_{N}
+ \int\cdots\int
\Psi_{m}^{\ast} \left(
\frac{\partial U_{nn}}{\partial Q_\alpha}
\right)_{\boldsymbol{R}_{0}} \Psi_{n}
d^4\boldsymbol{x}_{1}\cdots d^4\boldsymbol{x}_{N}
\nonumber\\
&=& \int\cdots\int
\Psi_{m}^{\ast} \left(
\frac{\partial U_{ne}}{\partial Q_\alpha}
\right)_{\boldsymbol{R}_{0}} \Psi_{n}
d^4\boldsymbol{x}_{1}\cdots d^4\boldsymbol{x}_{N},
\end{eqnarray}
where $U_{ne}$ denotes the sum of nuclear-electronic potentials,
$U_{nn}$ stands for the sum of nuclear-nuclear potentials,
and $Q_{\alpha}$ is a mass-weighted normal coordinate of mode $\alpha$.
Since $\left( \partial U_{ne} / \partial Q_\alpha \right)_{\boldsymbol{R}_{0}}$
can be written as the sum of one-electron potential derivatives:
\begin{equation}
\left(
\frac{\partial U_{ne}}{\partial Q_\alpha}
\right)_{\boldsymbol{R}_{0}}
= \sum_{i=1}^{N} v_\alpha(\boldsymbol{r}_i),
\end{equation}
$V^{mn}_\alpha$ is given by the integral of
a vibronic coupling density
$\eta_{\alpha}^{mn}(\boldsymbol{r})
= \rho^{mn}(\boldsymbol{r})\times v_{\alpha}(\boldsymbol{r})$
\cite{Sato2013_012010,Uejima2014_14244}.

\clearpage
%%%%%%%%%%%%%%%%%%%%%%%%%%%%%%%%%%%%%%%%%%%%%%%%%%%%%%%%%%%%%%%%%%%%%
\section{Optimized Structures}
%%%%%%%%%%%%%%%%%%%%%%%%%%%%%%%%%%%%%%%%%%%%%%%%%%%%%%%%%%%%%%%%%%%%%
\begin{figure}[!h]
\centering
\begin{tabular}{ccc}
\multicolumn{1}{l}{{\bf\large (a)}} &
\multicolumn{1}{l}{{\bf\large (b)}} &
\multicolumn{1}{l}{{\bf\large (c)}}\\
\includegraphics[scale=0.16]{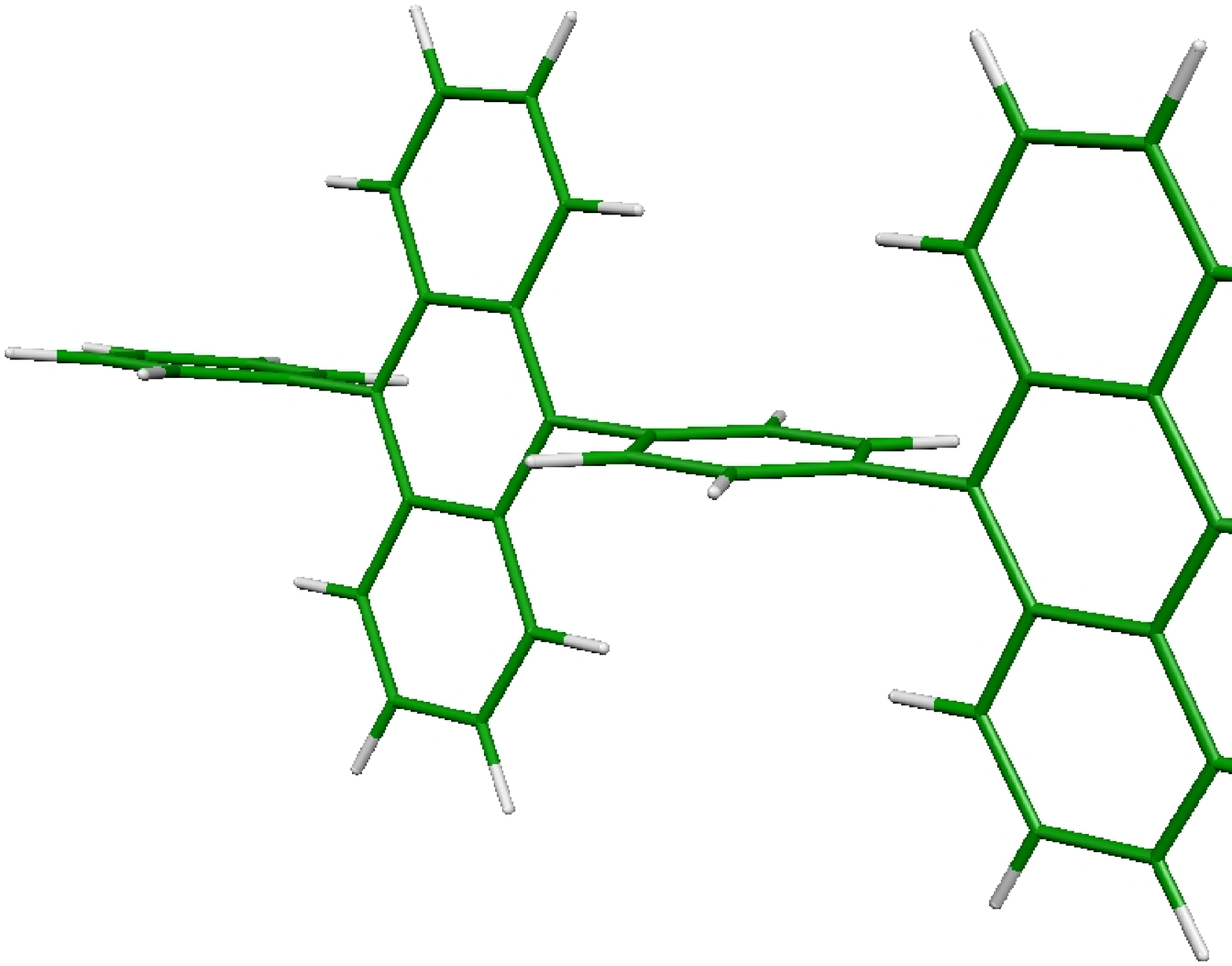} &
\includegraphics[scale=0.16]{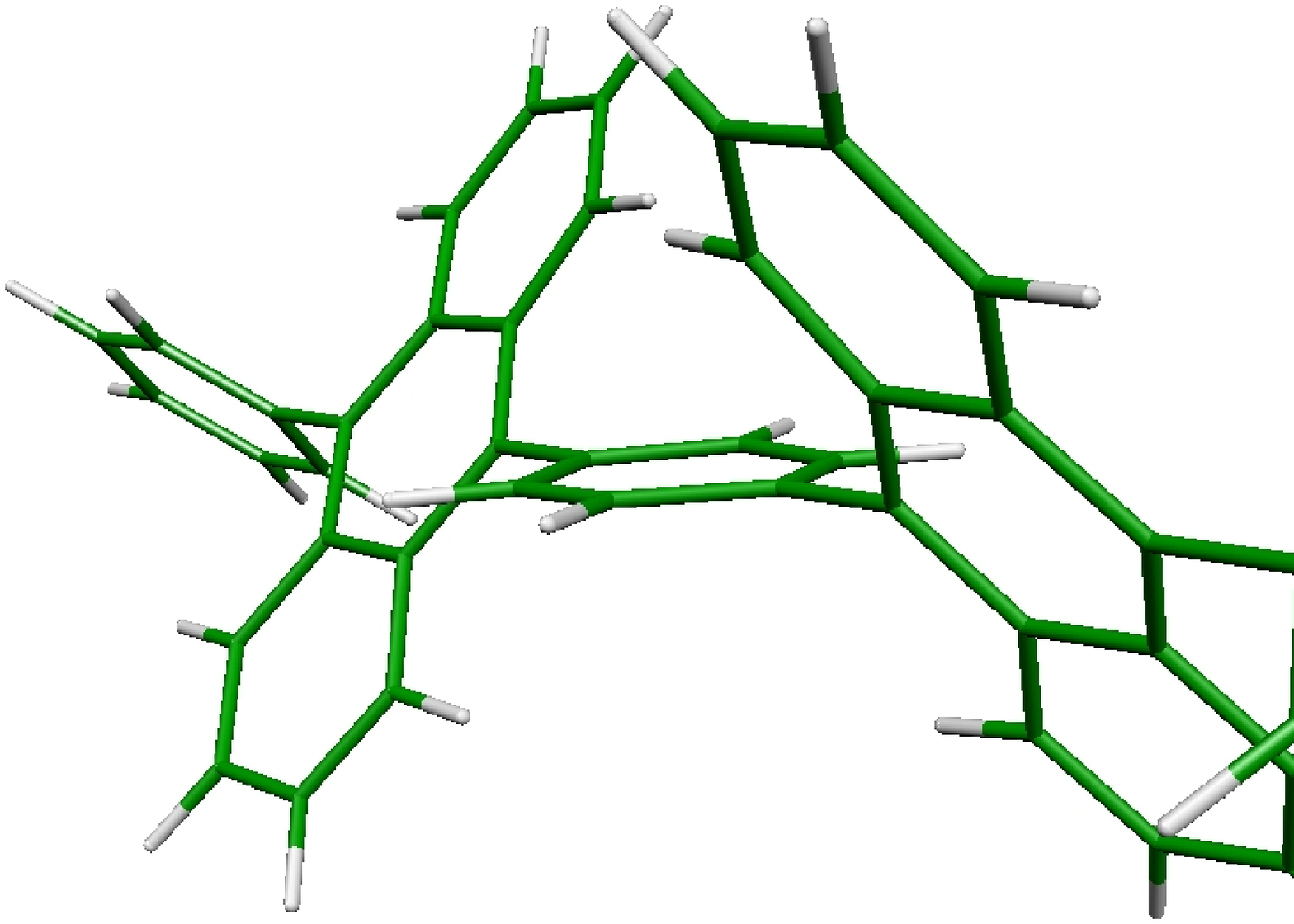} &
\includegraphics[scale=0.16]{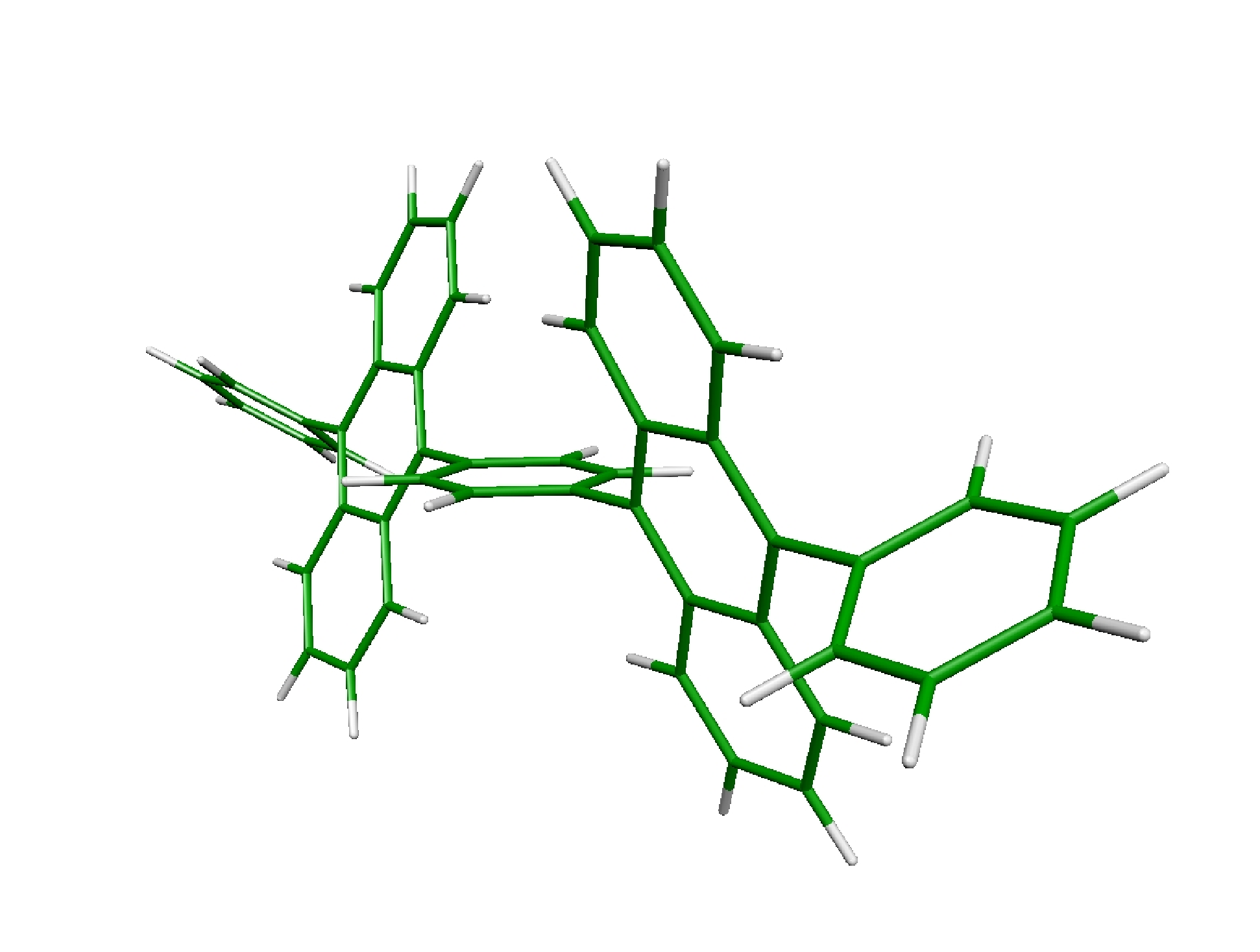} \\
 &
\multicolumn{1}{l}{{\bf\large (d)}} &
\multicolumn{1}{l}{{\bf\large (e)}}\\
 &
\includegraphics[scale=0.11]{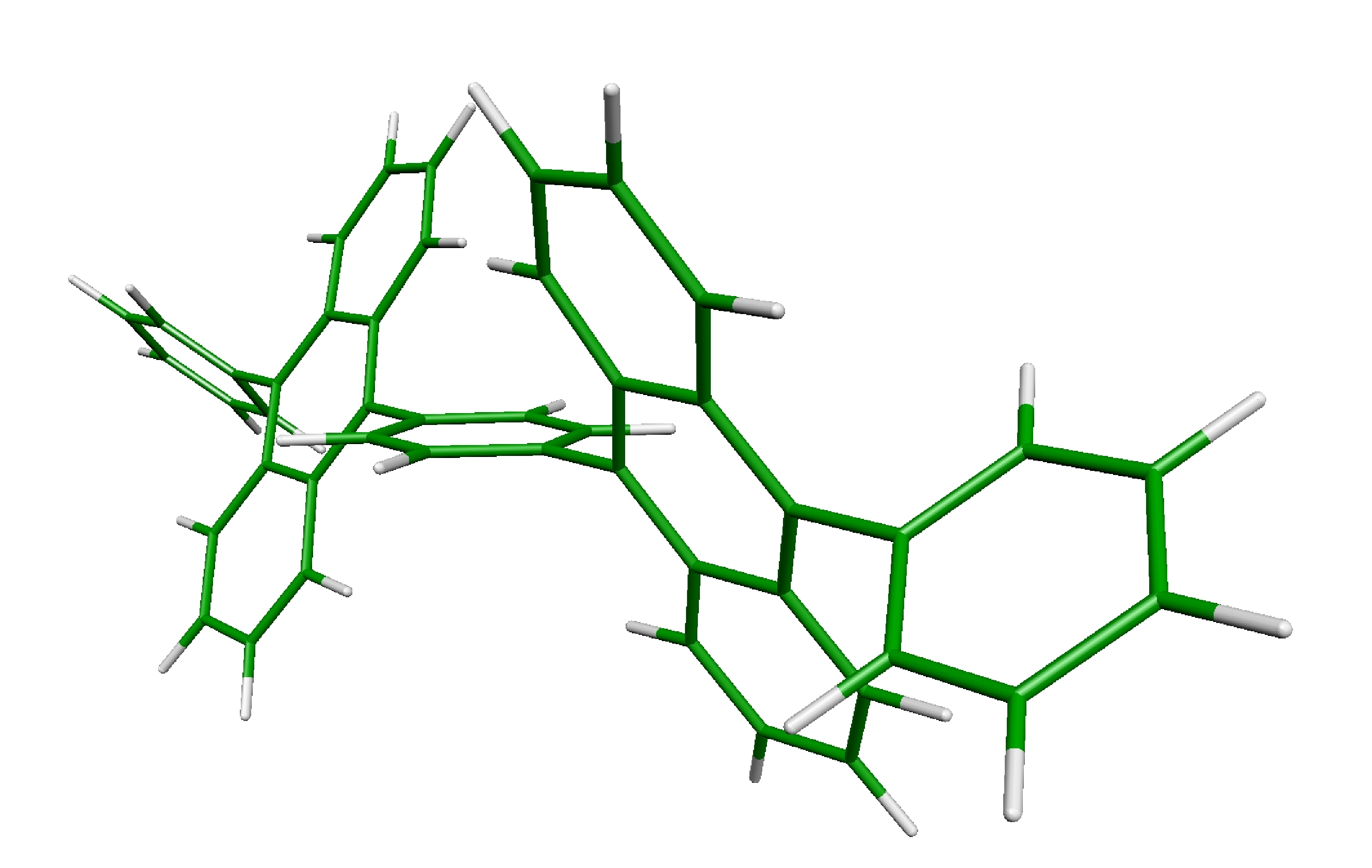} &
\includegraphics[scale=0.13]{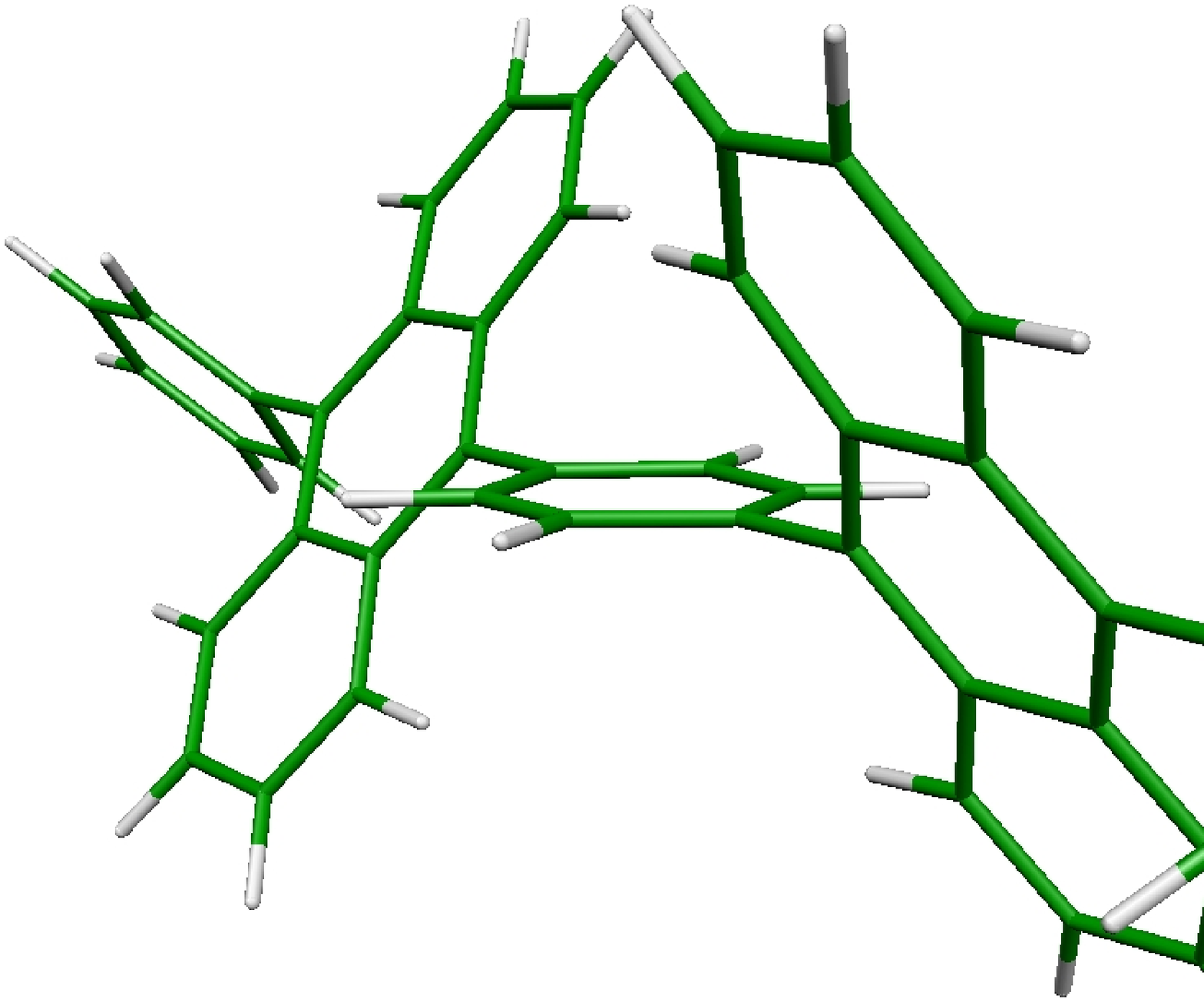} \\
\end{tabular}
\caption{Optimized structures of BD1 for
(a) S$_0$, (b) S$_1$, (c) S$_2$, (d) T$_3$, and (e) T$_4$.
\label{Fig:ADStructures}}
\end{figure}

\begin{figure}[!h]
\centering
\includegraphics[scale=0.4]{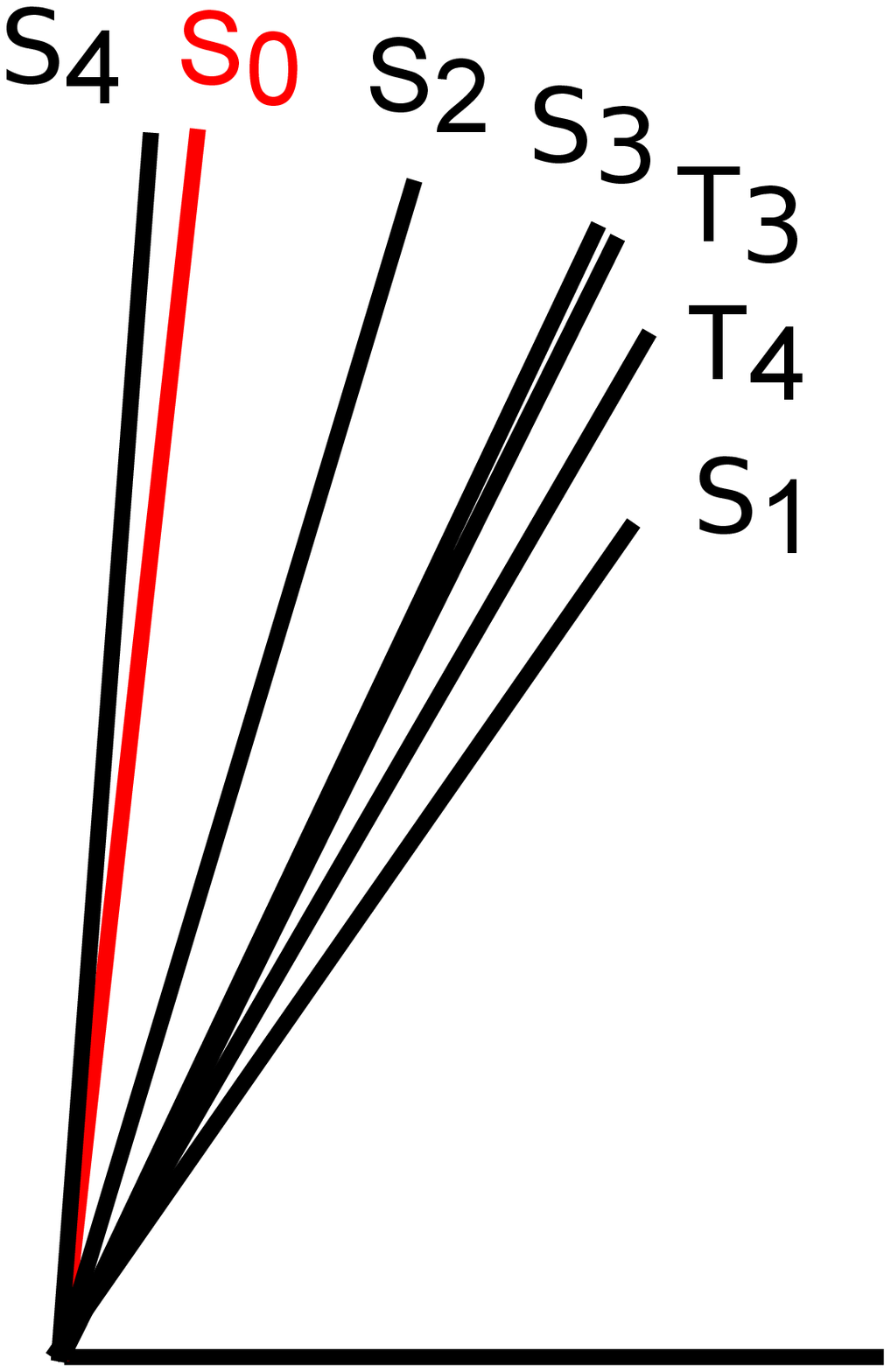}
\caption{Dihedral angles (C1--C2--C3--C4) between the anthracene and benzene moieties
for the geometry-optimized excited states.\label{Fig:Dihedral}}
\end{figure}

\clearpage
%%%%%%%%%%%%%%%%%%%%%%%%%%%%%%%%%%%%%%%%%%%%%%%%%%%%%%%%%%%%%%%%%%%%%
\section{Selection Rules}
%%%%%%%%%%%%%%%%%%%%%%%%%%%%%%%%%%%%%%%%%%%%%%%%%%%%%%%%%%%%%%%%%%%%%
\begin{table}[!h]
\centering
\caption{Character table of $D_2$.\label{Table:Character}}
\begin{tabular}{crrrrll}
\hline\hline
\bm{$D_2$} & $E$  & $C_2(z)$ & $C_2(y)$ & $C_2(x)$ & \multicolumn{2}{c}{$h = 4$}    \\
\hline
$A_1$ & $1$  & $1$      & $1$      & $1$      &            & $x^2$,$y^2$,$z^2$ \\
$B_1$      & $1$  & $1$      & $-1$     & $-1$     & $z$,$R_z$  & $xy$              \\
$B_2$      & $1$  & $-1$     & $1$      & $-1$     & $y$,$R_y$  & $zx$              \\
$B_3$      & $1$  & $-1$     & $-1$     & $1$      & $x$,$R_x$  & $yz$              \\
\hline\hline
\end{tabular}
\end{table}

\begin{table}[!h]
\centering
\caption{Direct products of irreducible representations in $D_2$.\label{Table:DirectProduct}}
\begin{tabular}{ccccc}
\hline\hline
\bm{$D_2$} & $A_1$ & $B_1$ & $B_2$ & $B_3$ \\
\hline
$A_1$      & $A_1$ & $B_1$ & $B_2$ & $B_3$ \\
$B_1$      &       & $A_1$ & $B_3$ & $B_2$ \\ 
$B_2$      &       &       & $A_1$ & $B_1$ \\ 
$B_3$      &       &       &       & $A_1$ \\ 
\hline\hline
\end{tabular}
\end{table}

\begin{table}[!h]
\centering
\caption{Selection rule of the transition dipole moment in $D_2$.
$-$ denotes a forbidden transition, and $\mu_i$ $(i=x,y,z)$ denotes a symmetry-allowed component 
of transition dipole moment $\langle m |\hat{\boldsymbol{\mu}}| n \rangle$.
\label{Table:SelectionRuleTDM}}
\begin{tabular}{ccccc}
\hline\hline
$m$/ $n$   & $A_1$ & $B_1$     & $B_2$     & $B_3$ \\
\hline
$A_1$      & $-$   & $\mu_{z}$ & $\mu_{y}$ & $\mu_{x}$ \\
$B_1$      &       & $-$       & $\mu_{x}$ & $\mu_{y}$ \\ 
$B_2$      &       &           & $-$       & $\mu_{z}$ \\ 
$B_3$      &       &           &           & $-$ \\ 
\hline\hline
\end{tabular}
\end{table}

\begin{table}[!h]
\centering
\caption{Selection rule of spin-orbit coupling in $D_2$.
$-$ denotes a forbidden transition, and $L_i$ $(i=x,y,z)$ denotes a symmetry-allowed component 
of orbital angular moment $\langle m |\hat{\boldsymbol{L}}| n \rangle$.
\label{Table:SelectionRuleSO}}
\begin{tabular}{ccccc}
\hline\hline
$m$/ $n$   & $A_1$ & $B_1$   & $B_2$   & $B_3$ \\
\hline
$A_1$      & $-$   & $L_{z}$ & $L_{y}$ & $L_{x}$ \\
$B_1$      &       & $-$     & $L_{x}$ & $L_{y}$ \\ 
$B_2$      &       &         & $-$     & $L_{z}$ \\ 
$B_3$      &       &         &         & $-$  \\ 
\hline\hline
\end{tabular}
\end{table}

\begin{table}[!h]
\centering
\caption{Selection rule of vibronic coupling in $D_2$.
$-$ denotes a forbidden transition, and $V_\Gamma$ $(\Gamma=A_1,B_1,B_2,B_3)$ denotes a symmetry-allowed component 
of vibronic coupling constant $\langle m |\hat{V}_{\Gamma}| n \rangle$, 
where $\hat{V}_{\Gamma}=(\partial H/\partial Q_{\Gamma})$ with normal coordinate $Q_{\Gamma}$.
\label{Table:SelectionRuleVC}}
\begin{tabular}{ccccc}
\hline\hline
$m$/ $n$   & $A_1$     & $B_1$     & $B_2$     & $B_3$ \\
\hline
$A_1$      & $V_{a_1}$ & $V_{b_1}$ & $V_{b_2}$ & $V_{b_3}$ \\
$B_1$      &           & $V_{a_1}$ & $V_{b_3}$ & $V_{b_2}$ \\ 
$B_2$      &           &           & $V_{a_1}$ & $V_{b_1}$ \\ 
$B_3$      &           &           &           & $V_{a_1}$ \\ 
\hline\hline
\end{tabular}
\end{table}

\clearpage
%%%%%%%%%%%%%%%%%%%%%%%%%%%%%%%%%%%%%%%%%%%%%%%%%%%%%%%%%%%%%%%%%%%%%
\section{Level Shifts of Triplet Excites States during Optimizations}
%%%%%%%%%%%%%%%%%%%%%%%%%%%%%%%%%%%%%%%%%%%%%%%%%%%%%%%%%%%%%%%%%%%%%
\begin{figure}[!h]
\centering
\begin{tabular}{c}
\includegraphics[scale=0.35]{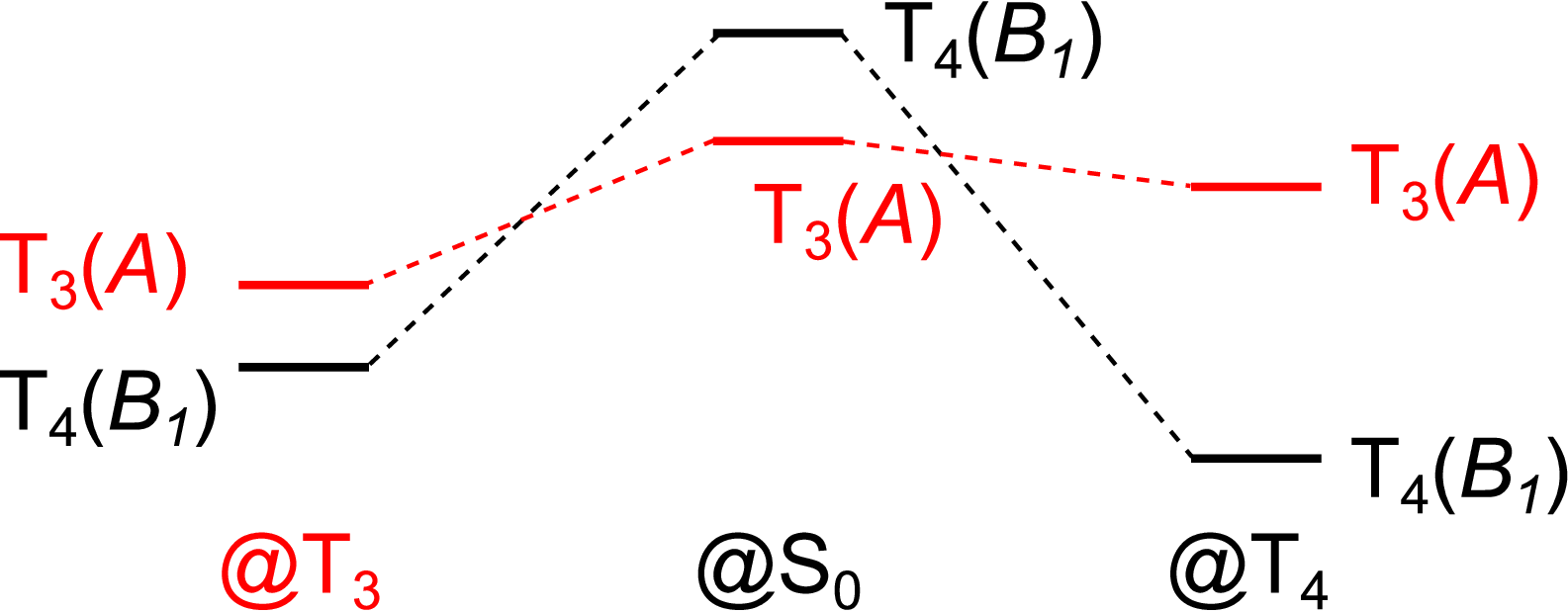} \\
\end{tabular}
\caption{Level shifts of triplet excited states of BD1
during geometry optimizations.}
\end{figure}

\clearpage
%%%%%%%%%%%%%%%%%%%%%%%%%%%%%%%%%%%%%%%%%%%%%%%%%%%%%%%%%%%%%%%%%%%%%
\section{Vibrational Modes with Strong Couplings}
%%%%%%%%%%%%%%%%%%%%%%%%%%%%%%%%%%%%%%%%%%%%%%%%%%%%%%%%%%%%%%%%%%%%%
\begin{figure}[!h]
\centering
\begin{tabular}{cc}
\multicolumn{1}{l}{{\bf\large (a)}} & \\
\includegraphics[scale=0.35]{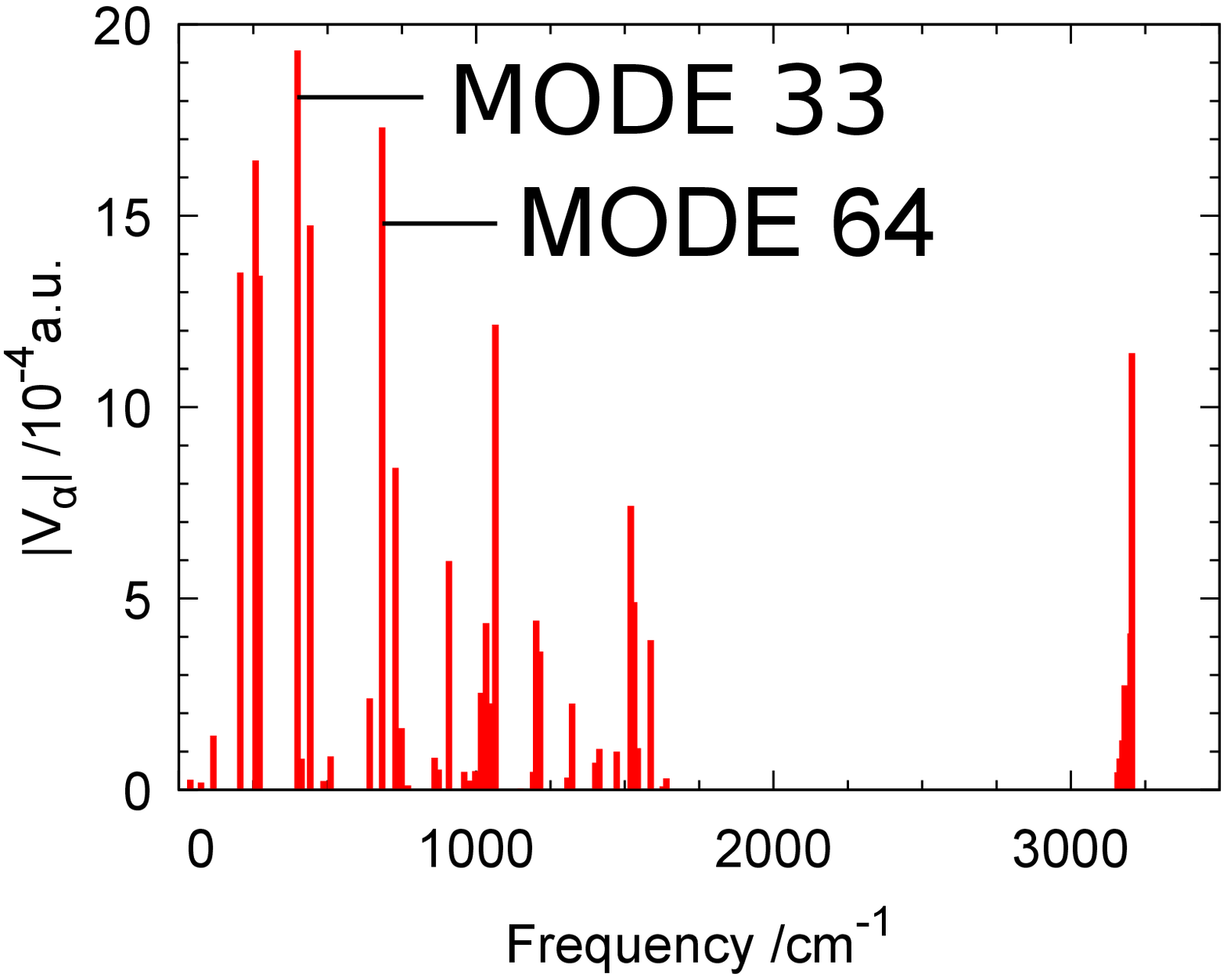}
& \\
\multicolumn{1}{l}{{\bf\large (b)}} &
\multicolumn{1}{l}{{\bf\large (c)}}\\
\includegraphics[scale=0.28]{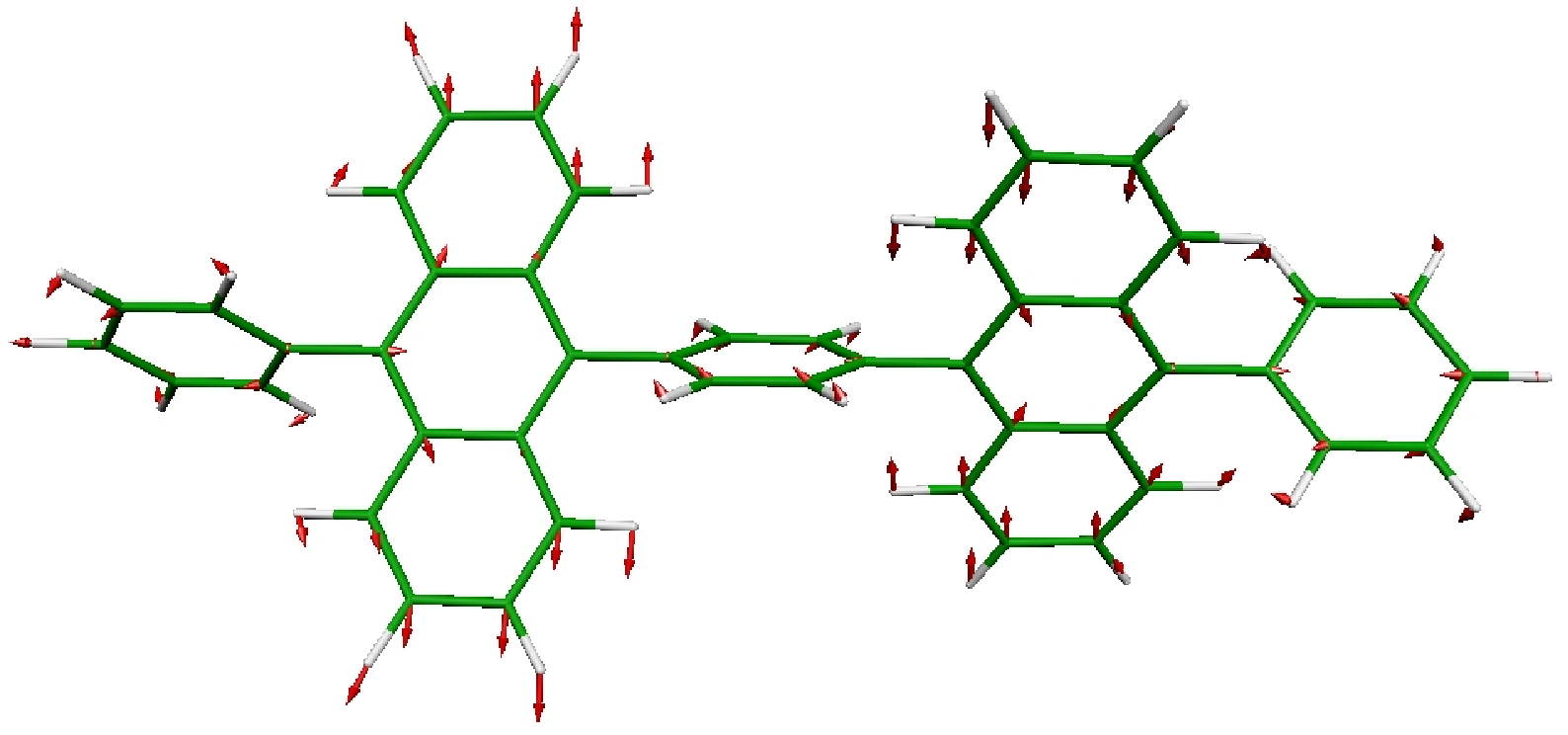} &
\includegraphics[scale=0.28]{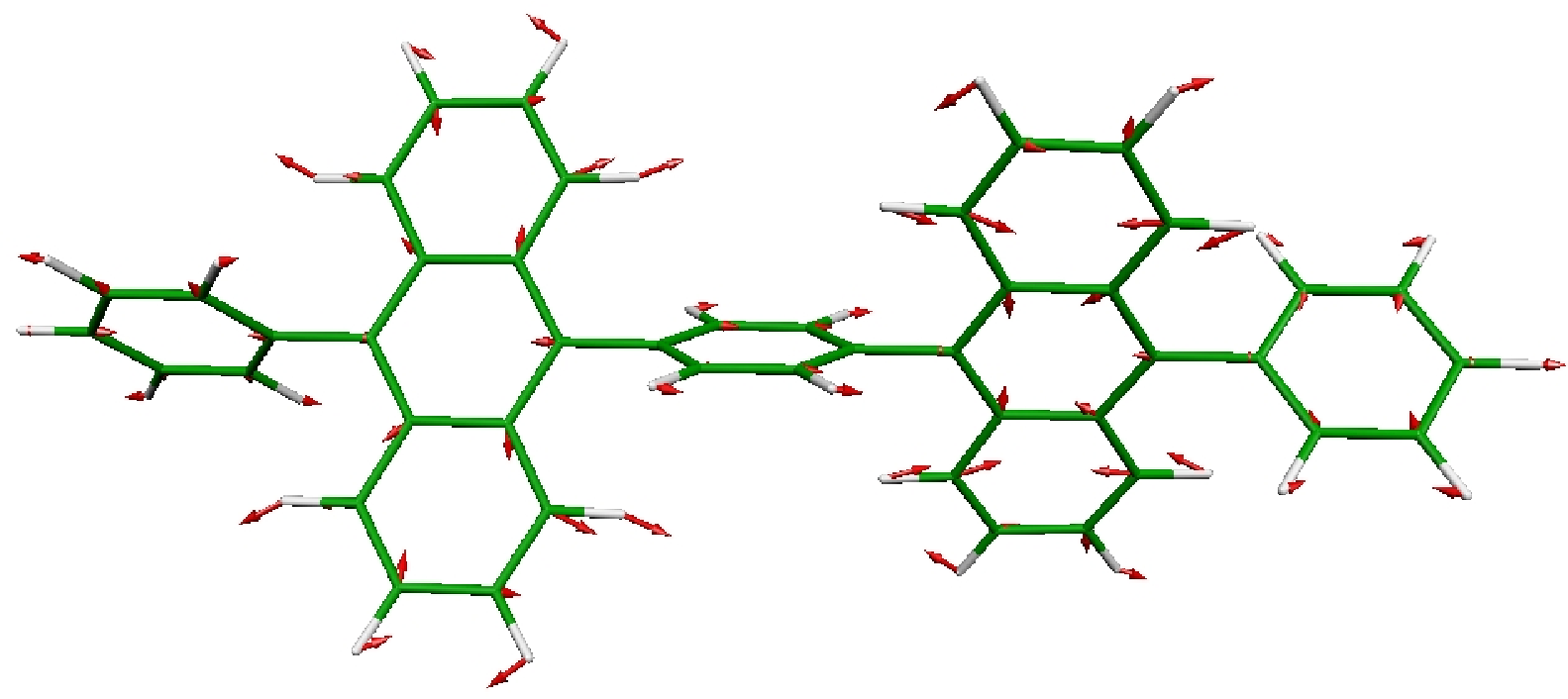} \\ 
\end{tabular}
\caption{(a) Off-diagonal VCCs between T$_3$ and T$_4$ at the T$_3$ optimized structure,
and strong coupling modes: (b) mode 33 and (c) mode 64. 
\label{Fig:VCCT3T4}}
\end{figure}
\clearpage
%%%%%%%%%%%%%%%%%%%%%%%%%%%%%%%%%%%%%%%%%%%%%%%%%%%%%%%%%%%%%%%%%%%%%
\section{Vibronic Coupling Density Analyses}
%%%%%%%%%%%%%%%%%%%%%%%%%%%%%%%%%%%%%%%%%%%%%%%%%%%%%%%%%%%%%%%%%%%%%
\begin{figure}[!h]
\centering
\begin{tabular}{ccc}
\multicolumn{1}{l}{{\bf\large (a)}} &
\multicolumn{1}{l}{{\bf\large (b)}} &
\multicolumn{1}{l}{{\bf\large (c)}}\\
\includegraphics[scale=0.145]{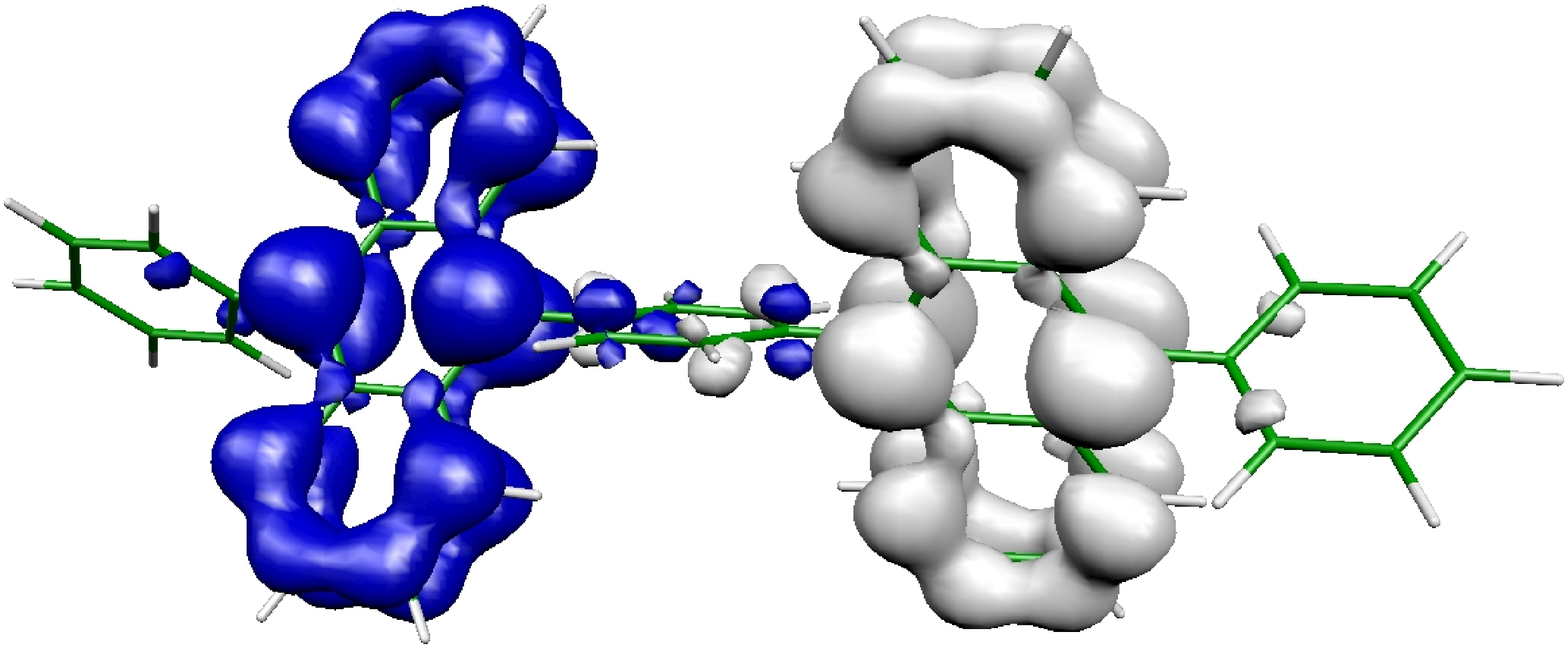} & 
\includegraphics[scale=0.145]{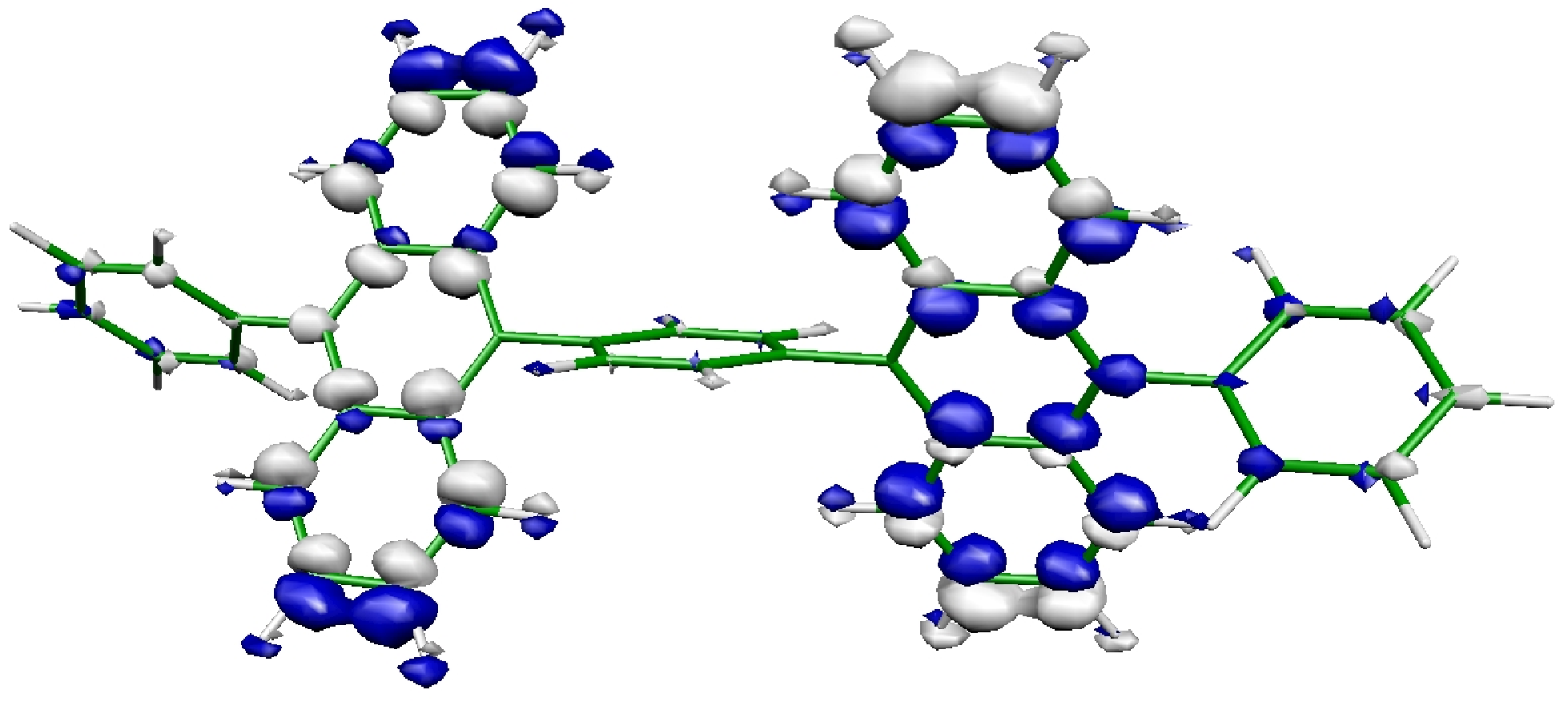} &
\includegraphics[scale=0.145]{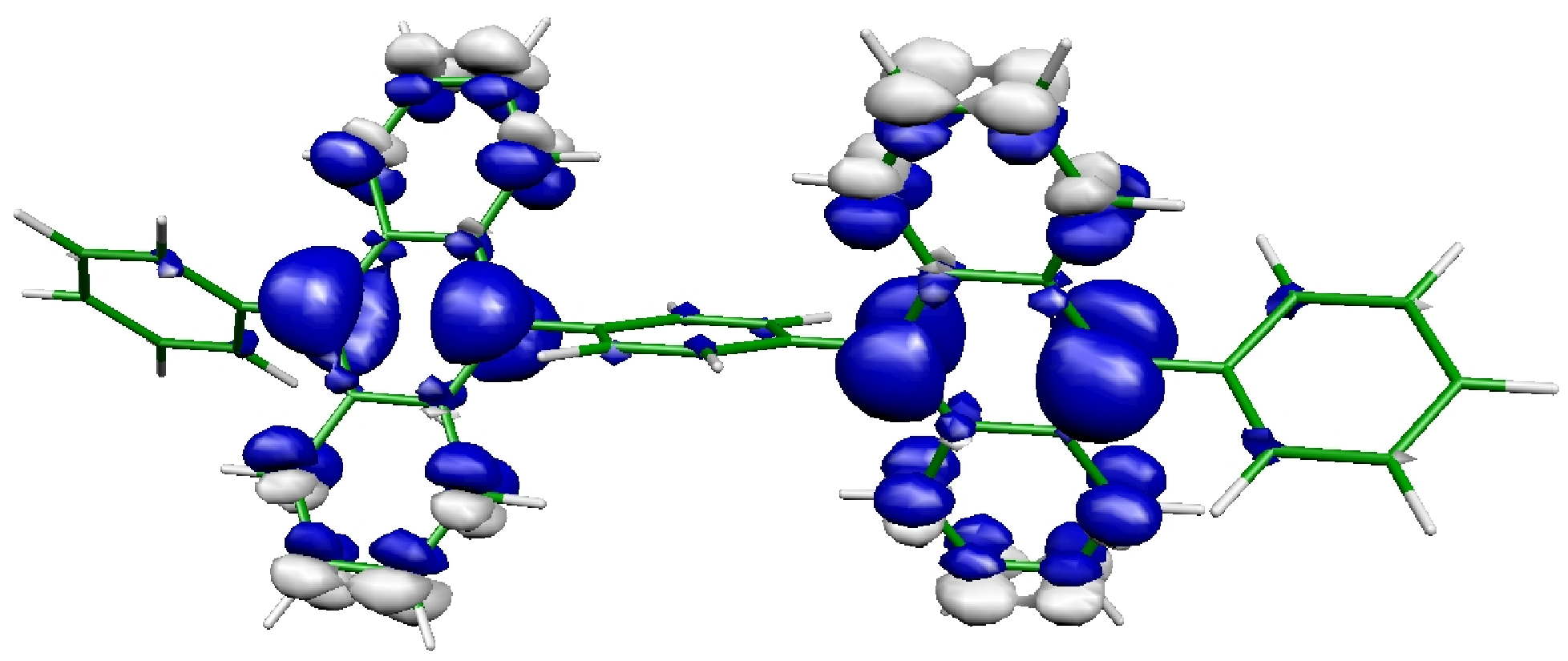} \\ 
 & & \\
 &
\multicolumn{1}{l}{{\bf\large (d)}} &
\multicolumn{1}{l}{{\bf\large (e)}}\\
 &
\includegraphics[scale=0.145]{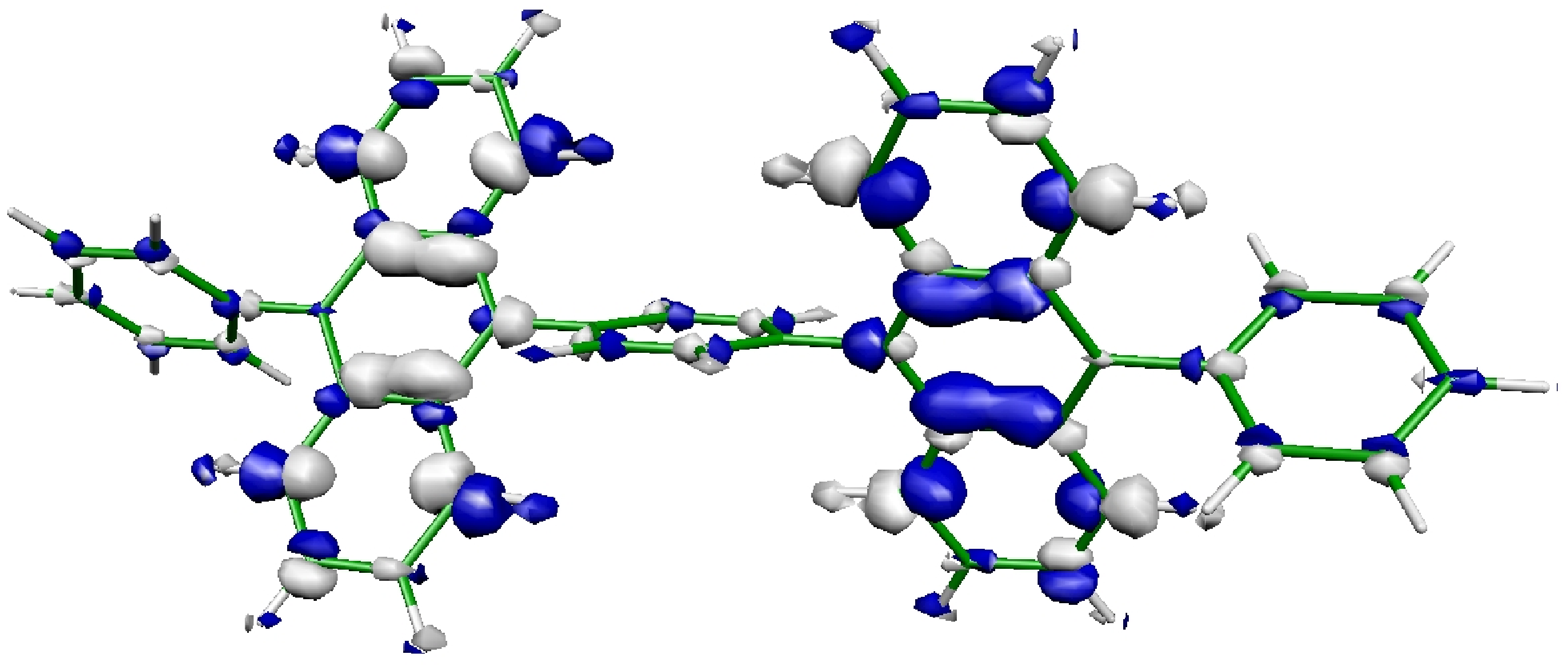} &
\includegraphics[scale=0.145]{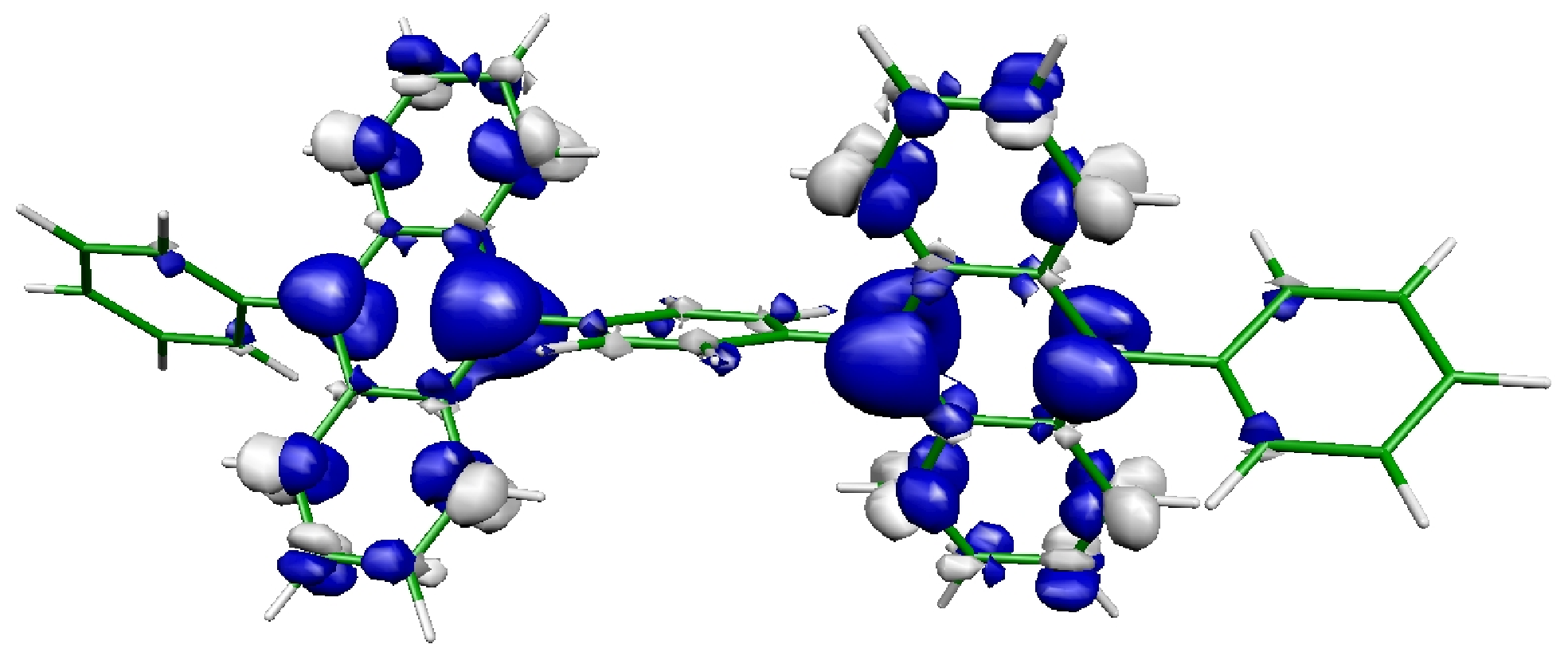} \\ 
\end{tabular}
\caption{Off-diagonal VCD analyses between T$_3$ and T$_4$ at the T$_3$ optimized structure: 
(a) overlap density $\rho_{34}$,
(b) potential derivative $v_{33}$,
(c) VCD $\eta_{34,33}$,
(d) potential derivative $v_{64}$, and
(e) VCD $\eta_{34,64}$.
The isosurface values for $\rho$, $v$, and $\eta$ are 1.0$\times$10$^{-3}$, 5.0$\times$10$^{-3}$ and 5.0$\times$10$^{-6}$ a.u.,
respectively.
\label{Fig:VCDT3T4}}
\end{figure}

\clearpage
%%%%%%%%%%%%%%%%%%%%%%%%%%%%%%%%%%%%%%%%%%%%%%%%%%%%%%%%%%%%%%%%%%%%%
\section{General Conditions for Disappearances of Overlap Densities
in Pseudo-Degenerate Systems}
%%%%%%%%%%%%%%%%%%%%%%%%%%%%%%%%%%%%%%%%%%%%%%%%%%%%%%%%%%%%%%%%%%%%%
In this section, we discuss the overlap densities in a pseudo-degenerate system,
(moiety $L$)-(linker)-(moiety $R$) system, in general, 
where moiety $L$ and moiety $R$ are the same fragments.
As the pseudo-degenerate frontier orbitals,
we consider the following forms:
\begin{eqnarray}
\psi_{\textrm{LU}}=\frac{1}{\sqrt{2}}(\phi_{\textrm{LU}}(L) - \phi_{\textrm{LU}}(R)),
&\quad&
\psi_{\textrm{NLU}}=\frac{1}{\sqrt{2}}(\phi_{\textrm{LU}}(L) + \phi_{\textrm{LU}}(R))
,
\label{Eq:LUNLU}
\\
\psi_{\textrm{HO}}=\frac{1}{\sqrt{2}}(\phi_{\textrm{HO}}(L) + \phi_{\textrm{HO}}(R)),
&\quad&
\psi_{\textrm{NHO}}=\frac{1}{\sqrt{2}}(\phi_{\textrm{HO}}(L) - \phi_{\textrm{HO}}(R))
,
\label{Eq:HONHO}
\end{eqnarray}
where $\phi_{\textrm{HO/LU}}(L/R)$ denotes a fragment
HOMO/LUMO
on moiety Left($L$)/Right($R$).

Since these frontier orbitals are localized on moieties $L$ and $R$,
the overlaps between the different moieties can be neglected:
$\phi_{i}(L)\phi_j(R)=0$ ($i,j=$HO, NHO, LU, and NLU).
Therefore, we can obtain the following relations of orbital overlap densities:
\begin{eqnarray}
& &|\psi_{\textrm{LU}}|^{2}
=
\frac{1}{2}(|\phi_{\textrm{LU}}(L)|^{2} + |\phi_{\textrm{LU}}(R)|^{2})
=
|\psi_{\textrm{NLU}}|^{2}
=:
r
,
\label{Eq:SquareLUNLU}
\\
& &
\psi_{\textrm{HO}}\psi_{\textrm{NLU}}
=
\frac{1}{2}(
  \phi_{\textrm{LU}}(L)\phi_{\textrm{HO}}(L)
  +  
  \phi_{\textrm{LU}}(R)\phi_{\textrm{HO}}(R))
=
\psi_{\textrm{NHO}}\psi_{\textrm{LU}}
=:
s
,
\label{Eq:OverlapHONLU-NHOLU}
\\
& &
|\psi_{\textrm{HO}}|^{2}
=
\frac{1}{2}(|\phi_{\textrm{HO}}(L)|^{2} + |\phi_{\textrm{HO}}(R)|^{2})
=
|\psi_{\textrm{NHO}}|^{2}
=:
t
,
\label{Eq:SquareHONHO}
\\
& &
\psi_{\textrm{HO}}\psi_{\textrm{LU}}
=
\frac{1}{2}(
  \phi_{\textrm{LU}}(L)\phi_{\textrm{HO}}(L)
  -  
  \phi_{\textrm{LU}}(R)\phi_{\textrm{HO}}(R))
=
\psi_{\textrm{NHO}}\psi_{\textrm{NLU}}
=:
u
.
\label{Eq:OverlapHOLU-NHONLU}
\end{eqnarray}
In addition, we define the orbital overlap densities
\begin{equation}
q
:=
\psi_{\textrm{HO}}\psi_{\textrm{NHO}}
=
\frac{1}{2}(
  |\phi_{\textrm{HO}}(L)|^{2} - |\phi_{\textrm{HO}}(R)|^{2})
,
\label{Eq:OverlapHONHO}
\end{equation}
\begin{equation}
p
:=
\psi_{\textrm{LU}}\psi_{\textrm{NLU}}
=
\frac{1}{2}(
  |\phi_{\textrm{LU}}(L)|^{2} - |\phi_{\textrm{LU}}(R)|^{2})
.
\label{Eq:OverlapLUNLU}
\end{equation}

We consider wave functions of the pseudo-degenerate system shown
in TABLE \ref{Table:ApproximateWavefunctions}.
The overlap densities of the pseudo-degenerate system are summarized
in TABLE \ref{Table:GeneralCondition}.
The pairs denoted by 0 in TABLE \ref{Table:GeneralCondition}
indicate reduced overlap density,
i.e.  suppressed radiative/non-radiative transition.
This approach can be extended for degenerate systems
and pseudo-degenerate systems with
more than three-fold degeneracy.
\begin{table}[!h]
\centering
\caption{Approximate wave functions of a pseudo-degenerate system, $c=1/\sqrt{2}$.
\label{Table:ApproximateWavefunctions}}
\begin{tabular}{lccc}
\hline\hline
Approximate form of wave function                                              & @T$_4$/@T$_3$ & @S$_2$/@S$_1$\\
\hline
$\Psi_a = c \Phi_{\text{HO}}^{\text{LU}} - c \Phi_{\text{NHO}}^{\text{NLU}}$   & T$_4$ &       \\
$\Psi_b = c \Phi_{\text{HO}}^{\text{NLU}} - c \Phi_{\text{NHO}}^{\text{LU}}$   & T$_3$ & S$_2$ \\
$\Psi_c = c \Phi_{\text{HO}}^{\text{LU}} + c \Phi_{\text{NHO}}^{\text{NLU}}$   & T$_1$ &       \\
$\Psi_d = \Phi_{\textrm{HO}}^{\textrm{LU}}$                                    &       & S$_1$ \\
$\Psi_e = c \Phi_{\text{HO}}^{\text{NLU}} + c \Phi_{\text{NHO}}^{\text{LU}}$   & T$_2$ & S$_3$ \\
$\Psi_f = \Phi_{\textrm{HO}}^{\textrm{NLU}}$                                   &       &       \\
$\Psi_g = \Phi_{\textrm{NHO}}^{\textrm{NLU}}$                                  &       & S$_4$ \\
$\Psi_h = \Phi_{\textrm{NHO}}^{\textrm{LU}}$                                   &       &       \\
$\Psi_0 = \Phi_{0} $                                                           &       & S$_0$ \\
\hline\hline
\end{tabular}
\end{table}
\begin{table}[!h]
\centering
\caption{Overlap densities in a pseudo-degenerate system. 
$\alpha=p-q$, $\beta=\rho_0 - t + r$, and $\gamma=p + q$.
\label{Table:GeneralCondition}}
\begin{tabular}{cccccccccc}
\hline\hline
         & $\Psi_a$ & $\Psi_b$ & $\Psi_c$ & $\Psi_d$ & $\Psi_e$ & $\Psi_f$ & $\Psi_g$ & $\Psi_h$ & $\Psi_0$ \\
\hline
$\Psi_a$ &          &$\alpha$  & 0        &$c\beta$  & 0        &$c\alpha$ &$-c\beta$ &$-c\alpha$& 0        \\
$\Psi_b$ &          &          & 0        &$c\alpha$ & 0        &$c\beta$  &$-c\alpha$ &$-c\beta$  & 0      \\
$\Psi_c$ &          &          &          &$c\beta$  &$\gamma$  &$c\gamma$ &$c\beta$  &$c\gamma$ &$2cu$     \\
$\Psi_d$ &          &          &          &          &$c\gamma$ &$p$       & 0        &$q$       &$u$       \\
$\Psi_e$ &          &          &          &          &          &$c\beta$  &$c\gamma$ &$c\beta$  &$2cs$     \\
$\Psi_f$ &          &          &          &          &          &          &$q$       & 0        &$s$       \\
$\Psi_g$ &          &          &          &          &          &          &          &$p$       &$u$         \\
$\Psi_h$ &          &          &          &          &          &          &          &          &$s$       \\
$\Psi_0$ &          &          &          &          &          &          &          &          &          \\
\hline\hline
\end{tabular}
\end{table}

\clearpage
%%%%%%%%%%%%%%%%%%%%%%%%%%%%%%%%%%%%%%%%%%%%%%%%%%%%%%%%%%%%%%%%%%%%%
\section{Imposition of $D_{2h}$-Symmetry Constraint on BD1}
%%%%%%%%%%%%%%%%%%%%%%%%%%%%%%%%%%%%%%%%%%%%%%%%%%%%%%%%%%%%%%%%%%%%%
\subsection{Singlet and Triplet Excited States}
%%%%%%%%%%%%%%%%%%%%%%%%%%%%%%%%%%%%%%%%%%%%%%%%%%%%%%%%%%%%%%%%%%%%%
\begin{table}[!h]
\centering
\caption{Excited states of BD1 at the optimized structure for S$_0$
with a $D_{2h}$-symmetry constraint.}
\begin{tabular}{ccccc}
\hline\hline
\multicolumn{2}{c}{State}&\multicolumn{2}{c}{Excitation Energy}&Oscillator Strength \\
      &                  & eV     & nm              & \\
\hline
T$_1$ & $^3B_{3u}$       & 2.2072 & 561.72          & 0.0000  \\
T$_2$ & $^3A_{g}$        & 2.2078 & 561.57          & 0.0000  \\
S$_1$ & $^1B_{3u}$       & 3.5520 & 349.05          & 0.5269  \\
S$_2$ & $^1A_{g}$        & 3.6186 & 342.63          & 0.0000  \\
T$_3$ & $^3B_{1g}$       & 3.7529 & 330.37          & 0.0000  \\
T$_4$ & $^3B_{2u}$       & 3.7533 & 330.34          & 0.0000  \\
\hline\hline
\end{tabular}
\end{table}

\clearpage
%%%%%%%%%%%%%%%%%%%%%%%%%%%%%%%%%%%%%%%%%%%%%%%%%%%%%%%%%%%%%%%%%%%%%
\subsection{Subduction from $D_{2h}$ to $D_2$}
%%%%%%%%%%%%%%%%%%%%%%%%%%%%%%%%%%%%%%%%%%%%%%%%%%%%%%%%%%%%%%%%%%%%%
\begin{center}
\begin{eqnarray}
A_{g}\downarrow D_2  &=& A \\
B_{1g}\downarrow D_2 &=& B_{1} \\
B_{2g}\downarrow D_2 &=& B_{2} \\
B_{3g}\downarrow D_2 &=& B_{3} \\
A_{u}\downarrow D_2  &=& A \\
B_{1u}\downarrow D_2 &=& B_{1} \\
B_{2u}\downarrow D_2 &=& B_{2} \\
B_{3u}\downarrow D_2 &=& B_{3} 
\end{eqnarray}
\end{center}

\clearpage
\bibliography{SM} %your .bib file
\bibliographystyle{rsc} %the RSC's .bst file